\documentstyle[12pt]{book}

\begin{document}

\author{Eugene Perchik}
\title{Methodology of Syntheses of Knowledge: Overcoming Incorrectness of
the Problems of Mathematical Modeling\\
(revised version, March 2005) }
\maketitle

\newpage

{\it Eugene Perchik. Methodology of syntheses of knowledge: overcoming
incorrectness of the problems of mathematical modeling}

{\it (revised version, March 2005)}

{\it \ / www.pelbook.narod.ru}

\bigskip

{\small J. Hadamard's ideas about the correct statement of the problems of
mathematical physics have been analyzed. In this connection various
interpretations of the directly related Banach theorem about the inverse
operator has been touched. The contemporary apparatus of mathematical
modeling is shown to be in a drastic contradiction with concepts of J.
Hadamard, S. Banach and a number of other outstanding scientists in the
sense that the priority is given to the realization of algorithms, which
actually imply that incorrectly stated problems are adequate to real
phenomena.}

{\small A new method is developed for solving problems traditionally
associated with the Fredholm integral equation of the first kind }$A\psi =f$%
{\small , }$x\in \left[ 0,1\right] ${\small . It is based on the
representation of the integration error in the form }$\delta f=\psi -\lambda
B\psi ${\small , }$x\in \left[ 0,1\right] ${\small , where }$B${\small \ is
the integral operator with limits }$-1,1${\small \ and Poisson kernel; }$%
\lambda ${\small \ is parameter. Incompletely continuous perturbation of
operator }$A${\small \ with }$I-\lambda B${\small , provided that }$\delta
f=0${\small , makes it possible to change the statement of the problem. This
involves (i) the extension of the problem }$\delta f=0${\small ; }$\delta
f=\mu A\psi -\mu f${\small \ (}$\mu ${\small \ is parameter) onto }$x\in %
\left[ -1,0\right) ${\small \ and (ii) the use of equations with similar
structure and the same function }$\psi ${\small , }$x\in \left[ 0,1\right] $%
{\small . The essence of this is the practical realization of the condition }%
$f+\delta f/\mu \in R\left( A\right) ${\small . A key point here is to
interrelate the components of the above systems of equations to enable their
mutual conversion. In the case when function }$\psi ${\small \ is harmonic
the problem is reduced to a Fredholm integral equation of the second kind
with properties favorable in computational respect.}

{\small The solution to this equation is given in the form of Fourier series
with coefficients depending on parameters of particular problem and also
parameter }$0<r<1${\small . The class of possible }$\psi ${\small \ may be
extended to }$L_{2}${\small \ by the limit transition }$r\rightarrow 1$%
{\small . In the second approach, the belonging }$\psi \in L_{2}${\small \
was assumed from the very beginning. Accordingly, condition }$\delta f=0$%
{\small \ needs to be addressed in terms of generalized functions. In
comparison with the first approach, this one is more formal; relatively
simple transformations result in second-order Fredholm integral equation
with properties most favorable for the numerical realization.}

{\small The general concept in this book is as follows. There is one and
only one function }$\psi =\psi _{\ast }${\small \ such that }$A\psi _{\ast
}=f${\small , but the problem to restore it from this equation with known }$%
A ${\small \ and }$f${\small \ is incorrect. Along with this, it is not
difficult to imagine a Fredholm integral equation of the second kind with
such a free term that }$\psi _{\ast }=\mu Z\psi _{\ast }+F${\small , }$x\in %
\left[ 0,1\right] ${\small . The essence of this text is to show how to
construct this equation starting from }$A\psi =f${\small \ . In other words,
the problem to find a function that satisfies the Fredholm integral equation
of the first kind is stated correctly.}

{\small The possibility is shown to extend the approach suggested for a wide
circle of problems that may be reduced to two-dimensional Fredholm integral
equations of the first kind; these are linear boundary-value and
initial-boundary-value problems with variable coefficients, non-canonical
domain of definition and other peculiarities complicating their solution.
The elaborated algorithm is shown to be directly applicable to them. Note
that this may be used for examining the above problems for solvability.}

{\small In discussing the statement of problems of mathematical physics,
considerable attention is paid to methodological aspects. Conclusions about
cause-and-effect relations are argued to be essentially illegitimate when
the solution of a problem is traced in the long run to a primitive renaming
of known and unknown functions of a corresponding direct problem. The aim of
this work is a constructive realization of J. Hadamard's opinions that
physically meaningful problems always have correct statements.}

{\small E-Mail: eperchik@bk.ru}

\newpage
\tableofcontents

\chapter{Introduction}

At the beginning, we should explain the title of the work and, in the first
place, the meaning of the employed notions. In this regard, we assume the
availability of information allowing us to formulate a mathematical model of
a certain phenomenon in a traditional way. Correspondingly, the
determination of unknown functions using the data of the problem is implied.
If the dependence of the solution to the problem on these data with respect
to the norm of the chosen space is continuous, such a problem, as a rule,
belongs to the domain of analysis or, in other words, its formulation is
direct.

However, the investigation of a concrete phenomenon in a variety of the
determining factors with the aim of obtaining, as a final result, of
qualitatively new information (the synthesis of knowledge) also envisages
the realization of problems in their inverse formulation,i.e., the
restoration of data using the hypothetically known solution: In other words,
the restoration of the cause using its consequence, which is usually
identified with the necessity of solving ill-posed problems.

The purpose of the present investigation consists in the justification of
the illegitimacy of this statement and, on the contrary, in a constructive
development of J. Hadamard's ideas of the existence of well-posed problems,
adequately describing real processes and phenomena. Note that the difference
between these two notions in the context of the book is unessential.
However, the term \textquotedblright process\textquotedblright\ accentuates
a time factor.

In the focus of the attention is a natural, to our mind, issue that, as an
example, can be explained by the evaluation of the integral 
\begin{equation}
\left( A\psi \right) \left( x\right) \equiv \int\limits_{0}^{1}k\left( x,\xi
\right) \psi \left( \xi \right) d\xi =f\left( x\right) ,\quad x\in \left[ 0,1%
\right] ,  \label{i1}
\end{equation}%
which amounts to the determination of the function $f\left( x\right) $ using
given $k\left( x,\xi \right) $ and $\psi \left( x\right) $ (from the space $%
L_{2}$). This procedure can be easily associated with a lot of physical, as
well as other, interpretations. Its realization, at least in the case of the
bounded integrand, does not pose any problem.

On the other hand, if the kernel $k\left( x,\xi \right) $, which is assumed
to be bounded, and the function $f\left( x\right) $, evaluated beforehand
from Eq. (\ref{i1}), are given, the function $\psi \left( x\right) $ is
objectively existent and unique. Thus, the questrion is whether it is
legitimate to restore this function by means of the solution of the Fredholm
integral equation of the first kind (\ref{i1}), just renaming the known and
unknown components in the formulation of the direct problem, i.e., by
assuming that the function $f\left( x\right) $ is given and $\psi \left(
x\right) $ is to be determined. And, generally speaking, what is the basis
to argue that mathematical formulations of the direct the inverse problems
can be absolutely identical?

The very fact of mechanical renaming of the known and unknown functions,
without any additional corrections, raises objections. Thus, we put forward
a thesis that an adequate approach to the formulation of inverse problems
should differ from the approach that became common. This position
predetermined the presence in the title of the work the notion of
methodology.

As a matter of fact, we hope to find reserves of the synthesis of the whole
complex of knowledge about the phenomenon by investigating it from different
sides, using formulations whose mathematical representations are not
identical. Although Eq. (\ref{i1}) is absolutely sufficient for the
evaluation of $f\left( x\right) $, the restoration of the function $\psi
\left( x\right) $ does not necessarily consists in the solution of the
Fredholm integral equation of the first kind, which is an ill-posed problem.

However, if there exists an alternative to the above-mentioned renaming of
the known and unknown components, one can assume that corresponding
formulations of the inverse problems may possess much more attractive
properties in a computational sense. From this point of view, the arguments
of Hadamard acquire a rather concrete meaning, stimulating a search of
correct and, at the same time, appropriate to the nature of the considered
phenomena formulations of problems of mathematical physics. A realization of
the outlined orientation seems to be possible in the context of the
following considerations.

The reasons for the difficulties related to the solution of ill-posed and
essentially mathematically senseless problems are, in principle, well
understood. In the Fredholm integral equation of the first kind (\ref{i1}),
there exists a mismatch between the function $f\left( x\right) $ and a
solution of the corresponding direct problem (the result of integration),
which is a result of errors in the determination of the data as well as of
rounding of digits in arithmetical operations.

As a consequence, considerable attention is paid to the phenomenon of (as it
is sometimes called) smoothing of information about the functions in the
process of their integration. At the same time, the data of the problem,
i.e., the free term $f\left( x\right) $ and also the kernel $k\left( x,\xi
\right) $, are usually determined experimentally, which inevitably incurs a
considerable error in Eq. (\ref{i1}). In this regard, we should point out
the dominance of the methodology of A. N. Tikhonov that is based on
objective incorrectness of the formulation of most problems of mathematical
simulation.

There appears a rather obvious, as it seems, question: Why not take into
account in practice the above-mentioned errors in the formulation of
problems, instead of merely bearing them in mind when identifying the
reasons for computational discrepancies? One can assume that an adequate
simulation of the error may contribute to a correct formulation of the
inverse problems.

Here, the adequacy implies, in the first place, the functional structure of
the representation of the error. In this regard, let us turn to the
procedure of integration (\ref{i1}). On the basis of general considerations,
it is logical to represent the loss of information about the function $\psi
\left( x\right) $ in the evaluation of $f\left( x\right) $ in the form 
\begin{equation}
\left( \delta f\right) \left( x\right) =\psi \left( x\right) -\lambda
\int\limits_{-1}^{1}h\left( x,\xi \right) \psi \left( \xi \right) d\xi
,\quad x\in \left[ 0,1\right] .  \label{i2}
\end{equation}%
Here, the function $\psi \left( x\right) $, $x\in \lbrack -1,0)$, the kernel 
$h\left( x,\xi \right) $ and the parameter $\lambda $ should satisfy the
requirement of the realization of the condition 
\begin{equation}
\left( \delta f\right) \left( x\right) =0  \label{i2.1}
\end{equation}
in the spaces $C\left[ 0,1\right] $ or $L_{2}\left( 0,1\right) $ for $\psi
\left( x\right) $, $x\in \lbrack 0,1]$ from a rather representative class.

Note that compared to the values of the sought and given functions the
considered error is really small. Therefore, in the case of the construction
of a stable algorithm of the evaluation of $\psi \left( x\right) $, its
exclusion by condition (\ref{i2.1}) should not considerably influence the
solution.

The structure of the error (\ref{i2}) under the condition (\ref{i2.1})
embodies the difference between the function $\psi \left( x\right) $,
subject to integration, and its approximate expression that, in turn,
appears as a result of the execution of an analogous procedure. One should
note the absence of any a priori premises of self-sufficiency of (\ref{i2})
in achieving the goal, namely, a correct formulation of the problem of the
determination of the function $\psi \left( x\right) $ from the data (\ref{i1}%
).

As a matter of fact, we put forward a hypothesis about the priority of a
qualitative side of the phenomenon of smoothing of information in modelling
the error of integration as well as, in general, about the expediency of the
suggested "measures" for the realization of a correct formulation of the
problem that is inverse to the procedure (\ref{i1}).

On the basis of (\ref{i2}) and (\ref{i2.1}), instead of the ill-posed
problem (\ref{i1}) for the determination of the function $\psi \left(
x\right) $, the following system of equations will be employed: 
\[
\mu \left( A\psi \right) \left( x\right) =\mu f\left( x\right) +\left(
\delta f\right) \left( x\right) ; 
\]%
\begin{equation}
\quad \left( \delta f\right) \left( x\right) =0,\quad x\in \left[ 0,1\right]
,  \label{i2.2}
\end{equation}%
where $\mu $ is a parameter analogous to $\lambda $.

We have worked out two versions of the solution of the formulated problem.
In the first version, the Fourier coefficients of the function $\psi \left(
x\right) \in \left[ 0,1\right] $ are represented in quadratures via the data
of the problem by means of the solution of the Fredholm integral equation of
the second kind that possesses rather favorable properties. This implies the
absence of singularities or oscillations of the kernel that are not caused
by $k\left( x,\xi \right) $ (i.e., those that are enforced by the employed
algorithm) as well as the absence of a small factor explicitly multiplying
the sought function $\psi \left( x\right) $.

In the second version, the computational procedure reduces to a consecutive
solution of two Fredholm integral equation of the second kind, i.e. of the
above-mentioned one and of another one that differs from the former one only
by the form of the free term. We also demonstrate a possibility of the
determination of the function $\psi \left( x\right) $ with the use of the
solution of just the first of these two equations.

The basis of the effectiveness of the performed transformations is formed by
the following factors:

1) An incompletely continuous perturbation of the operator $A$ in
combination, of course, with the condition (\ref{i1}) that led to the
derivation of the system (\ref{i2.1}) instead of (\ref{i2.2}).

2) The extension of (\ref{i2.2}) to $x\in \left[ -1,0\right) $, which allows
one to use a typical peculiarity of the solution of the Fredholm integral
equation of the second kind%
\begin{equation}
\psi \left( x\right) =\lambda \int\limits_{-1}^{1}h\left( x,\xi \right) \psi
\left( \xi \right) d\xi +\left\{ 
\begin{array}{c}
0,\quad x\in \left[ 0,1\right] ; \\ 
\kappa \left( x\right) ,\quad x\in \left[ -1,0\right) ,%
\end{array}%
\right.  \label{i2.3}
\end{equation}%
where $\kappa \left( x\right) $ is an undefined function stipulated by the
form of the free term.

3) The use of an equation, which is an analogue of (\ref{i2.3}), that
possesses the same solution on $x\in \left[ 0,1\right] $ and a free term
that goes to zero on the other part of the interval of definition $\left[
-1,0\right) $.

4) The choice of the kernel $h\left( x,\xi \right) $ in such a way that be
means of a linear change of the variables it could be transformed into the
canonical Poisson's kernel with a parameter $r$, which allows one to do the
following:

- determine the function $\psi \left( x\right) $ in the form of an
expression that explicitly depends on $r$ for the case when it is harmonic;

- by proceeding to the limit $r\rightarrow 1$, express the kernel and the
free term of the above-mentioned Fredholm integral equation of the second
kind via the data (\ref{i1});

- as a result, extend the class of admissible belonging of the function $%
\psi \left( x\right) $ to the whole space $L_{2}\left( 0,1\right) $ ( the
first version of the solution of the problem).

5) The consideration of the function $\psi \left( x\right) $ satisfying Eq. (%
\ref{i1}) as a generalized function, which resulted in a considerable
simplification of the procedure of meeting condition (\ref{i2.1}) (the
second version of the solution of the problem). Simultaneously, the
necessity of using the passage to the limit with respect to the parameter $r$
lost its relevance.

The objective of this work can also be explained by means of the following
example. Consider a beam (bar) supported at the ends and subject to a
transverse load: the problem consists in finding its deflection (in the
linear interpretation). Correspondingly, using the notation of (\ref{i1}),
we have:

$k\left( x,\xi \right) $ is the deflection at the cross-section with the
coordinate $x$ caused by a unit force applied in the cross-section with the
coordinate $\xi $;

$\psi \left( x\right) $ is the intensity of the distributed load;

$f\left( x\right) $ is the deflection whose determination by integration
according to (\ref{i1}) is successfully carried out by undergraduate
university students taking a course in the strength of materials.

However, the deflection of the beam is here to stay: it can be measured; and
the load does exist in reality. Therefore, fully justified is the
formulation of the inverse problem that consists in the determination of $%
\psi \left( x\right) $ from the given $k\left( x,\xi \right) $ and $f\left(
x\right) $. Such a problem is considered to be of an incomparably higher
degree of complexity, and the attempts of its solution is a preoccupation of
not the students, but rather of scientists and, in particular, of their
lecturers. In these attempts, the Fredholm integral equation of the first
kind (\ref{i1}) is used, whose solution, in reality, cannot be realized.
Moreover, even obtaining a palliative implied by this solution requires
application of great efforts.

At the same time, it is reasonable to suggest that the difficulties arise
due to the fact that the problem is ill-posed, as was explained above.
Really, there exists, on the other hand, a very convenient, from the point
of view of a numerical realization, object, namely, the Fredholm integral
equation of the second kind that can be represented in the form%
\begin{equation}
\psi \left( x\right) =\mu \int\limits_{0}^{1}K\left( x,\xi \right) \psi
\left( \xi \right) d\xi +F\left( x\right) ,\quad x\in \left[ 0,1\right] ,
\label{i2.4}
\end{equation}%
where the kernel $K\left( x,\xi \right) $ and the free term $F\left(
x\right) $ are given; the function $\psi \left( x\right) $ (the intensity of
the load) is subject to determination; $\mu $ is a parameter that has to be
chosen from the solvability condition.

However, the question arises: What are the reasons to believe that the
function $\psi \left( x\right) $, provided it is the same as that entering (%
\ref{i1}), should satisfy this equation, and what is understood under $%
K\left( x,\xi \right) $ and $F\left( x\right) $? On the other hand, if the
function $\psi \left( x\right) $ is assumed to be known and the kernel $%
K\left( x,\xi \right) $ is given even in an arbitrary form from the space $%
L_{2}$, one can always find a free term $F\left( x\right) $ allowing one to
satisfy Eq. (\ref{i2.4}). As a consequence, a Fredholm integral equation of
the second kind satisfied by the sought function $\psi \left( x\right) $
objectively exists.\footnote{%
One can assume that analogous arguments formed the basis of J. Hadamard's
statement.} Moreover, the number of such equations is not limited.

The construction of Eq. (\ref{i2.4}), simultaneously with Eq. (\ref{i1})
satisfied by the function $\psi \left( x\right) $, is, in fact, is the
objective of this work. In other words, it is devoted to the construction of
the free term $F\left( x\right) $ depending on the data (\ref{i1}) in such a
way that the function $\psi \left( x\right) $ and the solution of (\ref{i2.4}%
) coincide. Thus, equation (\ref{i2.4}) is a well-posed problem for the
determination of the function $\psi \left( x\right) $ satisfying (\ref{i1}).

However, a broad class of linear boundary-value and initial-boundary-value
problems of mathematical physics can be rather elementarily reduced to the
Fredholm integral equations of the first kind. To this end, considering, for
example, a problem described by the Laplace equation, one has to set $%
\partial _{x}^{2}u=\psi $ (or $\partial _{x}^{2}u+\beta u=\psi $, where $%
\beta $ is a constant).

By means of integration with respect to $x$ the function $u\left( x,y\right) 
$ and its derivatives are expressed via $\psi $. Integration of the
differential equation with respect to $y$ allows one to obtain a second
representation of $u$ via $\psi $. The one-dimensional functions of
integration in these representations are expressed via $\psi $ and satisfy
the boundary conditions. The elimination of $u$ from the representations of
the solution leads to a two-dimensional Fredholm integral equations of the
first kind with respect to the function $\psi \left( x,y\right) $.

The outlined scheme is practically indifferent to the type and order of
differential operators, the presence of variable coefficients, the
configuration of the boundary of the domain of the function and some other
factors that usually complicate the realization of numerical algorithms. The
above-mentioned method of the solution of the problem (\ref{i1}) is directly
extended to the determination of the function $\psi \left( x,y\right) $ (the
variable $y$ plays the role of a parameter). In this regard, there appears
an interesting possibility to check the solvability of the problems of
mathematical simulation that can be represented by partial differential
equations.

Perhaps, the motivation of the proposed investigation could be of certain
interest. The reason was the confusion caused by the absence in the
specialized literature of a clear statement of the universality of the
outlined method of the reduction of problems of mathematical physics to the
Fredholm integral equations of the first kind. However, a placement of the
whole lot of initial data in their kernels is nevertheless rather
attractive. Indeed, a conventional classification of problems according to
the complexity of their numerical realization is, in fact, depleted, and the
construction of an effective method of the determination of the functions
satisfying the stated type of equations comes to the foreground.

In Chapter 2, we analyze J. Hadamard's arguments concerning the issue of
correct formulation of problems for partial differential equations. Both
related and alternative positions on this issue of known specialists are
illuminated. We also discuss Banach's theorem on inverse operator that is
closely related, in a contextual sense, with the methodology of correctness.
Arguments are given that mathematical formulations of the direct and the
inverse problems should not be identical.

Chapter 3 contains an analytical review of the methodological approaches and
methods of the solution of ill-posed problems (mostly, of Fredholm integral
equations of the first kind) related to the concepts of A. N. Tikhonov and
V. M. Fridman. Pointed out are some expert opinions about rational use of
digital information and, on the whole, about priorities of the development
of computational mathematics.

The material of Chapter 4, in a sense, refracts principle difficulties,
accompanying the solution of ill-posed problems by the prism of fundamental
concepts of J. Hadamard and S. Banach. We present arguments for
inconsistency of the methodology of the solution of ill-posed problems.
General premises for a correct formulation of\ the problem of determination
of the function satisfying the Fredholm integral equations of the first kind
are given.

Chapter 5 is devoted to the construction of the method of the reduction of
the problems, usually associated with the Fredholm integral equation of the
first kind, to the solution of Fredholm integral equations of the second
kind. This section is, in a constructive sense, basic. Exactly here we
consecutively construct an algorithm that practically realizes the main
factors ensuring the efficiency of the transformations pointed out above.
The first version of the solution of the problem is presented.

In Chapter 6, we emphasize the main points of the carried out
transformations and also study a possibility of their variation. An
interpretation of the algorithm of the previous section is given from a
generalized point of view. Furthermore, the second version of the considered
problem is given. Its correct formulation in terms of the Fredholm integral
equation of the second kind is presented.

The material of Chapter 7 illustrates the universality of the technique of
the reduction of linear boundary-value and initial-boundary-value problems
to Fredholm integral equations of the first kind. An extension of the
suggested algorithm of the solution of the equations of this type to a
two-dimensional case is demonstrated. The issue of solvability of the
problems of mathematical simulation is touched on.

Chapter 8 develops the outlined orientation of the reduction of the problems
to the Fredholm integral equation of the first kind involving into the
sphere of transformations sufficiently nontrivial applications (including
factors of nonlinearity, singular perturbations and some other). The
presentation of the material has the form of sketches.

In Chapter Conclusions, we summarize the main points of the work from the
same position of priority importance of correct formulation of problems of
mathematical simulation for the efficiency of their numerical realization.

Mathematical techniques employed in the presentation of the material is
comparatively simple: basics of the classical theory of integral equations;
elements of functional analysis; general principles of formulations of
problems of mathematical physics and of methods of their solution. When
performing transformations, we often refer to the book by F. G. Tricomi, 
{\it Integral Equations} (Dover, New York, 1957).

The literature to each chapter is given in reference order. (Note that page
numbers refer to the Russian edition of a corresponding literature source.)
Chapters and sections (chapters and paragraphs of the literature sources)
are referred to, respectively, as Chapter 1, section 1.1, sections 1.1, 1.2.

The numbering of formulas in the text is dual: the first numeral refers to
the chapter number, whereas the second one refires to the formula number
inside the chapter.

The author is most grateful to I. Zhuravlev for pointing out a contradiction
in the transformations of Chapter 5. As a result, this section was
substantially revised. Some revisions were made also in other chapters.
However, the methodology of the work and the basis of the method of the
solution did not change.

I also thank I. Stepanov for making a web-site in the Internet, M.
Katchamanova for her assistance in the preparation of the manuscript, and S.
V. Kuplevakhsky for translating the Russian version of the manuscript (see: 
{\it www.pelbook.narod.ru}) into English.

\chapter{The issue of the correct formulation of problems of mathematical
physics}

\section{Hadamard's definition of correctness}

J. Hadamard has defined two conditions that should be satisfied by a
correctly formulated boundary-value (initial-boundary-value) problem for
partial differential equations: existence and uniqueness of the solution ( 
\cite{1.1}, p. 12).\footnote{%
For the first time, the concept of correct formulation was put forward by
Hadamard in his article of 1902.} At the same time, the third condition of
Hadamard's definition of correctness that concerns continuous dependence on
the data of the problem is well-known. Indeed, he paid serious attention to
the investigation of this issue with regard to Cauchy-Kovalevskaya's theorem
concerned with the solution of the differential equation 
\begin{equation}
\partial _{t}^{k}u=f\left( t,x_{1},x_{2},\ldots ,x_{n},\partial
_{t}u,\partial _{x_{1}}u,\partial _{x_{2}}u,\ldots ,\partial
_{x_{n}}^{k}u\right)  \label{1.1}
\end{equation}%
(a system of analogous equations), where $f$ is an analytical function of
its arguments in the vicinity of the origin of coordinates, with initial
conditions 
\[
u\left( 0,x_{1},x_{2},\ldots ,x_{n}\right) =\varphi _{0}\left(
x_{1},x_{2},\ldots ,x_{n}\right) ; 
\]%
\[
\partial _{t}^{s}u\left( 0,x_{1},x_{2},\ldots ,x_{n}\right) =\varphi
_{s}\left( x_{1},x_{2},\ldots ,x_{n}\right) , 
\]%
\begin{equation}
\quad s=1,\ldots ,k-1.  \label{1/2}
\end{equation}

As is pointed out by Hadamard, the consideration of the problem (\ref{1.1}),
(\ref{1/2}), named after Cauchy, raises three questions (\cite{1.1}, p. 17):

1) Does it admit a solution?

2) Is the solution unique? (In general, is the problem well-posed?)

3) Finally, how the solution can be derived?

Cauchy-Kovalevskaya's theorem (in its authors' interpretation) states that,
except for some special cases, the above-mentioned problem admits a unique
solution that is analytical at the origin of coordinates. Moreover, the
functions $\varphi _{0}$,..., $\varphi _{k-1}$ in (\ref{1/2}) can be not
only analytical but regular, i.e., continuous together with their
derivatives up to a certain order. A possibility of a uniform approximation
of $\varphi _{0}$,..., $\varphi _{k-1}$ by Taylor series expansions in
powers of $x_{1}$, ..., $x_{n}$ , retaining all operations on analytical
functions, including differentiation up to a corresponding order, is implied.

However, such an approach was strongly criticized by Hadamard. In his
opinion, the question is not how such an approximation affects the initial
data, but rather what is an effect on the solution? He emphasized the
non-equivalence of the notion of small perturbation for given Cauchy's
problem and of the solution to this problem (\cite{1.1}, p. 39). In this
regard, J. Hadamard presented his prominent example of a solution of the
differential equation 
\begin{equation}
\partial _{t}^{2}u+\partial _{x}^{2}u=0,  \label{1.3}
\end{equation}%
subject to the conditions 
\begin{equation}
u\left( x,0\right) =0;\quad \partial _{t}u\left( 0,x\right) =\alpha _{n}\sin
\left( nx\right) ,  \label{1.4}
\end{equation}%
where $\alpha _{n}$ is a rapidly decreasing function of $n$.

The expression on the right-hand side of (\ref{1.4}) can be arbitrarily
small. Nevertheless, the problem admits the solution 
\begin{equation}
u\left( x,t\right) =\frac{\alpha _{n}}{n}\sin \left( nx\right) \sinh \left(
nt\right) .  \label{1.5}
\end{equation}%
For $\alpha _{n}=1/n$ or $1/n^{\mu }$, or $e^{-\sqrt{n}}$, this solution is
rather large for any nonzero $t$, because of the prevailing growth of $%
e^{nt} $ and, correspondingly, of $\sinh \left( nt\right) $. Thus, the
function (\ref{1.5}) does not depend continuously on the initial data and,
as a result, the problem (\ref{1.3}), (\ref{1.4}) is ill-posed.

Concerning the regularity of the right-hand side of (\ref{1/2}), J. Hadamard
remarked: \textquotedblright ...actually, one of the most curious facts of
the theory is that equations, seemingly very close to each other, behave in
a completely different way\textquotedblright\ (\cite{1.1}, p. 29).

A large number of investigations devoted to the issue of the correct
formulation of Cauchy's problems. The authors of these investigations
concerned themselves with specification of corresponding classes of
differential equations and with minimization of requirements imposed on the
initial data (see \cite{1.2}). However, we are mostly interested in the
actual character of the dependence of the solution on the data of the
problem and, in this regard, the classic J. Hadamard's statement that
\textquotedblright an analytical problem is always well-posed in the
above-mentioned sense, when there exists a mechanical or physical
interpretation of the question\textquotedblright\ (\cite{1.1}, p. 38).

As was pointed out by V. Y. Arsenin and A. N. Tikhonov \cite{1.3}, the
latter questioned the legitimacy of studies of ill-posed problems, specified
by the authors as the following: the solution of integral equations of the
first kind; differentiation of approximately known functions; numerical
summation of Fourier series whose coefficients are approximately known in
the metric $l_{2}$; analytical continuation of functions; the solution of
inverse problems of gravimetry and of ill-defined systems of linear
algebraic equations; minimization of functionals for divergent sequences of
coordinate elements; some problems of linear programming and of optimal
control; the design of optimal systems and, in particular, the synthesis of
aerials. It is emphasized that this list is by no means complete, because
ill-posed problems appear in investigations of a broad spectrum of problems
of physics and engineering.

In his talk at the meeting of the Moscow Mathematical Society devoted to
Hadamard's memory, G. E. Shilov said the following \cite{1.4}:
\textquotedblright Our time has brought about corrections in Hadamard's
instructions, because it turned out that ctill-posed, according to Hadamard,
problems could have meaning (as, e.g., the problem of restoration of a
potential from scattering data). However, the studies of well-posed
problems, proclaimed by Hadamard, was a cementing means for the formation of
the whole theory\textquotedblright\ (functional analysis is implied). This
quotation is borrowed from a biographical sketch by E. M. Polishtuk and T.
O. Shaposhnikova \cite{1.5}, where it is also pointed out that in the course
of time J. Hadamard's opinion about the importance for practice of
exclusively well-posed problems was understood in a less absolute sense.

At the same time, rather sharp statements were made:

\textquotedblright And what is more, Hadamard put forward a statement that
ill-posed problems had no sense at all. Since (as can be seen from a modern
point of view) most applied problems, represented by equations of the first
kind, are ill-posed, this statement of the outstanding scientist,
apparently, strongly slowed down in 1920-1950's the development of the
theory, methods and practice of the solution of problems of this
class\textquotedblright\ (\cite{1.6}, p. 12).

\textquotedblright Until quite recently, it was thought that ill-posed
problems had no physical sense and that it was unreasonable to solve them.
However, there are many important applied problems of physics, engineering,
geology, astronomy, mechanics, etc., whose mathematical description is
adequate although they are ill-posed, which poses an actual problem of the
development of efficient methods of their solution\textquotedblright\ (\cite%
{1.6}, p. 225).\footnote{%
In the context of what follows, we draw attention to the \textquotedblright
adequate description\textquotedblright .}

\textquotedblright From the results of this work [of A. N. Tikhonov]
followed a limitation of the well-known notion of J. Hadamard \cite{1.1} of
a well-posed problem of mathematical physics, which was of indisputable
methodological interest, and inconsistency of Hadamard's thesis, wide-spread
among investigators, that any ill-posed problem of mathematical physics was
unphysical.\textquotedblright\ (\cite{1.7}, p. 3).

\textquotedblright For a long time, activities related to the analysis and
solution of problems called ill-posed used to be relegated (by famous
mathematicians too) to the domain of metaphysics\textquotedblright\ (\cite%
{1.8}, p. 126). \textquotedblright A prevailing number of mathematicians
(including Hadamard) expressed.their attitude towards this problem in the
following way: If a certain problem does not meet the requirements of
correctness, it is of no practical interest and, hence, does not need to be
solved\textquotedblright\ (\cite{1.8}, p.127) (I. G. Preobrazhenskii, the
author of the section \textquotedblright Ill-posed problems of mathematical
physics\textquotedblright ).

Note that the latter paper most distinctively reveals the style that causes
a principal objection. Thus, A. Poincar\'{e} is accused of inconsistency of
methodological views on the nature of causal relationship (\cite{1.9})
(\textquotedblright The Last Thoughts\textquotedblright ). Indeed, the text
does not contain any evidence that he makes a fetish of the problem of
restoration of the cause from the effect. On this basis, a conclusion is
made about the great scientist's misunderstanding of the essence of
instability of computation procedures inherent to ill-posed problems and, in
particular, to integral equations of the first kind.\footnote{%
In particular, the exact statement reads: \textquotedblright However, one
must remember that vagueness of philosophical positions of some scientists
in the West, even rather renown, results in the fact that, based on correct
starting points, they draw rather inaccurate conclusions, repeating old
mistakes of, for example, A. Poincar\'{e}, who writes: 'If two organisms are
identical, or simply similar, this similarity could not occur by chance, and
we can assert that they lived under the same conditions...' In other words,
the fact of possible incorrectness of the inverse problem is completely
ignored.\textquotedblright\ However, one would hardly mention Poincar\'{e}'s
mistakes... if modern \textquotedblright spontaneous supporters of the
principle of determinism did not repeat them\textquotedblright\ ( [8], p.
134).}

The adequacy of employed models to considered concrete processes is not even
touched on by the authors of (\cite{1.8}). Thus, a quite legitimate question
arises: How does one know that Poincar\'{e}, if necessary, could not find a
way of a mathematically correct formulation of the same physical problems?
Anyway, is there any contradiction in general arguments for the existence of
such a possibility, including the aspects of its constructive realization?

By the way, exactly Poincar\'{e} repeatedly mention Hadamard while
establishing a relationship between the correct formulation of problems and
a practical realization of employed models. We draw attention to an
expressive thesis: \textquotedblright If a physical problem reduces to an
analytical one, such as (\ref{1.3}), (\ref{1.4}), it will seem to us that it
is governed by a pure occasion (according to Poincar\'{e}, it means that
determinism is violated) and it does not obey any law\textquotedblright\ (%
\cite{1.1}, p. 43).

In light of the above, the arguments of I. Prigogine and I. Stengers \cite%
{1.10} are of interest: ''...one can speak of a 'physical law' of some
phenomenon only in the case when this phenomenon is 'coarse' with respect to
a limiting transition from a description with a finite accuracy to that with
an infinite accuracy and thus inaccessible to any observer, whoever he may
be'' (p. 9). ''Scientist in a hundred different ways expressed their
astonishment that a correct formulation of the question allows them to solve
any puzzle suggested by nature'' (p. 44).

Thus, underlined are: first, methodological importance of correct
formulation of problems of mathematical physics; second, a leading role of
the employed procedures and, finally, substantial influence of the quality
of their realization on the degree of complexity of obtaining the final
result. In other words, one and the same problem can be better or worse
formulated.

The above-mentioned statement or Hadamard's postulate, as it called by S. K.
Godunov (\cite{1.11}, p. 113), as a matter of fact, implies a possibility of
a \textquotedblright good\textquotedblright\ (correct) formulation of any
meaningful problem and, consequently, can be interpreted as having a global
orientation.\footnote{%
The above mentioned reference contains the following definition:
\par
A problem is called well-posed if is solvable for arbitrary initial (or
boundary) data belonging to a certain class, has a unique solution, and this
solution continuously depends on the initial data.
\par
A problem is called ill-posed either if it is not solvable for arbitrary
initial data or if it is impossible to choose such norms for the solution
and for the initial data that continuos dependence of the solution on the
data of the problem with respect to these norms be ensured.}

In this regard, one can establish an obvious relationship to D. Hilbert's
comments on his 20th problems that suggested a possibility of correct
formulation of arbitrary boundary-value problems of mathematical physics by
means of special requirements on boundary values of corresponding functions
(a type of continuity or piecewise differentiability up to a certain order)
and, by necessity, by giving an extended interpretation to the notion of the
solution (\cite{1.12}, pp. 54-55).

For the first time, the three conditions of the correctness of problems of
mathematical physics were clearly pointed out by D. Hilbert and R. Courant ( 
\cite{1.13}, pp. 199-200): existence, uniqueness and continuous dependence
of the solution on the data of the problem. Concerning the last, they say:
\textquotedblright ...it has crucial importance and is by no means
trivial... A mathematical problem can be considered adequate to the
description of real phenomena only in the case when a change of given data
in sufficiently narrow limits is matched by an alike small, i.e. restricted
by predetermined limits, change of the solution\textquotedblright .

V. A. Steklov's position is quite analogous (\cite{1.14}, p. 62):
\textquotedblright ...if differential equations with the above-mentioned
initial and boundary conditions are not constructed on erroneous grounds,
are not in direct contradiction to the reality, they must yield for each
problem a unique and completely definite solution...\textquotedblright .
Along the same lines, I. G. Petrovskii writes (\cite{1.15}, p. 87): "The
above-mentioned arguments for the correct formulation of Cauchy's problem
show that other boundary-value problems for partial differential equations
are of interest for natural science only in the case when there is, in a
sense, continuous dependence of the solution on boundary
conditions\textquotedblright .

S. L. Sobolev is less categorical (\cite{1.16}, p. 38): \textquotedblright
The solution to an ill-posed problem in most cases has no practical
value\textquotedblright . Of considerable interest is the opinion of V. S,
Vladimirov (\cite{1.17}, p. 69): \textquotedblright The issue of finding
correct formulations of problems of mathematical physics and methods of
their solution (exact or approximate) is the main content of the subject of
equations of mathematical physics\textquotedblright .

V. V. Novozhilov, in fact, drew attention to the potential of variation of
the formulation of the considered problem with the aim of the simplification
of the procedure of its numerical realization (\cite{1.18}, p. 352):
\textquotedblright The absence in the term \textquotedblright a mathematical
model\textquotedblright\ of the indication of its inevitable approximate
character leaves way for a formal mathematical approach to models,
disregarding those concrete problems for whose solution they were intended,
which is, unfortunately, wide-spread at present\textquotedblright .

\section{J. Hadamard's postulate and incorrectness of \textquotedblright
real\textquotedblright\ problems}

Thus, J. Hadamard and a number of other outstanding scientists thought that
any physically interpretable problem could be well-posed. However, a quite
opposite point of view dominates in modern publications. Indeed, a visibly
larger part of practically important problems considered therein are
incorrect. However, is the actual methodology of mathematical formulation of
these problems and, correspondingly, the results of its refraction with
respect to realities adequate?

Here we will not elaborate on something like general principles of the
construction of differential equations, and, generally speaking, it is
reasonable at the beginning to restrict the question to the following: What
arguments allow one to conclude that an ill-posed problem adequately
describes an observable phenomenon or a potentially real process? In this
regard, let us turn to the procedure of the solution of the Fredholm
integral equation of the first kind 
\begin{equation}
\int\limits_{0}^{1}k\left( x,\xi \right) \psi \left( \xi \right) d\xi
=f\left( x\right) ,\quad x\in \left[ 0,1\right] ,  \label{1.6}
\end{equation}%
which is a classical incorrect problem: the square summable kernel $k\left(
x,\xi \right) $ and the free term $f\left( x\right) $ are given; the
function $\psi \left( x\right) $ is to be restored.

Let us assume that the kernel is symmetric and closed, i.e. $k\left( x,\xi
\right) \equiv k\left( \xi ,x\right) $ and its eigenfunctions $\bar{\psi}%
_{n}\left( x\right) $, being nontrivial solutions of the integral equation 
\[
\psi \left( x\right) =\lambda \int\limits_{0}^{1}k\left( x,\xi \right) \psi
\left( \xi \right) d\xi =f\left( x\right) ,\quad x\in \left[ 0,1\right] 
\]%
with characteristic numbers $\lambda =\lambda _{n}$, $n=1,2,\ldots $, form a
complete in $L_{2}\left( 0,1\right) $ orthogonal system of elements. In this
case, the solution to Eq. (\ref{1.6}) exists and is unique under the
condition (see, e.g., \cite{1.19}, pp. 185-187) 
\begin{equation}
\sum_{n=1}^{\infty }\alpha _{n}^{2}\lambda _{n}^{2}<\infty ,\quad \alpha
_{n}=\int\limits_{0}^{1}f\left( x\right) \bar{\psi}_{n}\left( x\right) dx.
\label{1.7}
\end{equation}

If all the above-mentioned conditions are fulfilled, there is still the
third condition of correctness that, as it is known, is certainly not
satisfied by Eq. (\ref{1.6}). Numerous literature references clearly
illustrate an inadequately strong influence on the solution of small
perturbations of the data of the problem, in the first place of $f\left(
x\right) $. As a rule, this function is determined experimentally and
mismatch the kernel $k\left( x,\xi \right) $, in particular, with respect to
smoothness. Thus, Eq. (\ref{1.6}), strictly speaking, looses sense. At the
same time, a possibility of an equivalent description of the problems of
mathematical physics by means of integral equations of the first kind is
indisputably admitted at present, which is confirmed by their colossal list (%
\cite{1.6}, section 4.2).

Let us specify Eq. (\ref{1.6}): 
\begin{equation}
k\left( x,\xi \right) =\left\{ 
\begin{array}{c}
x\left( 1-\xi \right) ,\quad x\leq \xi \leq 1; \\ 
\xi \left( 1-x\right) ,\quad 0\leq \xi \leq x;%
\end{array}%
\right. \quad f\left( x\right) =\frac{1}{\left( m\pi \right) ^{2}}\sin
\left( m\pi x\right) ,  \label{1.8}
\end{equation}%
where $m$ is an integer. For this choice, $\lambda $$_{n}=\left( m\pi
\right) ^{2}$; $\bar{\psi}_{n}\left( x\right) =\sqrt{2}\sin \left( n\pi
x\right) $; $n=1,2,\ldots $ (\cite{1.20}, p. 149).

Since the kernel $k$ is symmetric and continuos, and all $\lambda $$_{n}>0$,
the use of Mercer's theorem \cite{1.19}, according to which 
\[
k\left( x,\xi \right) =\sum_{n=1}^{\infty }\frac{\bar{\psi}_{n}\left(
x\right) \bar{\psi}_{n}\left( \xi \right) }{\lambda _{n}}, 
\]%
and a representation of $\psi \left( x\right) $ as a series expansion in
terms of $\bar{\psi}_{n}\left( x\right) $ with undetermined coefficients
allows one to find the solution to Eq. (\ref{1.6}): 
\begin{equation}
\psi \left( x\right) =\sin \left( m\pi x\right) .  \label{1.9}
\end{equation}

However, the procedure of calculations turned out to be so simple owing to a
special choice of the data of the problem. If this is not the case or in the
case of the solution of Eq. (\ref{1.6}) with the kernel and the free term (%
\ref{1.8}) by means of one of numerical methods, the complexity of the
realization of a an approximation of sufficiently high order is practically
identical to the most general situation, characterized by an error in the
determination of $f\left( x\right) $ and $k\left( x,\xi \right) $.\footnote{%
Here, complexity implies an ill definition of the system of linear algebraic
equations obtained as a result of some sort of discretization.} As a matter
of fact, even if the data are objectively compatible, the incorrectness of
Eq. (\ref{1.6}) appears as a result of rounding off the digits in the
process of calculations.

The factor of the incorrectness of Eq. (\ref{1.6}) follows from a comparison
of the free term (\ref{1.8}) with the solution (\ref{1.9}). Indeed, by
increasing $m$, the function $f\left( x\right) $ may turn out to be
arbitrarily small, whereas the bounds of the values of $\psi \left( x\right) 
$ are unchanged. Correspondingly, any error in the calculations with $%
f\left( x\right) $ is projected onto the function $\psi \left( x\right) $
with the factor $m^{2}$. The mechanism of this phenomenon of the smoothing
of information about the function in the process of integration will be
repeatedly discussed in what follows.

However, let us return to the question of the relation of an incorrect
formulation to the reality. In this regard, we draw attention to the
following. By considering (\ref{1.6}) as the Fredholm integral equation of
the first kind , we mean the solution of the inverse problem (I). However,
equation (\ref{1.6}) can be used for the solution of the corresponding
direct problem (D): the determination of the function $f\left( x\right) $
from the data $k\left( x,\xi \right) $ and $\psi \left( x\right) $. This
procedure is correct and thus is radically simpler than the problem I. It is
sufficient to note the absence of any principal difference between the
evaluation of the integral (\ref{1.6}) in an analytical form and its
essentially numerical realization.

Here we want to draw attention to an issue that seems to be of substantial
importance. The problem D, as a rule, is transparent: in its categories, we
adequately model realistic current processes and phenomena by, which should
be emphasized, explicit means of linear superposition. Correspondingly, if,
for instance, $k\left( x,\xi \right) $ is a characteristic of the system and 
$\psi \left( x\right) $ is intensity of external influence, a resulting
effect in this or that subject sphere is to be elementarily summed up.

The situation is diametrically different for the problem I. One could hardly
point out any realistic process (phenomenon) for which it could be
formulated in mathematical terms directly on the basis of the subject
sphere. In other words, without any relation to the problem D, which
commonly implies a transformation of the latter into the problem I just by
means of renaming of known and unknown components.

It seems that the methodology, which states the adequacy of the problem I,
obtained by the above-mentioned renaming of the components, to the realities
on the basis of a high-quality information about a concrete problem D, is
profoundly deficient. Correspondingly, the opinion of experts who a priori
reject J. Hadamard's argument for the existence of correct formulations of
problems of mathematical physics should be considered unjustified.

Let us turn to the problem D that describes some realistic process
(phenomenon) (\ref{1.6}). For this process, the determination of $\psi
\left( x\right) $ from the data $k\left( x,\xi \right) $ and $f\left(
x\right) $, i.e. the formulation of the corresponding inverse problem that
will be denoted as I$^{^{\prime }}$, is, of course, reasonable. Suppose that
in this case Hadamard's argument holds, and, hence, the problem I$^{^{\prime
}}$ is correct. However, the problem I, the solution of the Fredholm
integral equation of the first kind (\ref{1.6}), is ill-posed by definition.

The conclusion is obvious: Mathematical formulations (representations,
expressions) of the problems I and I$^{^{\prime }}$ are non-identical. As a
result, the formulation of the problem I$^{^{\prime }}$ cannot be restricted
to readdressing the status of the unknown variable between the functions $f$
and $\psi $ in the problem D. Note in this regard that a development of the
methodology of the correct formulation of the problem that is inverse to D,
i.e. I$^{^{\prime }}$, is the main objective of the present investigation.

The above arguments seem to be rather convincing, however, at this stage of
our consideration, we can neither prove the correctness of Hadamard's
postulate (argument) in the general case nor illustrate its constructive
character with respect to separate classes of problems. One should also bear
in mind that, using special methods, the solution of the ill-posed problem I
(or what is understood under the solution), as a rule, can be obtained with
accuracy that is considered to be practically acceptable. In this regard,
the question arises: Should one aim at the correct formulation I$^{^{\prime
}}$, if the algorithm of the calculation of the function $\psi \left(
x\right) $ in the formulation of the problem I in some way realizes its
regularization? This implies a well-known deformation of the formulation I
with the use of a small parameter that yields the property of correct
solvability.

Thus, can the algorithm to a full extent, including the efficiency of
numerical realization, level off the principal difficulties inherent to the
incorrectness of the problem I in the form (\ref{1.6})? It is clear that the
answer is definitely negative: Otherwise, the deep-rooted differentiation
between ill-posed problems and well-posed ones would make no sense.

Furthermore, the indicated difference is of exceptional importance, because
correctness of the formulation is a criterion of a qualitative level,
whereas the efficiency of a method of the solution of the Fredholm integral
equation of the first kind can be estimated only in terms of quantitative
factors of a palliative property. The latter is caused by a direct
relationship between a degree of regularization and the deformation
(distortion) of the problem I.

What is, however, the actual difference in the interpretation of the
formulations I and I$^{^{\prime }}$? The answer to this question is
contained in sections 4.5, 5 and 6. At this stage, we only note that a
transformation of the formulation I into the formulation I$^{^{\prime }}$
will be realized by means of an incompletely continuos perturbation of the
integral operator of the problem (\ref{1.6}) that simulates the phenomenon
of smoothing of information.

\section{Banach's theorem on the inverse operator}

Let us quote (\cite{1.5}, p. 175): \textquotedblright First, Hadamard
defined the correctness of the problem by the conditions of solvability and
uniqueness and strongly insisted on continuous dependence of the solution on
the initial data only in the consideration of Cauchy's problem. In the book
'The theory of partial differential equations', published in Peking a year
after his death, he wrote: 'This third condition that we introduced in
\textquotedblright Lectures on Cauchy's problem...\textquotedblright\ but
did not consider as part of well-posed problems, was added, quite justified,
by Hilbert and Courant \cite{1.13}. Here, we accept their point of view.'"

E. M. Polishuk and T. O. Shaposhnikova made the following comment on this
text \cite{1.5}, pp. 175-176]: \textquotedblright From a mathematical point
of view, the question of the necessity of the requirement of the continuity
of the solution with respect to the data seems to be rather delicate. As a
matter of fact, according to Banach's well-know theorem on closed graph,
unique solvability of a linear problem leads to boundedness of the inverse
operator and, thus, continuous dependence of the solution on the right-hand
sides.\textquotedblright\ It is pointed out that variations of the
coefficients of differential equations and of the boundary of the considered
domain can also influence the solution of the problem; hence, the use of the
three conditions of the correctness is preferable.

At the same time, Banach's theorem on the inverse operator (\cite{1.21}, p.
34), being a consequence of the above-mentioned one, is more closely related
to the considered issue. Its formulation, given by A. I. Kolmogorov and S.
V. Fomin, is the following (\cite{1.22}, pp. 259-260): Let $A$ be a linear
bounded operator that maps a Banach space $B_{1}$ in a one-to-one fashion
onto a Banach space $B_{2}$. Then the inverse operator $A^{-1}$ is unique.

In addition, L. A. Lyusternik and V. I. Sobolev (\cite{1.23}, pp. 159-161)
emphasized that a one-to-one mapping of the whole Banach space $B_{1}$ onto
the whole Banach space $B_{2}$ is implied. Besides, a situation is discussed
when \textquotedblright ...an operator, being the inverse of a bounded
operator, although linear, turn out to be defined not on the whole space $%
B_{2}$ but only on a certain linear manifold and unbounded on this
manifold\textquotedblright .

The formulation of the same theorem in (\cite{1.24}, p. 60) reads: If a
linear bounded operator $A$ that maps a Banach space $B_{1}$ onto a Banach
space $B_{2}$ has an inverse $A^{-1}$, then $A^{-1}$ is bounded. It is
pointed out that this statement becomes invalid if one gives up the
requirement of completeness of one of the spaces. There is also a
clarification: The existence and uniqueness of the solution of the equation $%
A\psi =f$ with an arbitrary right-hand side from $B_{2}$ leads to continuous
dependence of the solution $\psi =A^{-1}f$ on $f$.

S. Banach himself made the following statement: If a linear operation
realizes a one-to-one transformation of $B_{1}$ onto $B_{2}$, the
transformation is mutually continuos. At the same time, in the formulation
of the theorem on the closed graph, he pointed out that the transformation $%
B_{1}$ is realized onto the whole space $B_{2}$.

L. V. Kantorovich and G. P. Akilov made a refinement concerning a mapping
under the specified conditions onto a closed subspace of the Banach space $%
B_{2}$ (\cite{1.25}, p. 454). The essence is that a closed subspace of a
Banach space is itself a Banach space.

S. G. Mikhlin gave a proof of the theorem (\cite{1.26}, p. 507): For the
linear problem $A\psi =f$ to be well-posed in a pair of Banach spaces $B_{1}$%
, $B_{2}$, it is necessary and sufficient that the operator $A^{-1}$ exist,
be bounded and map the whole space $B_{2}$ onto $B_{1}$. At the same time, a
clear distinction is made between the category of the existence and
uniqueness of the solution of the boundary-value problem and its correctness
as a whole, which implies, as a result, continuous dependence on the data
(the third condition according to Hadamard). The following definition is
given: \textquotedblright A boundary-value problem is called well-posed in a
pair of Banach spaces $B_{1}$, $B_{2}$ if its solution is unique in $B_{1}$
and exists for any data from $B_{2}$, and if an arbitrarily small change of
the solution in the norm $B_{2}$ corresponds to a sufficiently small change
of the initial data in the norm $B_{1}$\textquotedblright\ (p. 204).

The author pointed out that the problem might turn out to be well-posed in
one pair of spaces and ill-posed in another one. Besides, the fact that the
Fredholm integral equation of the first kind (\ref{1.6}) is ill-posed
follows from the contradiction: If the problem is well-posed, there exists a
bounded operator $A^{-1}$ and, hence, the identical operator $I=A^{-1}A$ is
completely continuos in the corresponding infinite dimensional space, which
contradicts the fundamentals of the general theory \cite{1.24}. Mikhlin also
quite encouragingly pointed out the approach of an approximate solution of
ill-posed problems headed by A. N. Tikhonov.

In an analogous, as to its content, course (\cite{1.27}, pp. 169-170),
Mikhlin reiterated the above-mentioned formulations. However, Tikhonov is
not mentioned at all, whereas the discussion of Eq. (\ref{1.6}) found a
rather interesting continuation (p. 171). It is shown that the problem of
its solution becomes well-posed if the pair of spaces $B_{1}$, $B_{2}$ is
replaced with such one that the operator $A$ is no longer completely
continuos. The general considerations are illustrated by the following
example. Let $k\left( x,\xi \right) $ and $f\left( x\right) $ satisfy the
conditions of section 2.2, including (\ref{1.7}). It turns out that if one
retains $L_{2}\left( 0,1\right) $ as $B_{1}$ and for $B_{2}$ also takes a
Hilbert space of functions normalized according to (\ref{1.7}), the solution
of Eq. (\ref{1.6}) becomes a well-posed problem: the operator $A$ is
incompletely continuos and $A^{-1}$ is bounded.

In this regard, one can point out that the operator $A$ is restrictively
invertible not only when it acts from $B_{1}$ onto the whole Hilbert space $%
B_{2}$. It is sufficient that the operator $A$ be bounded from below and
that its range $R\left( A\right) $ be dense everywhere in $B_{2}$. At the
same time, $R\left( A\right) $ is not necessarily closed (\cite{1.28}, p.
34).

A decade later, Mikhlin, in fact, gave up the investigations related to the
issue of correctness \cite{1.29}: \textquotedblright The author adheres to
the classical point of view, according to which the problem being solved by
mathematical methods should be considered as well-posed. Of course, there
are other opinions (p. 7)... Thus, we neglect the so-called incorrigible
errors related to the formulation of the above-mentioned problem as a
problem of natural science or of social studies (measurement errors,
insufficient accuracy of basic hypotheses, etc.)\textquotedblright\ (p. 17).

M. M. Lavrentiev and L. Y. Saveliev characterized investigations of the
issue of the solvability of Eq. (\ref{1.6}) on the basis of considerations
of the type of \cite{1.27} as trivial, because it is difficult to imagine
that for experimentally determined $f\left( x\right) $ the corresponding
error may prove to be small in the norm of the space $B_{2}$ (\cite{1.30},
p. 217). At the same time, it is pointed out that, generally speaking, for
any operator equation, one can choose pairs of spaces such that the problem
of its solution will be well-posed.

G. M. Vainikko and A. Y. Veretennikov draw attention to the complexity of
the description of such spaces. Thus, even the Volterra integral equation of
the first kind 
\[
\int\limits_{0}^{x}k\left( x,\xi \right) \psi \left( \xi \right) d\xi
=f\left( x\right) ,\quad x\in \left[ 0,1\right] , 
\]%
which admits the regularization 
\[
\psi \left( x\right) +\int\limits_{0}^{x}\partial _{x}k\left( x,\xi \right)
\psi \left( \xi \right) d\xi =f^{\prime }\left( x\right) ,\quad x\in \left[
0,1\right] 
\]%
and is elementarily solvable by quadratures, for reasons of the norm for $%
\psi \left( x\right) $, as a rule, has to be considered as an ill-posed
problem (\cite{1.31}, p. 6).

As regards the pair of spaces that realize the conditions of the correct
formulation, an original remark of K. I. Babenko is of interest (\cite{1.32}%
, p. 304): \textquotedblright Hadamard's well-known example (\ref{1.3}), (%
\ref{1.4}) that yields the solution of Cauchy's problem of the type (\ref%
{1.5}) by no means tells of the absence of continuous dependence on the
initial data, as it is usually interpreted. It rather tells of the fact that
small changes of the initial data may result in leaving the totality of the
initial data for which the solution of Cauchy's problem
exists.\textquotedblright\ 

By the way, R. Richtmyer demonstrated the correctness of the procedure of a
numerical realization of a rather complicated problem of the above-mentioned
type with the representation of sought functions by two-dimensional power
series and with the use of special methods of suppression of errors of
arithmetical operations (\cite{1.33}, section 17.B).

In the course of V. A. Trenogin (\cite{1.34}, p. 225), the following two
theorems are given:

Let $E_{1}$ and $E_{2}$ be infinite dimensional normed spaces, with $E_{2}$
being complete. If $A$ is a completely continuos linear operator from $E_{1}$
into $E_{2}$, different from a finite dimensional one, its range $R\left(
A\right) $ is not a closed manifold in $E_{2}$.

Let $A$ be a completely continuos operator from an infinite dimensional
normed space $E_{1}$ into a normed space $E_{2}$, with the inverse operator $%
A^{-1}$ existing on $R\left( A\right) $. Then $A^{-1}$ is bounded on $%
R\left( A\right) $.

\section{The premises of the realization of the conditions of correctness}

Let us assume that $f=f_{\ast }\left( x\right) $ is an exact result of
integration of the function $\psi \left( x\right) \in L_{2}\left( 0,1\right) 
$ and of the symmetric closed kernel $k\left( x,\xi \right) $ by means of
the formula (\ref{1.6}):%
\begin{equation}
\left( A\psi \right) \left( x\right) =f_{\ast }\left( x\right) ,\quad x\in 
\left[ 0,1\right] .  \label{1.10}
\end{equation}

However, in general, the right-hand side of this equation is actually the
following:%
\begin{equation}
f\left( x\right) =f_{\ast }\left( x\right) -\left( \delta f^{\prime }\right)
\left( x\right) ,  \label{1.11}
\end{equation}%
where $\delta f^{\prime }$ is an admissible error.\footnote{%
Here, the prime is used to match the notation of section 4.5 and thereafter.}
Quite naturally, $f$ does not belong to the range of the operator $A$
defined by the condition (\ref{1.7}). In what follows, the space of
functions for which this condition holds will be denoted $l_{2}^{\prime }$.
In contrast to the usual $l_{2}$, the property of belonging to $%
l_{2}^{\prime }$ depends both on the function $f\left( x\right) $ and on the
operator $A$.

Thus, $l_{2}^{\prime }$ is the space of functions obtained as a result of
integration according to (\ref{1.6}) of the given kernel $k\left( x,\xi
\right) $ and of the whole set $\psi \left( x\right) $ from $L_{2}$. As a
matter of fact, $l_{2}^{\prime }$ and $R\left( A\right) $ coincide in the
considered case. At the same time, the notion of the space $l_{2}^{\prime }$
characterizes to a greater extent the form of the normalizing functional (%
\ref{1.7}). Between $l_{2}^{\prime }$ and such an abstraction as as the
range of the operator $A$ is Picard's theorem that provides the condition of
the solvability of Eq. (\ref{1.6}) \cite{1.20}. Moreover, it may prove to be
useful to compare $l_{2}^{\prime }$ with the space $l_{2}$ whose close
relationship with $L_{2}$ is established by the Riesz-Fischer theorem \cite%
{1.19}.

On the contrary, the free term of Eq. (\ref{1.10}) $f_{\ast }\left( x\right)
\in l_{2}^{\prime }$, and, at the same time the only verification of the
condition (\ref{1.7}) may prove to be infeasible because of the accumulation
of errors of the calculations. Specific \textquotedblright
diffusion\textquotedblright\ of the space $l_{2}^{\prime }$ is caused by the
structure of its normalizing functional. In this sense, the $L_{2}\left(
0,1\right) $ is much more tangible for the function $f\left( x\right) $.
Nevertheless, the use of it incurs rather negative consequences.

Indeed, in this case $R\left( A\right) $ does not belong to the closed space 
$L_{2}\left( 0,1\right) $, the operators $A$ and $A^{-1}$ become,
respectively, completely continuos and unbounded. As a consequence, the
procedure of a numerical realization of equation (\ref{1.6}), in fact, turn
out to be beyond the sphere of the application of Banach's fundamental
theorem on the inverse operator. Isn't it a too high price to pay for
seemingly ephemeral clarity in the formulation of the problem under the
conditions of mapping inside the space $L_{2}$?

We draw attention to a known point of view that a choice of appropriate
spaces for the solutions to problems of mathematical physics should be done
on the basis of practical applications, which can hardly be disputed. As the
same time, a wide-spread opinion that, for example, a sociologist should
formulate a problem to be solved by mathematical methods with a
specification of appropriate spaces for its data. This, as a rule, admits
variety, which is a prerequisite for an increase in the efficiency of
procedures of numerical realization.

Are there any prospects to overcome the above-mentioned complexity in mating
the free term of Eq. (\ref{1.6}) with the adequate space $l_{2}^{\prime }$?
In this regard, let us turn to Eq. (\ref{1.10}) that by virtue of (\ref{1.11}%
) takes the form 
\begin{equation}
\left( A\psi \right) \left( x\right) =f\left( x\right) +\left( \delta
f^{\prime }\right) \left( x\right) ,\quad x\in \left[ 0,1\right] .
\label{1.12}
\end{equation}

However, there is a chance of a reduction of the given function $f\left(
x\right) $ to $f_{\ast }\in l_{2}^{\prime }$ by means of adaptive
simulations of the error $\left( \delta f^{\prime }\right) \left( x\right) $%
. Indeed, it can be interpreted as the smoothing of information by the
procedure of integration. From this point of view, it seems to be reasonable
to represent $\delta f^{\prime }$ as a difference between the explicit form
of the sought function $\psi \left( x\right) $ and an integral over this
function whose kernel would not impose any additional restrictions on the
formulation of the problem. Simultaneously, given that the error of
integration by means of (\ref{1.6}) is objectively small, there appears a
condition of the form 
\begin{equation}
\left\Vert \delta f\right\Vert _{L_{2}\left( 0,1\right) }=0.  \label{1/13}
\end{equation}%
Thus, instead of traditional determination of the function $\psi \left(
x\right) $ by means of the solution of the Fredholm equation of the first
kind (\ref{1.6}), we suggest to employ the perturbation $\delta f^{\prime }$
which leads to the problem (\ref{1.12}), (\ref{1/13}). In this way, a
prerequisite is formed for ensuring $f+\delta f^{\prime }\in R\left(
A\right) $.\footnote{%
A practical realization of the outlined orientation is a key aspect of the
constructive part of the present consideration (see sections 4.5, 5, 6).} As
will be shown, by use of some additional considerations, the determination
of the function $\psi \left( x\right) $ that satisfies Eq. (\ref{1.6}) can
be reduced to the solution of a well-posed problem.

Note that, in the case of a considerable mismatch between $R\left( A\right) $
and the actually known function $f\left( x\right) $, condition (\ref{1/13})
can hardly be regarded as feasible. Nevertheless, the outlined approach
still applies interpreting, figuratively, the reduction of the free term of
Eq. (\ref{1.6}) to a form which makes it solvable.

\chapter{The existing approaches to the solution of ill-posed problems}

\section{A. N. Tikhonov's methodology}

The consideration of this subsection is based on the material of the
monograph by A. N. Tikhonov and V. Y. Arsenin \cite{2.1} that is, literally,
pierced by the concept of the adequacy of incorrect formulations and, in
particular, of integral equations of the first kind to problems of
mathematical physics. As an illustration, we show that the solution to the
Fredholm integral equation of the first kind 
\begin{equation}
\left( A\psi \right) \left( x\right) \equiv \int\limits_{a}^{b}k\left( x,\xi
\right) \psi \left( \xi \right) d\xi =f\left( x\right) ,\quad x\in \left[ a,b%
\right] ,  \label{2.1}
\end{equation}%
with $k\left( x,\xi \right) $ and $\partial _{x}k\left( x,\xi \right) $
being continuous with respect to $x$, can undergo arbitrarily considerable
changes both in the metric $C$ and $L_{2}$ for small in $L_{2}\left(
a,b\right) $ variations of the right-hand side in the form $\varepsilon \sin
\left( \omega \xi \right) $.

The situation with the perturbation of the kernel $k\left( x,\xi \right) $
is, in fact, analogous. In this regard, the authors pose the question: What
should be understood by the solution of Eq. (\ref{2.1}) when $k$ and $f$ are
known approximately? In their opinion, a problem of this type should be
considered \textquotedblright underspecified\textquotedblright , and,
correspondingly, a choice of possible solutions should be made taking into
account \textquotedblright usually available" additional qualitative or
quantitative information about the function $\psi \left( x\right) $. In this
regard, we draw attention to N. G. Preobrazhenskii's considerations
concerning a system of linear algebraic equations, obtained by the
discretization of (\ref{2.1}) (\cite{2.2}, p. 130):

''An analysis shows that choosing sufficiently high order of an
approximation, we transform [the above-mentioned problem] into an
arbitrarily ill-defined one... Under these conditions, it is necessary to
add to the algorithm some a priori nontrivial information, only by the use
of which we can expect to filter out veiling false variants and single out
the solution, closest to the sought one. any purely mathematical tricks that
do not employ additional a priori data are equivalent to an attempt to
construct an informational perpetuum mobile producing information from
nothing.''

The so-called method of the selection of the solution to ill-posed problems
is based on a priori quantitative information. It is shown that if a
compactum $M$ of a metric space $E_{1}$ is mapped in a one-to-one and
continuous manner onto a set $F$ of a metric space $E_{2}$, the inverse map $%
F$ onto $M$ is also continuous. Correspondingly, an assumption that the
solution, in particular, to Eq. (\ref{2.1}) belongs to the compactum $M$
allows us to consider the operator $A^{-1}$ to be continuous on the set $%
F=AM $.

A practical realization is reduced to an approximation of $M$ by a series
with parameters that change within certain limits (for $M$ to represent a
closed set of a finite dimensional space) and should be determined from the
condition of the minimum of the error of closure of (\ref{2.1}). Note the
absence of any more or less general recommendation with respect to the
choice of $M$.

In light of the above, M. M. Lavrentiev has formulated the notion of
correctness according to Tikhonov for an equation of the type (\ref{2.1}) 
\cite{2.3}:

1) It is a priori known that the solution $\psi _{*}$ to the considered
equation exists and belongs to a set $M$ of the space $B_1$.

2) The solution $\psi _{*}$ is unique on the set $M$.

3) The operator $A^{-1}$ is continuous on the set $AM$ of the space $B_2$.

If $M$ is a compactum (this case is called ''usual'') the last condition
becomes a consequence of the first two conditions.

Those problems in which the operator $A^{-1}$ is unbounded on the set $%
AE_{1} $ and the set of possible solutions $E_{1}$ is not a compactum are
called substantially ill-posed. For such problems, Tikhonov has put forward
an idea of a regularizing operator $G$, in a sense close to $A^{-1}$, whose
value domain for the map from $E_{2}$ into $E_{1}$ admits matching to the
right-hand side of (\ref{2.1}), known approximately. Moreover, $G$ must
contain a regularization parameter $\alpha $ that depends on the accuracy of
the initial information.

The operator $G\left( f,\alpha \right) $ is called a regularizing operator
for Eq. (\ref{2.1}) if it possesses the following properties:

1) It is defined for any $\alpha >0$ and $f\in E_2$.

2) For $A\psi _{*}=f_{*}$, where $\psi _{*}$ and $f_{*}$ are corresponding
exact expressions, there exists such $\alpha \left( \delta \right) $ that
for any $0<\epsilon \leq \rho _{E_1}\left( \psi _{_{*}},\psi _\alpha \right) 
$ there is $\delta \left( \epsilon \right) \geq \rho _{E_2}\left( \psi
_{_{*}},\psi _\alpha \right) $. Here, $\psi _\alpha =G\left( f,\alpha
\right) $.

It is implied that there is a possibility of a choice of $\alpha \left(
\delta \right) $ such that for $\delta \rightarrow 0$ the regularized
solution $\psi _{\alpha }\rightarrow \psi _{\ast }$, i.e., $\epsilon
\rightarrow 0$. At the same time, it is pointed out that the construction of
the dependence $\alpha \left( \delta \right) $, for which the operator $%
G\left( f,\alpha \left( \delta \right) \right) $ is a regularizing one, is
algorithmically complicated for classes of practically important problems.
There are a lot of publications of Tikhonov's followers devoted to the
resolution of this difficulties, which will be discussed below.

Namely in \cite{2.1}, the construction of $G\left( f,\alpha \right) $ is
carried out by the use of techniques of calculus of variations that reduce
the evaluation of $\psi \left( x\right) $ to the minimization of the
functional 
\begin{equation}
\Phi ^{\alpha }\left[ f,\psi \right] =\rho _{E_{1}}^{2}\left( A\psi
,f\right) ^{2}+\alpha \Omega \left[ \psi \right] .  \label{2.2}
\end{equation}

For Eq. (\ref{2.1}), its stabilizing component is recommended to be taken in
the form 
\begin{equation}
\Omega \left[ \psi \right] =\int\limits_{a}^{b}\left\{ p_{0}\left( x\right)
\psi ^{2}\left( x\right) +p_{1}\left( x\right) \left[ \partial _{x}\psi
\left( x\right) \right] ^{2}\right\} dx,  \label{2.3}
\end{equation}%
where $p_{0}$, $p_{1}\geq 0$ are given functions.

In the case of a symmetric kernel $k\left( x,\xi \right) $, the procedure of
the minimization of (\ref{2.2}) is equivalent to the solution of the
integrodifferential equation 
\begin{equation}
\alpha \left\{ p_{0}\left( x\right) \psi \left( x\right) -\partial _{x}\left[
p_{1}\left( x\right) \partial _{x}\psi \left( x\right) \right] \right\}
+\left( A\psi \right) \left( x\right) =f\left( x\right) ,\quad x\in \left[
a,b\right] ,  \label{2.4}
\end{equation}%
under the conditions 
\begin{equation}
\left. p_{1}\left( x\right) \partial _{x}\psi \left( x\right) \upsilon
\left( x\right) \right\vert _{a}^{b}=0.  \label{2.5}
\end{equation}

Here, $\upsilon \left( x\right) $ is an arbitrary variation of $\psi \left(
x\right) $ in the class of admissible functions.

In the opinion of the authors of \cite{2.4}, an overwhelming majority of
inverse problems are ill-posed, and attempts to solve them, in view of their
great practical importance, were being undertaken for a long period.
\textquotedblright But only as a result... of the appearance of fundamental
publications of academician A. N. Tikhonov, the modern theory of the
solution of inverse problems, based on the notion of a regularizing
algorithm, was constructed\textquotedblright\ (p. 7). In what follows, the
authors construct the procedure of a numerical realization of the Fredholm
integral equations of the first kind, related to the interpretation of
astrophysical observations, by means of the selection of the compactum of
possible solutions in the class of monotonically bounded functions.

As is pointed out by O. A. Liskovets \cite{2.5}, \textquotedblright ...the
correctness according to Tikhonov is achieved at the expense of the
reduction of the admissible manifold of solutions to the class of
correctness\textquotedblright\ (p. 13). The following quotation from the
above-mentioned monograph is also of interest: \textquotedblright In
contrast to a previously prevailing opinion that all the problems describing
physical reality are ill-posed, according to the modern point of view any
realistic problem can be regularized, i.e., it has at least one
regularizer\textquotedblright\ (p. 14).

Here is V. A. Morozov's conclusion (\cite{2.6}, p. 9): \textquotedblright A.
N. Tikhonov's method of regularization turned out to be simple in practice,
because it did not require actual knowledge of the compactum $M$ that
contained the sought solution to Eq. (\ref{2.1})... The main difficulty of
the application of this method consists in the formulation of algorithmic
principles of the selection of the parameter of regularization $\alpha $%
\textquotedblright . According to his own monograph (\cite{2.7}, p. 4),
\textquotedblright The importance of A. N. Tikhonov's paper \cite{2.8} can
hardly be overestimated. It served as impetus for a number of publications
by other investigators in different fields of mathematical analysis and
natural science: spectroscopy, electron microscopy, identification and
automatic regulation, gravimetry, optics, nuclear physics, plasma physics,
meteorology, automation of scientific research and some other spheres of
science and engineering\textquotedblright .

V. V. Voevodin's opinion (\cite{2.9}, p. 43) is as follows:
\textquotedblright The success of the application of the regularization
method to the solutions of unstable systems of algebraic equations is
explained to a large extent by the fact that A. N. Tikhonov and his
followers did not restrict themselves to an investigation of separate
fragments of this complicated problem but considered the whole complex
related issues. This, in the first place, concerns a clear formulation of
the problem itself, the construction of a stable with respect to
perturbation of the input data algorithm of its solution, the development of
an efficient numerical method, estimates of a deviation of the actually
evaluated object from the sought one taking into account a perturbation of
the input data and errors of rounding\textquotedblright .

A quotation from the preface to the collected volume by A. A. Samarsky and
A. G. Sveshnikov \cite{2.9} reads: \textquotedblright A clarification of
Andrey Nikolaevich Tikhonov of the role of ill-posed problems in classic
mathematics and its applications (inverse problems) is of fundamental
importance for the who;e modern mathematics. He proposed a principally new
approach to this class of problems and developed methods of the construction
of their stable solutions based on the principle of
regularization\textquotedblright .

\section{A brief review of the development of the outlined concepts}

The results of investigations devoted to the determination of the
regularization parameter $\alpha $ are summarized in \cite{2.10}. Based on
the assumption that errors in the determination of the free term $f\left(
x\right) $ and the kernel $k\left( x,\xi \right) $ of Eq. (\ref{2.1}) are
known, one uses different methods of the minimization of the error of
closure of the type 
\[
\left\Vert \tilde{A}\psi _{\alpha }-\tilde{f}\right\Vert _{F}=\mu \left\Vert
\delta \tilde{f}\right\Vert _{F},\quad \mu \in \left( 0,1\right) . 
\]%
The evaluation of the parameter $\alpha $ as a root of the corresponding
equation does not pose any problem. However, a choice of $\mu $ is, in fact,
related to considerable uncertainty. The main obstacle is that a reliable
estimate of the error stipulated by the "measure of incompatibility" of the
concretely considered equation $\tilde{A}\psi =\tilde{f}$ is rather
questionable.

Considerable efforts were undertaken to reduce the volume of information
necessary for the evaluation of the parameter $\alpha $. A noticeable step
in this direction was made by A. N. Tikhonov and V. B. Glasko who suggested
a criterion of the minimization of the functional $\left\Vert \alpha d\psi
_{\alpha }/d\alpha \right\Vert $ with respect to $\alpha >0$ \cite{2.11}
(see also \cite{2.1}, section 2.7). However, its theoretical justification
proved to be possible only for rather narrow classes of problems. A number
of methods of the determination of $\alpha $ is related to the use of
solutions to Eq. (\ref{2.1}) for a special form of the functions $f\left(
x\right) $.

In \cite{2.10}, the status of the studies of estimates of the accuracy of
methods of the solution of the integral equation (\ref{2.1}) is also
illuminated. If $\psi \left( x\right) $ belongs to a compactum, any serious
complications, as a rule, do not arise, and the main interest is focused on
the algorithm of regularization. If $p_{1}\equiv 0$ in (\ref{2.3}) and the
parameter $\alpha $ is finite, Eq. (\ref{2.4}) becomes a Fredholm integral
equation of the second kind, to which, under the assumption that the error
in the determination of $k\left( x,\xi \right) $ and $f\left( x\right) $ is
known, the whole general theory of approximate methods of L. V. Kantorovich
applies (\cite{2.12}, section 14.1).\footnote{%
Note that the definition of the type of the equation, i.e. "the second
kind", in this case, because of the presence of $\alpha $, is purely formal.
This important issue will be repeatedly discussed in what follows.}

At the same time, as shown by V. A.Vinokurov \cite{2.13}, when a priori
information about the solution to Eq. (\ref{2.1}) is missing, the estimate
of the error of the evaluation of $\psi \left( x\right) $ by means of
regularization is impossible in principle. Justified is only a formulation
of the question of the convergence of the procedure of computation or of a
possibility of the regularization of the corresponding problem.

In this regard, we note the arguments of A. B. Bakushinskii and A. V.
Goncharskii (\cite{2.14}, p. 13): \textquotedblright Unfortunately, in the
general case, it is impossible to estimate the measure of closeness of $%
G\left( f,\alpha \right) $ to $A^{-1}\left( f_{\ast }\right) $ without
additional information about the solution to Eq. (\ref{2.1}). This is a
characteristic feature of ill-posed problems. In the general case, a
regularization algorithm ensures only asymptotic convergence of an
approximate solution to the exact one for $\delta \rightarrow 0$%
\textquotedblright .

The name of M. M. Lavrentiev is associated with a particular case of a
practical realization of A. N. Tikhonov's method consisting in the reduction
of the problem (\ref{2.4}), (\ref{2.5}) to the solution of the Fredholm
integral equation of the second kind 
\begin{equation}
\alpha \psi \left( x\right) +\int\limits_{a}^{b}k\left( x,\xi \right) \psi
\left( \xi \right) d\xi =f\left( x\right) ,\quad x\in \left[ a,b\right] ,
\label{2.6}
\end{equation}%
where $\alpha >0$ is a small parameter.

It is shown that $\left\| \psi _\alpha -\psi _{*}\right\| \rightarrow 0$ for 
$\delta \rightarrow 0$, $\gamma \rightarrow 0$ and $\left( \delta +\gamma
\right) /\alpha \left( \delta ,\gamma \right) \rightarrow 0$. Here, $\gamma $
is an error in the determination of the kernel $k\left( x,\xi \right) $,
analogous to $\delta $ (see section 3.1).

V. K. Ivanov's method \cite{2.15} allows one to find the so-called
quasisolution minimizing the error of closure of (\ref{2.1}) for a class of
functions $\psi \left( x\right) \in M_{R}$, where $M_{R}\in E_{1}$ is a
compactum. The quasisolution to (\ref{2.1}) on such a compactum has the form 
\begin{equation}
\psi \left( x\right) =\sum_{n=1}^{\infty }c_{n}\left( \lambda +\lambda
_{n}\right) \bar{\psi}_{n}\left( x\right) ,\quad x\in \left[ a,b\right] .
\label{2.7}
\end{equation}%
Here 
\[
c_{n}=\int\limits_{a}^{b}f\left( x\right) \bar{\psi}_{n}\left( x\right) dx; 
\]%
$\lambda _{n}$ and $\bar{\psi}_{n}\left( x\right) $ are, respectively,
characteristic numbers and eigenfunctions of the kernel $k\left( x,\xi
\right) $; the parameter $\lambda =0$ and represents a positive root of the
equation 
\begin{equation}
\sum_{n=1}^{\infty }\left( \frac{c_{n}\lambda \lambda _{n}}{\lambda +\lambda
_{n}}\right) ^{2}=R^{2}  \label{2.8}
\end{equation}%
under the conditions, respectively, 
\begin{equation}
\sum_{n=1}^{\infty }c_{n}^{2}\lambda _{n}^{2}\leq R^{2};\quad
\sum_{n=1}^{\infty }c_{n}^{2}\lambda _{n}^{2}>R^{2}.  \label{2.9}
\end{equation}

Special methods of regularization are developed for the situations when
considerable volume of information of statistical character (spectral
densities, mathematical expectations, etc.) about the solution to an
equation of the type (\ref{2.1}) is available. Thus, V. N. Vapnik \cite{2.16}
rather constructively employed the specifics of problems concerned with
recognition of images, related to nonuniqueness and, as a result, to extreme
behavior of the sought functions. We point out a definition in the
above-mentioned monograph (p. 8) that, apparently, was implied by many
authors but did not receive such a clear formulation:

''The problem of the restoration of dependencies from empirical data was
and, probably, will always be central in applied analysis. This problem is
nothing but mathematical interpretation of one of the main problems of
natural science: How to find the existent regularity from random facts.''

\section{V. M. Fridman's approach}

Let $k\left( x,\xi \right) $ be symmetric, positive definite kernel and Eq. (%
\ref{2.1}) be solvable. Then, as shown by V. M. Fridman \cite{2.17}, a
sequence of functions determined by iteration 
\begin{equation}
\psi _{n+1}\left( x\right) =\psi _{n}\left( x\right) +\lambda \left[ f\left(
x\right) -\int\limits_{a}^{b}k\left( x,\xi \right) \psi _{n}\left( \xi
\right) d\xi \right] ,\quad n=0,1,\ldots ,  \label{2.10}
\end{equation}%
converges in $L_{2}\left( a,b\right) $ to the solution of Eq. (\ref{2.1})
for an arbitrary choice of the initial approximation $\psi _{0}\left(
x\right) \in L_{2}\left( a,b\right) $ and $0<\lambda <2\lambda _{1}$, where $%
\lambda _{1}$ is the smallest characteristic number of the kernel $k\left(
x,\xi \right) $.

M. A. Krasnoselskii \cite{2.18} extended this result to an arbitrary
solvable equation of the type (\ref{2.1}) with a linear bounded operator $A$
in a Hilbert space $H$. A theorem on the convergence of successive
approximations 
\begin{equation}  \label{2.11}
\psi _{n+1}=\left( I-\nu A_1\right) \psi _n+\nu f_1
\end{equation}
to the solution is proved. Here, $A_1=A^{*}A$; $f_1\left( x\right) =\left(
A^{*}f\right) \left( x\right) $; $I$ is the identity operator; $A^{*}$ is
the conjugate operator to $A$; $0<\nu <2/\left\| A_1\right\| $; $\psi
_0\left( x\right) \in H$.

Note that in the case of the integral operator (\ref{2.1}) 
\[
A_{1\bullet }=\int\limits_a^bk_1\left( x,\xi \right) _{\bullet }d\xi , 
\]
where 
\[
k_1\left( x,\xi \right) =\int\limits_a^bk\left( \zeta ,x\right) k\left(
\zeta ,\xi \right) d\zeta . 
\]

A number of procedures are known that improve convergence of iterations
according to Fridman (see \cite{2.10}). For example, under the conditions
that are specified with respect to the procedure (\ref{2.10}), 
\begin{equation}  \label{2.12}
\psi _{n+1}\left( x\right) =\frac 1{m+1}\sum_{n=0}^m\varphi _n\left(
x\right) ,
\end{equation}
where $\varphi _0\left( x\right) \in L_2\left( a,b\right) $; 
\[
\varphi _n\left( x\right) =\varphi _{n-1}\left( x\right) +f\left( x\right)
-\int\limits_a^bk\left( x,\xi \right) \varphi _{n-1}\left( \xi \right) d\xi
,\quad n=1,2,\ldots . 
\]

G. M. Vainikko and A. Y. Veretennikov \cite{2.19} studied an iteration
algorithm of an implicit type: 
\begin{equation}  \label{2.13}
\alpha \psi _{n+1}\left( x\right) +\int\limits_a^bk\left( x,\xi \right) \psi
_{n+1}\left( \xi \right) d\xi =\alpha \psi _n\left( x\right) +f\left(
x\right) ,\quad n=0,1,\ldots ,
\end{equation}
where $\psi _0\left( x\right) \in L_2\left( a,b\right) $; the parameter $%
\alpha $$>0$.

Note that in contrast to the regularization of the type (\ref{2.6}), based
on the smallness of $\alpha $, the considered approach is characterized by
multiple iteration with, on the contrary, sufficiently large value of this
parameter. Moreover, one of the merits of the procedures (\ref{2.10})-(\ref%
{2.13}) is a possibility of a constructive application of an a posteriori
estimate of the error to accomplish the iteration.

In the simplest case, one finds the number $n$ for which for the first time 
\[
\left\Vert \psi _{n+1}-\psi _{n}\right\Vert _{L_{2}\left( a,b\right) }\leq
c_{1}\delta +c_{2}\gamma , 
\]%
where $\delta $ and $\gamma $ are errors in the determination of $f\left(
x\right) $ and $k\left( x,\xi \right) $, respectively; $c_{1}$, $c_{2}$ are
constants meeting a number of requirements to ensure the stability of the
procedures of computation. The influence of errors, small in a probabilistic
sense, on the convergence of successive approximations is also investigated.

The authors of \cite{2.20} gave arguments for usefulness of the combination
of the regularization of the equation of the type (\ref{2.1}), whose
parameter is the number of iterations, with algorithms of the saddle-point
type. This approach has its origin in the publication by V. M. Fridman \cite%
{2.21} and is realized, in particular, according to the scheme 
\begin{equation}
\psi _{n+1}=\psi _{n}-\beta _{n}A^{\ast }\left( A\psi _{n}-f\right) ,
\label{2.14}
\end{equation}%
where 
\[
\beta _{n}=\frac{\left\Vert A^{\ast }\left( A\psi _{n}-f\right) \right\Vert
^{2}}{\left\Vert AA^{\ast }\left( A\psi _{n}-f\right) \right\Vert ^{2}}, 
\]%
which is adequate to the choice of the step of the descent from the
condition of the minimum of the error of closure 
\[
\Delta _{n+1}=\left\Vert A\psi _{n+1}-f\right\Vert _{L_{2}\left( a,b\right)
}. 
\]

\section{Inverse problems for differential equations of mathematical physics}

The monograph by O. M. Alifanov, E. A. Artyukhin and S. V. Rumyantsev \cite%
{2.20} reflects established approaches in this field. In the procedure of
mathematical formulation of the problems, structural and parametric
identification is emphasized, which implies, respectively, a qualitative
description of the considered processes by means of differential operators
and allotting quantitative information to the model.

Interpretation of physical processes in terms of causality is also given.
The cause includes boundary and initial conditions with their parameters,
coefficients of the differential equations and also the domain of the
problem. The effect reflects the status of the investigated object and
represents, mostly, fields of physical quantities of different types.

The restoration of the cause from the information about physical fields is
considered as an inverse problem. A key consideration is as follows (p. 11):
\textquotedblright A violation of a natural causal relation that takes place
in the formulation of the inverse problem can lead to its mathematical
incorrectness, such as, in most cases, instability of the solution.
Therefore, inverse problems constitute a typical example of ill-posed
problems\textquotedblright .

In connection with the sought function, the following types of inverse
problems of the identification of physical processes for partial
differential equations are singled out:

1) Retrospective problems: the determination of the prehistory of a certain
state of the problem.

2) Boundary problems: the restoration of boundary conditions or of the
parameters contained therein.

3) Coefficient problems: the restoration of the coefficients of the
equations.

4) Geometrical problems: the determination of geometrical characteristics of
the contour of the domain or of the coordinates of points inside.

A principal difference between inverse problems of identification and those
of regulation is pointed out, concerning the width of classes of possible
solutions. Whereas in the former case their increase leads to complications
in the numerical realization, in the latter case, on the contrary, this is a
favorable factor. By the way, the algorithmic means \cite{2.20} are almost
completely based on the methods of the solution of integral equations of the
first kind, to which the considered problems of heat exchange are reduced.

In the formulation of inverse problems of mathematical physics, the proof of
corresponding theorems of existence and uniqueness is of primary importance.
In this regard, a general approach, outlined schematically by A. L. Buchgeim
(\cite{2.22}, pp.133-134) can be mentioned. Thus, the following equations
are considered: 
\begin{equation}
Pu=f;\quad Qf=g,  \label{2.15}
\end{equation}%
where $P$ is an operator of the direct problem; $Q$ is an \textquotedblright
information\textquotedblright\ operator describing the law of the change of
the right-hand side; $g$ is given, whereas $u$ and $f$ are the sought
elements of corresponding functional spaces.

The application of the operator $Q$ to the first equation (\ref{2.15})
yields $QPu=g$, which is equivalent to 
\[
PQu=\left[ P,Q\right] u+g, 
\]%
where $\left[ P,Q\right] =PQ-QP$ is the commutator of the operators $P$ and $%
Q$.

The meaning of the commutation lies in the fact that, as a rule, there is no
information, except for (\ref{2.15}), about the function $f$. Therefore, it
is easier to study the operator on the solution of the direct problem $u$
that satisfies some manifold of boundary conditions. It is important that in
typical applications the operator $Q$ does not \textquotedblright
spoil\textquotedblright\ the part of boundary conditions that reflects the
domain of the operator $P$. As a result, one gets a specific factorization
of the inverse problem (\ref{2.15}) as a product of two direct problems,
induced by the operators $P$ and $Q$ under the condition that the commutator
is, in a sense, \textquotedblright subordinate\textquotedblright\ to them.

In the trivial case $\left[ P,Q\right] =0$, the initial problem decomposes
into two simpler ones: $Pv=g$; $Qu=v$. For the description of properties of
the employed operators, a priori estimates are used.

Of interest is also a quotation from the introduction to the monograph by R.
Lattes and J.-L. Lions \cite{2.23}: \textquotedblright In this book, we
suggest a method of quasiinversion, intended for the numerical solution of
some classes of ill-posed, according to Hadamard, boundary value problems.
Practical and theoretical importance of such problems is being more and more
realized by investigators\textquotedblright . And further: "The main idea of
the method of quasiinversion (universal in numerical analysis!) consists in
an appropriate change of operators entering the problem. This change is done
by the introduction of additional differential terms that are

i) sufficiently ''small'' (they can be set equal to zero);

ii) ''degenerate on the boundary'' (to prevent, for example, the appearance
of complicated boundary conditions and of such conditions that may contain
unknown, sought variables)''.

In particular, the ill-posed problem of thermal conductivity 
\begin{equation}
\partial _{t}u-\partial _{x}^{2}u=0,  \label{2.16}
\end{equation}%
\[
u\left( 0,t\right) =u\left( 1,t\right) =0;\quad u\left( x,T\right) =\zeta
\left( x\right) , 
\]%
where $\zeta \left( x\right) $ is an unknown function, is replaced by the
following, with a small parameter $\epsilon $: 
\begin{equation}
\partial _{t}u-\partial _{x}^{2}u-\epsilon \partial _{x}^{4}u=0;
\label{2.17}
\end{equation}%
\[
u=\partial _{x}^{2}u=0,\quad x=0;\quad x=1;\quad u\left( x,T\right) =\zeta
\left( x\right) . 
\]

The authors point out (p. 36): ''In a numerical realization, it is natural
to choose $\epsilon $ as the smallest possible one. However, in problems of
the considered type, one should expect numerical instability for $\epsilon
\rightarrow 0$. Therefore one can expect at most that for any problem there
exists a certain optimal value of $\epsilon $ equal to $\epsilon _0$''. The
absence of convergence ''in a usual sense'' of the solution of the problem (%
\ref{2.17}) to the exact one for $\epsilon \rightarrow 0$ was pointed out by
A. N. Tikhonov and V. Y. Arsenin (\cite{2.1}, p. 52).

\section{Alternative viewpoints and developments}

In Y. I. Liubich's opinion, any more or less general theory of integral
equations of the first kind is absent, and only in some cases it is possible
to use special methods. An example is given by known Abel's equation (\cite%
{2.24}, p. 83).

K. I. Babenko's remark (\cite{2.25}, p. 310) is rather typical:
\textquotedblright Although from the point of view of the loss of
information algorithms are not estimated, it seems to us that this is an
important characteristic and it should be taken into
account\textquotedblright . In what follows, the lack of optimality of the
traditional approach to a numerical realization of ill-posed problems is
concretely demonstrated.

A profound analysis of methodological aspects of this sphere is given by R.
P. Fedorenko (\cite{2.26}, sections 40, 41). In particular, he failed to
establish the value of the regularization parameter $\alpha $ by minimizing
the functional (\ref{2.2}), because for small values the sought function
began to oscillate, whereas with it increase the value of $\Phi ^{\alpha }$
considerably exceeded the admissible one. The author arrived at the
conclusion that reason lay in the inadequacy of the theory \cite{2.1} to
problems of control, characterized by discontinuity of solutions.

By studying the problem (\ref{2.16}), Fedorenko brought up the following
consideration: \textquotedblright All the methods of the solution of
ill-posed problems more or less consist in preventing the appearance in the
sought solution of higher harmonics with large or even simply finite
coefficients. But what is \textquotedblright high
frequency\textquotedblright ? Beginning with what number $n$ should we
consider the function $\sin \left( n\pi x\right) $ redundant, only spoiling
the solution? This, of cause, depends on $T$\textquotedblright . It is
implied that a hypothetically known solution of the corresponding direct
problem can be expanded into a Fourier series 
\[
u\left( x\right) =u\left( x,0\right) =\sum_{n=1}^{\infty }\sin \left( n\pi
x\right) . 
\]

It is shown that the use in \cite{2.23} of the value $T=0.1$ and the errors
in $L_{2}\left( 0,1\right) $ of the satisfaction of the last condition (\ref%
{2.16}), with $\delta $ of order $10^{-3}$, imposes the restriction $n=2$.
In this context, the method of P. Lattes and G.-L. Lions came under
criticism. These authors, while solving the problem (\ref{2.17}) on a grid
with a step of $\Delta x=0.02$, obtained an absolutely unacceptable
component $u_{0}$, namely, $10^{8}\sin \left( 6\pi x\right) $. This occurred
for $\delta $ at the level of $0.05$, under the conditions when $\left\vert
\zeta \left( x\right) \right\vert \leq 1\ldots $ .

Note also the remark (\cite{2.26}, p. 360) that, aside the fact of the
boundedness of the regularizing operator $G$ (see section 3.1), its norm $%
\left\Vert G\right\Vert $ is an exceptionally important characteristic whose
value directly influences a relation between the accuracy of the given
function $\zeta $ and the solution $u_{0}=G\zeta $.\footnote{%
By the way, in most specialized publication this issue is not accentuated.}

Indeed, let us consider Eq. (\ref{2.6}), written in the canonical form 
\begin{equation}  \label{2.18}
\psi \left( x\right) =-\frac 1\alpha \int\limits_a^bk\left( x,\xi \right)
\psi \left( \xi \right) d\xi +\frac 1\alpha f\left( x\right) ,\quad x\in 
\left[ a,b\right] .
\end{equation}

Let $a=0$, $b=1$, and the kernel $k\left( x,\xi \right) $ be determined by
the expression (\ref{1.8}). In this case, for $\alpha ^{-1}\neq \left( n\pi
\right) ^{2}$, its solution is \cite{2.27}: 
\[
\psi \left( x\right) =\frac{1}{\alpha }f\left( x\right) -\frac{1}{\alpha }%
\sum_{n=1}^{\infty }\frac{c_{n}}{1+\alpha \left( n\pi \right) ^{2}}\sin
\left( n\pi x\right) , 
\]
\[
c_{n}=\int\limits_{0}^{1}f\left( x\right) \sin \left( n\pi x\right) dx. 
\]

It is not difficult to notice that for small values of $\alpha $ the error
in the determination of the function $f\left( x\right) $ can considerably
distort $\psi \left( x\right) $ [see also a footnote concerning the solution
of Eq. (\ref{2.18}) in section 3.2].

In a constructive aspect, Fedorenko recommends to use traditional
formulations of inverse problems of differential or variational character
with an application of additional conditions that rationally restrict
classes of possible solutions. As the main factor to achieve the desired
efficiency, a comprehensive analysis of qualitative peculiarities of
solutions to the considered problems, involving elements of numerical
simulations, is suggested.

What are the values of the regularization parameter $\alpha $, typical of
computational practice? The authors of \cite{2.28} point out that for
problems of restoration of time-dependent density of thermal flux on the
surface from the results of temperature measurements at internal points of
the samples the corresponding range is rather representative: $%
10^{-7}-10^{-4}$. The editors of the above-mentioned book have a different
point of view: ''One can give a lot of examples of solutions to inverse
problems thermal conductivity, when the range of acceptable values of $%
\alpha $ turn out to be rather narrow'' (p. 141).

The main technique of a numerical realization \cite{2.28} is interpreted by
the authors as a complement to the method of least squares by a procedure
that smooths oscillations of the solutions in high order approximations. In
this regard, they point out a relationship between Tikhonov's regularization
and algorithms of singular expansions and ridge regression (or damping) that
are widely used for the suppression of the instability of the method of
least squares \cite{2.29}.

In a number of publications, one can see an orientation towards
regularization of Eq. (\ref{2.1}) without the distortion of the original
operator along the lines of (\ref{2.4}) or (\ref{2.6}). Thus, A. P. Petrov 
\cite{2.30} suggested a formulation of the problem with $f\left( x\right)
\in R\left( A\right) $ by means of the representation $f=A\psi +\tilde{\omega%
}$, where $\tilde{\omega}$ is a random process reflecting errors of the data
and of the calculations. At the same time, the author failed to use his
formally achieved correctness to construct an efficient algorithm of a
numerical realization. It seems that the reason lies in the insufficiency of
the structure of $\tilde{\omega}$ from the point of view of adaptive
compensation of the error of closure of the satisfaction of (\ref{2.1}).

A. V. Khovanskii \cite{2.31} put forward arguments for the regularization of
the algorithm of the solution of Eq. (\ref{2.1}), not the operator $A$
(which is the basis of the theory of \cite{2.1}). The following quotation is
of interest: \textquotedblright What is more, Tikhonov's regularization
contains in an inseparable form two completely different notions, accuracy
and stability, and there is a transformation of one into another.
Nevertheless, there exists for a long time an idea of the predetermination
of the operator \cite{2.32}, although only in the context of conjugate
gradients and in a multiplicative form\textquotedblright .

However, the method of conjugate gradients is, in fact, Fridman's iterations
of the type (\ref{2.14}). Note that nonlinearities contained therein
facilitate the smoothing of a well-known slow-down of the convergence of the
procedure (\ref{2.10}) with approaching the solution to Eq. (\ref{2.1}).
This effect was demonstrated by A. D. Myshkis \cite{2.33} with the help of
the representation of the components of (\ref{2.10}) by series in terms of
the eigenfunctions of the kernel $k\left( x,\xi \right) $. This leads to the
relations 
\[
c_{n+1,m}=\left( 1-\lambda /\lambda _{m}\right) c_{n,m}+\lambda f_{m},\quad
m=1,2,\ldots , 
\]%
where $c_{n,m}$ and $f_{m}$ are coefficients of the above-mentioned
expansion of $\psi _{n}\left( x\right) $ and $f\left( x\right) $,
respectively.

When the number of the terms in the representation of the solution
increases, which seemingly had to improve the accuracy, the coefficient of
convergence $1-\lambda /\lambda _{m}$ approaches unity and, as a result of
the accumulation of errors, the iterations become "counterproductive".

Note an effective method of the suppression of instability of the procedure
of a numerical realization of the Fredholm integral equation of the second
kind 
\begin{equation}
\psi \left( x\right) =\lambda \int\limits_{a}^{b}k\left( x,\xi \right) \psi
\left( \xi \right) d\xi +f\left( x\right) ,\quad x\in \left[ a,b\right] ,
\label{2.19}
\end{equation}%
\textquotedblright positioned on the spectrum\textquotedblright , i.e., in
the case when $\lambda =\lambda _{n}$ with $\lambda _{n}$ being a
characteristic number, proposed by P. I. Perlin (\cite{2.34}, pp. 105-107).

This problem is ill-posed both with respect to the uniqueness of the
solution and as a result of the degeneracy of the system of linear algebraic
equations obtained by discretization. Nevertheless, a perturbation of the
right-hand side of $f\left( x\right) $ by a zero (within the limits of the
accuracy of calculations) component 
\[
-\bar{\psi}_{n}^{\prime }\left( x\right) \int\limits_{a}^{b}f\left( x\right) 
\bar{\psi}_{n}^{\prime }\left( x\right) dx, 
\]%
where $\bar{\psi}_{n}^{\prime }\left( x\right) $ is a normalized
eigenfunction of the kernel conjugate to $k\left( \xi ,x\right) $, allows
one to improve radically the situation.

The essence lies in the fact that, theoretically, the solution to Eq. (\ref%
{2.19}) is expanded in a power series of $\lambda $. Provided that
computational procedures, matching this situation, are identical, one can
compensate for the errors.

\section{A comparison between the main concepts of A. N. Tikhonov and V. M.
Fridman}

A. N. Tikhonov's original suggestion (1943) admitting of the consideration
of ill-posed problems by an a priori restriction on the class of possible
solutions is a kind of refraction of general methodology of investigations
of the issues of existence and uniqueness into the sphere of numerical
analysis. Note that A. N. Tikhonov's proof of the well-known theorem on the
uniqueness of the solution of the inverse problem of thermal conductivity in
an infinite $n$-dimensional domain under an additional condition of the type 
$\left\vert \partial _{x}^{n}u\right\vert \leq M$ dates back to 1935. A
vivid illustration of these considerations is provided by the algorithm of
the search for a quasisolution (\ref{2.7})-(\ref{2.9}).

Behind A. N. Tikhonov's method of regularization (1963), there is a global
idea of a limiting transition to the exact solution with respect to a small
parameter of the problem, which is unambiguously pointed out in (\cite{2.1},
p. 56): ''Note that regularizing operators, dependent on a parameter, have
been employed in mathematics since Newton's times. Thus, the classic problem
of an approximate calculation of the derivative $u^{\prime }\left( x\right) $
by means of approximate (in the metric $C$) values $u\left( x\right) $ can
be solved with the help of the operator 
\[
G\left( u,\alpha \right) =\frac{u\left( x+\alpha \right) -u\left( \alpha
\right) }{\alpha }". 
\]

Then, instead of the exact value of the function $u\left( x\right) $, an
approximate one $u_{\delta }\left( x\right) =u\left( x\right) +\Delta
u\left( x\right) $ with $\left\vert \Delta u\left( x\right) \right\vert \leq
\delta $ is substituted. On the basis of these calculations, one makes the
statement: \textquotedblright If $\alpha =\delta /\eta \left( \delta \right) 
$, where $\eta \left( \delta \right) \rightarrow 0$ for $\delta \rightarrow
0 $, then $\left( 2\delta /\alpha \right) =2\eta \left( \delta \right)
\rightarrow 0$ for $\delta \rightarrow 0$. Thus, for $\alpha =\alpha
_{1}\left( \delta \right) =\delta /\eta \left( \delta \right) $, $G\left(
u_{\delta },\alpha _{1}\left( \delta \right) \right) \rightarrow u^{\prime
}\left( x\right) $\textquotedblright .

It should be noted that, using the methodology of a small parameter,
Tikhonov obtained fundamental results in the field of investigations of
differential equations with a singular perturbation of the type 
\[
\epsilon \dot{u}=f\left( u,v,t\right) ;\quad \dot{v}=g\left( u,v,t\right) , 
\]%
where $\epsilon $ is a small parameter; $f\left( u,v,t\right) $ is a
nonlinear function (1948-1952)\footnote{%
See the review by A. B. Vasileva [35].}.

The solution of the system of equations does not depend continuously on the
parameter $\epsilon $. Proceeding to the limit $\epsilon \rightarrow 0$
creates a new object of investigations with completely different properties.
In the first place, it implies the issue of the so-called violation of the
stability of the root of the equation $f\left( u,v,t\right) =0$.
Nevertheless, Tikhonov managed to develop a rather constructive theory that
served as a basis for a number of productive approaches of both fundamental
and applied character. The importance of Tikhonov's achievements in the
sphere of system analysis is analyzed in detail by N. N. Moiseev (\cite{2.36}%
, section 5).

However, properties of the integral equation (\ref{2.6}) for $\alpha =0$
also change radically. In this regard, generally speaking, a certain analogy
emerges. One can suggest that Tikhonov undertook an attempt to use the
techniques of his theory of singular perturbations for the solution of
ill-posed problems.

This suggestion is supported by the following quotation from the monograph
by S. A. Lomov (\cite{2.37}, p. 12): ''Now it is becoming clear how to
isolate in singularly perturbed differential equations small terms that can
be neglected. It turned out that one needed additional information about the
solution to do this.''

Note J. Hadamard's remark that an extension of methods of the theory of
ordinary differential equations to problems of mathematical physics should
be done with great care (\cite{2.38}, p. 38). At the same time, at the turn
of the 1950s, the theory of singular perturbations became an efficient tool
in investigations of complicated problems of partial differential equations
(publications by M. I. Vishik and L. A. Liusternik, O.A. Olejnik, K. O.
Fiedrichs, and others). By the way, explaining the conceptual basis of their
method of quasiinversion, R. Lattes and G.-L.\ Lions (\cite{2.23}, p. 11)%
\footnote{%
Ideological closeness of quasiinversion and Tikhonov regularisation was
pointed out by M. M. Lavrentiev [23, p. 5].} refer to these authors and A.
N. Tikhonov.

Simultaneously, they pointed out that Tikhonov's priority publication on the
method of regularization \cite{2.8} (see also \cite{2.39}) was preceded by
D. L. Phillips' article \cite{2.40}, whose results with respect to integral
equations were analogous. In the monograph by F. Natterer \cite{2.41} this
regularization figures as Tikhonov-Phillips' method. V. A. Morozov estimated
the achievements of the latter author in a much more restrained manner (\cite%
{2.6}, p. 10): \textquotedblright Some recommendations on the use of this
method are contained in the publications by L. V. Kantorovich \cite{2.42}
and D. L. Phillips \cite{2.40}. There is no theoretical justification of
this approach in the above-mentioned publications\textquotedblright .

The chronological reference to the most important results in the field of
the construction of stable algorithms for the solution of integral equations
of the first kind (\cite{2.10}, p. 234) gives the following information:
''1962, Phillips's publication \cite{2.40}, where he suggested a variational
method of conditional minimization of the functional (with the use of
restrictions on the smoothness of the solution) and put forward the idea ...
of a choice of the regularization parameter $\alpha $''.

Turning to V. M. Fridman's achievements, note that it is rather difficult to
evaluate the premises that form the basis of the iteration procedure (\ref%
{2.10}). At the first sight, such a computational method has a lot of
analogs. However, its adequacy, in a sense, to the object of investigation,
the ill-posed problem of the solution the Fredholm integral equation of the
first kind, turned out to be rather unexpected.

Later on, with the aim to improve convergence, Fridman also employed the
nonlinear algorithm (\ref{2.14}). In our opinion, different ways of the
determination of the number of the final iteration and of the increase of
the rate of global convergence (see \cite{2.14,2.19,2.20}), despite their
actuality for practical application, should be interpreted as a technical
complement to Fridman's methodology.

Nowadays, the algorithm of conjugated gradients is considered to be nearly
the most efficient one for the solution of large ill-defined sparse systems
of linear algebraic equations, obtained by the reduction of, apparently,
most problems of numerical simulations \cite{2.32,2.43,2.44}. As is pointed
out by J. Ortega \cite{2.32}, this method was proposed by M. P. Hestens and
E. L. Stiefel (1952). However, for certain reasons, it was not employed for
a long time. It attracted considerable interest at the turn of the 1970s,
when one realized the actual sphere of its applications, the potential of
the above-mentioned predetermination and adaptivity with respect to
paralleling of computational operations in combination with the architecture
of modern computers.

Thus, the priority of the method of conjugated gradients ensured its
refraction to a class of problems of linear algebra, characterized by the
instability of the numerical realization, that is, in fact, ill-posed. In
this regard, we emphasize that Fridman's \textquotedblright methods of the
saddle-point type\textquotedblright\ \cite{2.21} can be interpreted as
somewhat simplified representatives of the family of the methods of
conjugated gradients (\cite{2.20}, section 2.1; \cite{2.43}, section 7.1).
It seems that V. M. Fridman, who was the first to use systematically
iterations for the solution of ill-posed problems, essentially foresaw the
development of computational mathematics that followed.

In light of the above, the position of M. A. Krasnoselskii and the
co-authors is worth noting \cite{2.18}. The role of V. M. Fridman in the
development of the iteration procedure (\ref{2.11}), which is an analog of (%
\ref{2.10}), is described as follows: \textquotedblright A transition to the
equation $\left[ \psi =\left( I-\nu A_{1}\right) \psi +\nu f_{1}\right] $
was pointed out for some cases by I. P. Natanson \cite{2.45}. For Fredholm
integral equations of the second kind, it was already employed by G. Wiarda 
\cite{2.46}. For integral equations of the first kind, it was, essentially,
employed in the publication by V. M. Fridman \cite{2.17}\textquotedblright\
(p. 73). There is no comment on a qualitative difference between the objects
of the investigation.

The nontriviality of Fridman's approach is noted in the remark of Natanson 
\cite{2.45}: \textquotedblright Our method does not apply to the solution of
the integral equation of the first kind. This could be expected, because the
use of the method implies complete arbitrariness of the free term of the
equation $A\psi =f$, whereas Eq. (\ref{2.1}) is solvable not for all $%
f\left( x\right) $\textquotedblright . In what follows, the author gives an
extended proof of the degeneracy of the corresponding discrete problem.

The gradient algorithm of V. M. Fridman \cite{2.21} is mentioned by the
authors of \cite{2.18} exclusively in the context of the equation $A\psi =f$%
, where both the operators $A$ and $A^{-1}$ are bounded (p. 115). We quote
the abstract to Fridman's paper \cite{2.21}: \textquotedblright We present a
new proof of the convergence of methods of the saddle-point type for a
linear operator equation. We do not assume, unlike L. V. Kantorovich \cite%
{2.47}, M. A. Krasnoselskii and S. G. Krejn \cite{2.48}, that zero is an
isolated point of the spectrum of the operator\textquotedblright .\footnote{%
This is equivalent to the boundedness of the operator $A^{-1}$.}

\section{Ill-defined finite-dimensional problems and issues of discretization%
}

In this subsection, $A\psi =f$ denotes a system of linear algebraic
equations. The conditionality number of the matrix $A$ (see, e.g., \cite%
{2.49}) 
\[
cond\left( A\right) =\max\limits_{\psi }\frac{\left\| A\psi \right\| }{%
\left\| \psi \right\| }/\min_{\psi }\frac{\left\| A\psi \right\| }{\left\|
\psi \right\| }, 
\]
where $\psi $ is a manifold of vectors of the Euclidean space, represents a
raising coefficient between a relative error of the data and the solution.
At the same time, cond$\left( A\right) $ characterizes the measure of
closeness of $A$ to a degenerate matrix, for which the solution of the
corresponding system of algebraic equations does not exist or is nonunique.

An algorithm of the solution of a degenerate system of linear algebraic
equations, based on the method of least squares, is presented in the book by
A. N. Malyshev \cite{2.50}. First, the matrix $A$ is transformed to a
two-diagonal one by means of a special transformation, and one finds its
eigenvalues that are subdivided into two groups, $\sigma _{1}$, $\sigma _{2}$%
,..., $\sigma _{n}$ and $\sigma _{n+1}$,..., such that $\sigma _{n}/\left(
\sigma _{n}-\sigma _{n+1}\right) $ is not very large. Then, with the help of
a rather laborious procedure of the exhaustion of the second group of the
eigenvalues, one constructs a matrix $A_{n}$ that is stably invertible
beginning with a certain value $n$. The accuracy of the thus obtained
generalized solution $\tilde{\psi}$ is determined by the error of closure $%
\left\Vert A\tilde{\psi}-f\right\Vert /\left\Vert A\tilde{\psi}\right\Vert $%
, using heuristic considerations.

It seems that in a methodological sense this scheme reminds of B. K.
Ivanov's algorithm \cite{2.15} that reflects computational relations (\ref%
{2.7})-(\ref{2.9}).

L. Hageman and D. Young \cite{2.43} studied the approach of
predetermination, employed for the solution of systems of linear algebraic
equations, close to degenerate ones, to accelerate by the method of
conjugated gradients iterations of the type 
\[
\psi _{n+1}=P\psi _n+g, 
\]
where $P=I-Q^{-1}A$; $g=Q^{-1}f$. It is assumed that this procedure can be
symmetrized in the sense that there exists a non-degenerate matrix $W$ such
that the matrix $W\left( I-P\right) W^{-1}$ is symmetric and positive
definite.

By use of $W$, the initial problem can be reduced to the solution of much
better defined systems of algebraic equations $B\varphi =q$, where 
\[
B=W\left( I-P\right) W^{-1};\quad \varphi =W\psi ;\quad q=Wg. 
\]

Formally, a choice of the predeterminer does not pose problems. However, in
practice, one has to resolve a contradiction between the conditions imposed
on the matrix $W$: ''closeness'' to $A^{-1}$ to reduce the number of
iterations; a ''rapid'' calculation of a product of the type $W\psi $ \cite%
{2.51}. In the above-mentioned publication, I. E. Kaporin analyzes different
approaches to the construction of predeterminers for systems of linear
algebraic equations of a general type. An analogous issue, in the
interpretation of J. Ortega \cite{2.32}, is oriented mainly towards sparse
matrices.

The complexity of problems of linear algebra that arise in the realization
of modern methods of investigations in the field of the mechanics of a
continuous medium are characterized as follows \cite{2.51}: ''The matrices
of corresponding systems are rather large (up to a hundred thousand nonzero
elements), rather densely filled (up to hundreds or even thousands of
nonzero elements in each line), have no diagonal predominance, are not $M$%
-matrices and are rather ill-defined. In general, one can expect only
symmetry and positive definiteness of the matrix of the system''.\footnote{%
The non-diagonal elements of an $M$-matrix are non-positive, and all the
elements of its inverse are non-negative.}

Note that, for example, in seismic tomography \cite{2.44}, one has to be
satisfied with a numerical realization of discrete analogs of integral
equations of the first kind, because their kernels cannot be represented
analytically and parameters of the considered models are determined with the
help of natural experiments.

In light of the above, the considerations of R. W. Hamming (\cite{2.52}, p.
360) may seem to be archaic: ''A system of linear equations is said to be
ill-defined, if, roughly speaking, the equations are almost linearly
dependent. Many efforts were made to investigate the problem of the solution
of ill-defined systems. However, one may pose the question: Is it necessary
to solve such systems in practical situations? In what physical situation
may the solutions prove to be useful, if they depend in such a substantial
manner on the coefficients of the systems? Usually, the following is true:
Instead of the solution, one is looking for a system of almost linearly
independent equations. In light of this information, the problem can be
better understood and is usually reformulated again in a more satisfactory
way. It is rather probable that ill-defined systems of equations, provided
that round-off and measurement errors are eliminated, are actually linearly
dependent and thus do not reflect the physical situation''.

Note that the renowned practitioner adheres to the position of correctness
according to Hadamard. Let us quote P. S. Guter's preface to \cite{2.52}:
\textquotedblright The name of R. W. Hamming, a renowned American scientist,
former President of the Computer Association, Head of the Mathematical
Service of Bell Telephone Laboratories, and his works in the field of
computational mathematics and the theory of information are rather
well-known and do not need special recommendations. ... The book 'Numerical
Methods for Scientists and Engineers' is without any doubt an outstanding
phenomenon in mathematical literature\textquotedblright .

Of special interest is Hamming's opinion about the priority of computational
procedures (\cite{2.52}, p. 90): \textquotedblright It is often believed
that the main problems of numerical analysis are concentrated on
interpolation, but this is not the case. They are mostly related to such
operations as integration, differentiation, finding zeros, maximization,
etc., in those cases when all we have or can compute are some nodes of
functions that are usually known not exactly, but approximately, because
they are spoiled by the round-off error\textquotedblright .

Thus, the problem should be posed correctly despite an inevitable error in
the data. It is obvious that such a position witnesses the preference of
algorithmic efficiency to the quality of initial information. Interpolation,
mentioned in the above quotation, implies approximate representation of the
latter for the performance of computer operations by means of a
finite-dimensional approximation.

However, in computational mathematics, alternative concepts are rather
wide-spread, which is reflected in K. I. Babenko's remark \cite{2.25}: ''In
some spheres of numerical analysis, the theory of approximation serves as
the foundation for the building of the numerical algorithm'' (p. 138).
''Information, inputted into the algorithm, is characterized, in the first
place, by its volume... All other characteristics, such as, e.g., accuracy,
are its derivatives and do not present a true picture of the input'' (p.
281).

Here, information is understood in the sense of Kolmogorov's theory of $%
\epsilon $-entropy that identifies it with the length of a given table or an
alphabet, whose words are manipulated by the algorithm. Correspondingly, the
issue of numerical analysis is interpreted in terms of, figuratively, the
deficiency in the search for necessary words and of the deletion of tables
in the course of operations.

Nevertheless, R. W. Hamming's point of view on the relation between the
method of investigations and the employed information is actively developed
by a group of specialists with J. Traub and G. Wasilkovski at the head. The
authors of \cite{2.53} point out (pp. 9, 6): ''In this book, we construct a
general mathematical theory of optimal reduction of uncertainty. We
interested in the two main questions: 1) Is it possible to reduce
uncertainty to a given level? 2) What will it cost? The aim of the theory of
informational complexity is to provide a unified approach to investigations
of optimal algorithms and their complexity for the problems that involve
incomplete, imprecise or paid information and to employ the general theory
to concrete problems from different fields''.

Here, complexity implies the number of arithmetic operations, the time of
their realization, computer memory resources, etc. By the way, the
interpretation of the notion of information \cite{2.52,2.53} correlates with
the expressive statement of R. Bellman and S. Dreyfus (\cite{2.54}, p. 342):
\textquotedblright Fortunately, in some cases, there is a very simple way to
overcome this difficulty. Instead of trying to study information as the
\textquotedblright smile of Cheshire Cat\textquotedblright , we consider the
actual physical process, where information is used to work out solutions.%
\footnote{%
The smile of Cheshire Cat, according to L. Carrols "Alice in th Miracle
Land", existed separately from this cat (editor's note to [54]).} The value
of information can then be measured by the efficiency of the solutions.

Thus, the usefulness of information depends on its application, which is the
most reasonable concept!''

It should be noted that the procedure of finite-dimensional approximation of
problems of mathematical physics is, of course, also very important, which
is accentuated by Babenko. Indeed, the obtained discrete model can turn out
to be incorrect, and the employed algorithms of the numerical realization
may prove to be divergent even in the solution of rather ordinary problems.
An example of instability of a finite-difference scheme is given by S. K.
Godunov and V. S. Ryabenkii (\cite{2.55}, section 4.9).

Babenko also emphasized the absence of any general methods of the
construction of finite-dimensional analogs (\cite{2.25}, p. 622):
\textquotedblright ...the provision of an approximation alone is
insufficient\textquotedblright ... one has to ensure that the discrete
problem \textquotedblright retains the type of the original continuous
problem\textquotedblright . In his opinion, to achieve the above goal,
\textquotedblright a detailed investigation in each concrete case is
required, which is the most nontrivial part of work\textquotedblright .

\section{The crisis of the technology of numerical simulations}

Of considerable interest is, in fact, a program statement of O. M.
Belotserkovskii and V. V. Stchennikov in the preface to \cite{2.56}:

''A rapid development of computers, especially during the last 10-15 years,
with a special acuteness posed the problem of the construction of a
principally new technology of the solution of problems by computers. ...
Historically, the problems of numerical simulations (in this notion, we
include the actual mathematical simulations related to a numerical
experiment), being rather advanced already in the ''precomputer'' period and
rapidly developing during the next periods, turned out to be the most
conservative component of the modern technology of the solution of problems
on the computer. Using, probably, redundant from the point of view of a
mathematician expressiveness of the description, one can characterize the
existent situation by two stable tendencies:

- an increase of the complexity of mathematical models;

- construction of rather sophisticated mathematical methods.

Both the tendencies inevitably lead to a technological deadlock, because
they create complications in the solution of the problem of the construction
of software-hardware means of the support of the whole technological chain.
... Without any pretension to profoundness and importance of the analogy, we
dare say that the present situation in numerical simulations is similar to
that in mechanics before the appearance of main ideas and concepts of
quantum mechanics''.

In the introductory article \cite{2.56} the same authors emphasize the
phenomenon of the accumulation of the round-off error in the numerical
realization of algorithms that include up to $10^{12}$ operations and the
absence of real means to estimate the error of solutions to, in particular,
evolution problems. In their opinion, \textquotedblright ...the following
conclusion is quite justified: a priori, any evolution problem for large
times is numerically (or computationally) ill-posed in the sense of the
absence of a practically important solution...

In the case, when a priori or a posteriori information about the error of an
approximate solution is absent, it is impossible to claim that the solution
exists. This conclusion fairly agrees with A. N. Tikhonov's theorem that
states that the problem with the data on the operator and the right-hand
side has no solution in the manifold of approximate numbers''.

Belotserkovskii and Stchennikov regard as constructive the idea that
discrete models of the considered problems should be assembled with the aim
of increasing the accuracy of information by means of special superposition.
They also suggest to search for the solution in the class of function with a
bounded variation, with would endow the difference operator of the problem
with smoothing properties.

As is well-known, N. N. Yanenko paid considerable attention to the
methodology of mathematical simulations (see \cite{2.2}). His concept of
overcoming the above-mentioned crisis is explained by O. M. Belotserkovskii (%
\cite{2.57}, p. 106):

''An investigation of finite-difference schemes, approximating different
classes of equations of mathematical physics, led N. N. Yanenko to an
extension of the notion of the scheme. For the first time, he begins to
consider the finite-difference scheme as an independent object of the
investigation, as a mathematical model, adequate to this or that physical
model. This fundamental concept is based on profound understanding of the
fundamentals of differential and integral calculus.

Indeed, physical and mathematical models, described by differential,
integral or integrodifferential equations, are obtained from discrete models
by means of averaging and passing to the limit with respect to certain
parameters. This is the case, for example, in the model of a continuous
medium, where for a sufficiently large number of elements in the unit volume
one comes to the notion of the continuous medium by averaging and passing to
the limit with respect to the volume. In this regard, one can interpret a
finite-difference scheme as an independent mathematical model with certain
properties''.

Note the fundamental, as it seems, considerations of Yanenko \cite{2.2}:
\textquotedblright The objects of modern mathematics, whose theoretical
\textquotedblright nucleus\textquotedblright\ comprises topology, geometry,
algebra and functional analysis, are ideal logical constructions forming a
certain operational system. We will call them ideal objects, which
underlines, on the one hand, their practical inaccessibility and, on the
other hand, their excellent operational properties that allow one to make
operations without loss of information. Ideal objects of mathematics are
essentially infinite and require an infinite number of
operations\textquotedblright\ (p. 12).

\textquotedblright The development of the experimental foundation and the
tool of investigations, the computer, increased interest in such objects as
computer numbers, programs, finite automata. In this regard, the definition
of mathematics as studies of the infinite, accepted in the 20th century,
should be replaced by another one, more correctly reflecting its essence,
i.e., as studies of the relationship between the finite and the
infinite\textquotedblright\ (p. 18).

Of interest is the following extract from (\cite{2.58}, p. 89):

"Let us make the following remark about the meaning of mathematically
ill-posed problems. In the old literature [I. G. Petrovskii, {\it Lectures
on Partial Differential Equations} (Fizmatgiz, Moscow, 1961) (In Russian)],
the above-mentioned lesser value of ill-posed problems was even interpreted
as their total senselessness. Nowadays it is accepted that this is not the
case. ... Nevertheless, the fact is, of course, that ill-posed problems are
substantially sensitive to small errors. A misunderstanding of this fact may
lead to paradoxes."

We think that on the basis of the above one can come to a very important
conclusion: In their construction of the conceptual basis of mathematical
simulation, the leading specialists were guided by the concept of
inapplicability of Banach's theorem on the inverse operator. Note that N.
Dunford and J. Schwartz considered this theorem as one of the three
principles of linear functional analysis, characterized as being rather
fruitful (\cite{2.59}, p. 61).\footnote{%
The other two are the principle of linear boundedness and the Hahn-Banach
theorem.}

A quotation from K. Maurin's manual (\cite{2.60}, p. 51) reads:
\textquotedblright This theorem [on the closed graph], in the last years,
has gained itself a reputation of being the most important theorem of
functional analysis, if this one is considered from the point of view of
applications\textquotedblright .

An attempt to renew the above-mentioned fundamentals in the context of the
accentuation of peculiarities of computational mathematics was made by A. V.
Chechkin \cite{2.61}, who suggested a division of sections of mathematics
into classical and non-classical ones, respectively: \textquotedblright
arithmetics, mathematical analysis, algebra, geometry, probability theory,
etc.; mathematical logic, the theory of information and statistics, the
theory of fuzzy sets, the theory of algorithms and recursive functions,
methods of computational mathematics, the theory of finite-difference
schemes, the theory of cubic formulas, methods of the solution of incorrect
problems, etc.\textquotedblright\ (p. 8). As a criterion, the authors choose
the fact of availability of absolutely complete or partial information about
the considered objects (points, functions etc.).

Let us quote the abstract of section (\cite{2.61}, p. 78):
\textquotedblright We define and study a new type of mappings that
generalize classical notions. Classical mappings realize correspondence
between the points of a set. This implies that the points are known with
absolute precision. The new mappings, termed ultramappings, realize
correspondence between pieces of information about points of sets. The main
construction of the ultramappings, termed ultraoperators, allows one to
obtain separate information about the image point from separate information
about the inverse image point.

Ultracontinuity of ultraoperators is defined, which is a broad
generalization of the notion of the stability of methods. It is found that,
for an arbitrary base operator, one can construct an ultracontinuous
operator over it. A class of ultracontinuous operators, termed Tikhonov's
operators, is singled out. For these operators, the base operators are not
continuous''. Furthermore, ''they are related to A. N. Tikhonov's ideas and
methods of the solution of incorrect mathematical problems''.

Returning to the question of adequate discretization, we quote the abstract
of the monograph by A. A. Dezin \cite{2.62}: \textquotedblright It is
devoted to the description of the basic structures of multidimensional
analysis and to the consideration of internally defined discrete problems of
analysis and mathematical physics. It implies not merely an approximation of
a given continuous object, but the construction its analog, starting from
the notion allowing for discrete interpretation\textquotedblright .

Arguments for contradiction to physical sense of differential models of
certain classes of problems of the mechanics of a continuous medium are
given by M. A. Zak \cite{2.63}. In this regard, he developed a general
approach, wholly based on the concepts of theoretical mechanics with a
special interpretation of Gauss' principle of least action.

The position of C. Truesdell is alternative. He thinks that continuum
mechanics of a deformed body \textquotedblright is, in essence, not only
subtler, more beautiful, majestic than a rather sparse particular case,
called \textquotedblright analytical mechanics\textquotedblright , but it is
much more suitable for the simulation of real bodies\textquotedblright\ (%
\cite{2.64}, p. 10).

\chapter{Comments on the material of the previous sections and some general
considerations}

\section{The correctness of the formulation of problems of mathematical
physics}

The conditions of correctness, formulated by J. Hadamard at the turn of the
20th century (see \cite{3.1}) and insistently advocated by him thereafter, \
primarily attract us by their ever-increasing importance for practical
applications. These conditions deal with the conceptual basis of numerical
simulation of physically meaningful problems, which, in fact, is disputed by
nobody. At the same time, nowadays, the prevailing opinion is that
Hadamard's concepts are principally invalid.

Implied is the basic statement that the properties of existence and
uniqueness, considered by Hadamard as inherent to mathematical models of
real processes, lead to the correctness of the formulation of adequate
boundary-value (initial-boundary-value) problems, which implies the
stability of the employed algorithms of a numerical realization. A
particular consequence is that the Fredholm integral equation of the first
kind is simply unsuitable for "application" in the problems of mathematical
simulation.

A natural course of investigations with the aim to confirm or disprove the
hypothesis, or, maybe, a prophecy, of Hadamard, seemingly had to be
conducted from the position of variability of formulations of the considered
problems, which was not the case. The main reason is, apparently, a
formulation of the belief in a special mission of computational means of
numerical simulations that lightheartedly neglected even one of the main
principles of functional analysis, i.e., Banach's theorem on inverse
operator (\cite{3.2,3.3},section 9, and \cite{3.4}).

One can hardly explain the absence in special literature of a consistently
introduced thesis that it is necessary to coordinate constructive matching
of the formulation of problems of mathematical physics with algorithms of
their numerical realization. The roots of this situation seem to be in
systemic character of the giant computer-supply complex oriented at
commercial efficiency at the expense of high costs of provided services.

As a result, the alternative school of A. N. Tikhonov builds up the
criticism of J. Hadamard ideas according to the following scheme:

- the solution of the Fredholm integral equation of the first kind 
\begin{equation}
\left( A\psi \right) \left( x\right) \equiv \int\limits_{0}^{1}k\left( x,\xi
\right) \psi \left( \xi \right) d\xi =f\left( x\right) ,\quad x\in \left[ 0,1%
\right]  \label{3.1}
\end{equation}%
is, in general, an ill-posed problem (which is undisputable);

- integral equations of this type are adequate to a variety of real
phenomena, which is actually supported by a rather transparent
interpretation of corresponding direct problems (calculations of $f$ from
given $k$ and $\psi $).

However, what are the grounds for the formulation of the problem, inverse to
the calculation of $f\left( x\right) $, by means of mechanical renaming the
given and the sought functions in (\ref{3.1})? The fact that the procedure
of the restoration of $\psi \left( x\right) $ for given $f\left( x\right) $
and $k\left( x,\xi \right) $ is computationally incorrect not imply any
consequences.

The reproaches to Hadamard, whose typical elements are reproduced in section
2.1, can be summarized as follows: The great scientist slowed down the
progress of science by refusing to admit that ill-posed problems were
adequate to a variety of real processes (see \cite{3.3,3.4,3.5}). Indeed,
the principles formulated by Hadamard do not allow for ill-posed problems,
but this by no means imply their invalidity. In contrast to Hadamard who put
forward convincing arguments in support of his concept and, one dares say,
relied on postulates of mathematical religion, the "science of ill-posed
problems" itself could not provide any argument for the very justification
of its existence.

Among supporters of studies of problems of mathematical physics exclusively
in the correct formulation are: A. Poincar\'{e}, D. Hilbert, V. A. Steklov,
I. G. Petrovsky, I. Prigogine \cite{3.6,3.7,3.8,3.9,3.10}. On the other
hand, the role of the three absolutely independent conditions of the
correctness (existence, uniqueness and continuous dependence on the data of
the problem), introduced by R. Courant and D. Hilbert \cite{3.11}, can
hardly be called positive.

The potential of the fact that the third condition is a corollary of the
previous ones could facilitate the activation of research related to correct
formulation of problems of mathematical physics. When considering the
Fredholm integral equation of the first kind (\ref{3.1}), one had to be more
careful with respect to a possibility of performing corresponding
transformations involving $f\left( x\right) \in R\left( A\right) $, as
opposed, figuratively, to a surrogate of continuous inversion with the use
of the regularization parameter $\alpha $.

\section{A relationship to the theorem on the inverse operator}

The above-mentioned fact that the third condition of the correctness has the
character of a corollary results from Banach's theorem on the inverse
operator \cite{3.12} whose optimistic meaning consists in the following: If
the solution to Eq. (\ref{3.1}), with $D\left( A\right) =B_{1}$ and $R\left(
A\right) =B_{2}$, where $B_{1}$, $B_{2}$ are Banach spaces, exists and is
unique, the inverse operator $A^{-1}$ from $B_{2}$ into $B_{1}$ is bounded
(see section 2.3).

Correspondingly, the procedure of evaluation of the function%
\begin{equation}
\psi \left( x\right) =\sum_{n=1}^{\infty }a_{n}\lambda _{n}\bar{\psi}%
_{n}\left( x\right)  \label{3.2}
\end{equation}%
(this formula follows from the Hilbert-Schmidt theorem \cite{3.13}),
satisfying Eq. (\ref{3.1}) in $L_{2}\left( 0,1\right) $, must be stable with
respect to small perturbations of $k\left( x,\xi \right) $ and $f\left(
x\right) $ under the condition $B_{2}=l_{2}^{\prime }$. In what follows we
assume that such a function exists, the kernel $k\left( x,\xi \right) $ is
symmetric and closed: we use the notation of sections 2.2 and 2.4. Thus, $%
l_{2}^{\prime }$ is a Hilbert space of functions normalized according to (%
\ref{1.7}).

It should be noted that the properties of the Fredholm integral equation of
the first kind with a symmetric kernel can be easily extended to the case
when $k\left( x,\xi \right) $ is an arbitrary function from the space $L_{2}$
(\cite{3.13}, pp. 188-194).

However, both fulfillment in the course of calculations and a verification
of the condition $f\in l_{2}^{\prime }$ are practically infeasible.
Therefore, such spaces are called \textquotedblright
inconvenient\textquotedblright\ (see \cite{3.14,3.15}). Hence we are in a
principle dilemma as to the choice of the methodology of the investigation:

- an urge to overcome the difficulties resulting from the use of the space $%
l_{2}^{\prime }$ related to the boundedness of the operator $A^{-1}$;

- the loss of this property in exchange for a possibility of studying
mathematical models in ''convenient'' spaces.

With the beginning of large-scale applications of computational methods to
mathematical investigations, the second way became dominant.

Instructive is the dynamics of the point of view of S. G. Mikhlin, reflected
in his courses of mathematical physics and the theory of errors of 1968,
1977 and 1988 \cite{3.16,3.17,3.18}. At the beginning, the author considers
Eq. (\ref{3.1}) under the traditional assumption that the operator $A$ is
completely continuos. In this case, the inverse operator $A^{-1}$ is
unbounded. As a result, the problem has no solution in the usual sense and
one has to turn to the methodology of A. N. Tikhonov.

Later, Mikhlin drew attention to the fact that if the Fredholm integral
equation of the first kind (\ref{3.1}) is interpreted from the point of view
of a mapping from the space $L_{2}$ into $l_{2}^{\prime }$, the operator $A$
is no longer completely continuos, the operator $A^{-1}$ is bounded, and the
problem of the determination of the function $\psi \left( x\right) $ becomes
well-posed. Simultaneously, the completeness of the conditions of
correctness is restored, whereas the third condition was initially singled
out by the author.

Thus, the use of the pair of spaces $L_{2}\left( 0,1\right) -l_{2}^{\prime }$
in a sense transfers the canonical ill-posed problem to the mainstream of
fundamentals of functional analysis. Note the fact that Mikhlin did not
devalue the importance of his arguments by reasoning in terms of
\textquotedblright convenient - inconvenient\textquotedblright\ or
\textquotedblright bad\textquotedblright\ and \textquotedblright
good\textquotedblright\ spaces.

Such a position apparently incurred criticism: In his concluding monograph,
Mikhlin somewhat irritably readdresses actual formulation of problems of
mathematical physics to specialists in applied sciences, including
sociologists, who are interested in their solution. Simultaneously, the
author has found it reasonable not to consider infinite dimensional models
with inherent aspects of incorrectness.

There is a well-known opinion of A. M. Lyapunov that, being posed in the
framework of initial premises, a problem of mechanics or physics should be
solved afterwards by means of rigorous methods. Here, implied is a problem
"... that is posed completely definitively from the point of view of
mathematics" ( \cite{3.19}, p. 26). In other words, this means a well-posed
problem.

At the same time, why not consider the procedure of the formulation of
problems of mathematical physics as an additional reserve of increasing the
efficiency of employed techniques of numerical realization? Moreover, maybe
rigidly predetermined formulations of problems themselves prose artificial
complications of computational character under the conditions when physical
considerations admit a small, in a sense, variation? In our opinion, the
formulation of problems of mathematical physics and the algorithm of its
numerical realization are essentially interrelated categories.

\section{The methodology of the solution of ill-posed problems}

Ill-posed problems of mathematical physics are deceptively transparent from
the point of view of the interpretation of considered processes. This is
stipulated, in reality, by their adequacy to spaces that in the
computational sense are practically infeasible. If the data of such problems
are specified in their natural classes of functions, the corresponding
formulations loose a mathematical sense because of their insolubility.

In such a nontrivial situation, of crucial importance is, of course, a role
of general methodological concepts. In other words, one has to be guided by
a certain system of global principles. From this point of view, if
Hadamard's insistence on the correct formulation of problems describing
physical phenomena \cite{3.1} still can be interpreted as a kind of
hypothesis, in fact related Banach's theorem on the inverse operator is a
universally accepted element of the foundation of modern mathematics \cite%
{3.20}.

Nevertheless, there appeared a notion of correctness according to Tikhonov
that played up a version of a search for the solution of the problem (\ref%
{3.1}) in a reduced class of functions \cite{3.14}. Any general
recommendations for finding such a class on the basis of reasonable
information were not worked out.

A shaky conceptual basis led to the failure of the idea of a limiting
transition with respect to a small parameter in the solution of a family of
problems that mimicked ill-posed ones (the method of regularization \cite%
{3.2}). The reason, apparently, lies in the same inadequacy of the use of
functional spaces. Given that $l_{2}^{\prime }$ is characterized by an
infinite number of features that depend on the operator $A$ (a superposition
of products of squared values that consist of characteristic numbers,
integrals over free terms and eigenfunctions), whereas $L_{2}$ is
characterized by only one (an integral over the squared function), is it
possible, even on a purely heuristic basis, to expect to overcome this
cardinal disagreement with the help of the regularization parameter $\alpha $%
?

The situation in the sphere of activity of numerous followers of Tikhonov
looks rather deplorable. Actually, the efforts are concentrated on a
mathematical object with a small factor $\alpha $, formed on the basis of (%
\ref{3.1}): 
\begin{equation}
\alpha \psi \left( x\right) +\int\limits_{0}^{1}k\left( x,\xi \right) \psi
\left( \xi \right) d\xi =f\left( x\right) ,\quad x\in \left[ 0,1\right] .
\label{3.3}
\end{equation}%
This is called the Fredholm integral equation of the second kind, without
any mentioning of its insufficiency in this respect. Despite a large number
of investigations devoted to the determination of the regularization
parameter $\alpha $, any more or less constructive algorithms are absent.
The main reason seems to be the inconsistency of the idea that implies a
possibility of efficient matching between the solution and the data of
ill-posed problems (see, e.g., \cite{3.2,3.21,3.22}).

As a matter of fact, one has to be satisfied only by a comparison of
solutions to (\ref{3.3}) obtained in the range of the decrease of $\alpha $.
One can assume that because of great labor input of numerical realization
for small values of the regularization parameter, a large-scale application
of Tikhonov methodology to the practice of scientific investigations
incurred considerable economic damage. As regards attempts to investigate
the Fredholm integral equation in functional spaces of its correct
solvability, they were isolated and were not accompanied by constructive
implementation \cite{3.23}.

V. M. Fridman, whose papers \cite{3.24,3.25} are considered in section 3.3,
approached the solution of (\ref{3.1}) regardless of its applicability to
modeling of concrete processes. From the point of view of our consideration,
the iterative algorithms of Fridman may be of interest, because they allow
one to achieve maximal possible efficiency in the framework of the chosen
object of investigation, which is indirectly confirmed by their simplicity
and brevity. In other words, it is hardly possible to obtain anything more
from the traditional interpretation of Eq. (\ref{3.1}). Despite formally
existing convergence, by approaching the solution, the determined
corrections become small against the background of the values of the sought
function: 
\[
\psi _{n+1}\left( x\right) =\psi _{n}\left( x\right) +\lambda \left[ f\left(
x\right) -\left( A\psi _{n}\right) \left( x\right) \right] . 
\]

In the absence of a timely halt of such a procedure, computational
\textquotedblright noise\textquotedblright\ from operations with numbers
that differ by order of magnitude can radically distort the solution \cite%
{3.5,3.15}. It becomes obvious that the Fredholm integral equation of the
first kind, by virtue of its nature, contains an inherent defect that
principally disagrees with pithy formulation of the problem of the
determination of the function $\psi \left( x\right) $ from the kernel and
the free term of (\ref{3.1}).

In section 3.5, we have given the argument of K. I. Babenko \cite{3.26} for
the necessity to take into account the fact of the loss of information when
evaluating comparative efficiency of computational algorithms. This argument
seems to be even more important at the stage of the formulation of the
problem. Since calculations of $f\left( x\right) $ from (\ref{3.1})
objectively delete the information on the function $\psi \left( x\right) $,
its restoration in the framework of the traditional approach quite naturally
reduces to an ill-posed problem.

If we hypothetically assume that for the determination of the function $\psi
\left( x\right) $ satisfying (\ref{3.1}) one can find a different equation
that contains this function not only under the sign of integration but also
in an explicit form, all the problems will be removed. Such an appearance of 
$\psi \left( x\right) $ can be viewed in the context of modeling of
computational errors including also the integral component (which yields
"zero" in the sum).

\section{Methodological concepts of numerical simulations}

The predetermined method of conjugate gradients is considered to be one of
the most efficient methods for the solution of ill-posed systems of linear
algebraic equations that appear as a result of discretization of different
problems of mathematical physics \cite{3.27}. The predeterminer, a
non-degenerate matrix, allows one to reduce the procedure of numerical
realization to a sequence of algebraic problems with desired favorable
properties. On the other hand, however, the number of necessary iterations
and the difficulty intermediate calculations increase (section 3.7).

One of the key problems of computational mathematics is the development of
the conceptual basis for a relationship between a representation of the data
and the efficiency of the employed algorithms. In this regard, the ideas of
K. I. Babenko \cite{3.26}, completely based on a qualitative interpretation
of the notion of information can be estimated as rather pessimistic. Indeed,
almost all computational operations of this guide are accompanied by a
\textquotedblright colossal\textquotedblright\ loss of information, whereas
rare exceptions correspond only to a special representation of initial
tables, which, as a rule, is not realized in practice.

The position of R. W. Hamming \cite{3.28}, who can be characterized as a
direct follower of the ideas of J. Hadamard in the field of computational
mathematics, is alternative. In his opinion, methods of numerical
realization must be adapted to the available information. As regards
principal difficulties, such as the incorrectness of the formulation, the
main attention should be concentrated on a modification of mathematical
models. The arguments of P. Bellman and S. Dreyfus for the expediency of the
evaluation of the quality of information on the basis of its efficiency
indices \cite{3.29} are also rather attractive.

O. M. Belotserkovsky and V. V. Shennikov \cite{3.30} stated a crisis in the
sphere of numerical simulations resulting from the complexity of both the
formulations of practical problems and the techniques of their numerical
realization (section 3.8). As a reason, they have pointed out an
inapplicability of methods of \textquotedblright domestic\textquotedblright\
mathematics to situations, when owing to the accumulation of round-off
errors actually any algorithm becomes computationally incorrect. As a matter
of fact, the authors proposed to develop more intensively approaches in the
style of Tikhonov, without any mentioning of the alternative way, i.e.,
matching the formulations of considered problems with Banach's theorem on
the inverse operator.

Note that generations of specialists in different fields of mathematical
physics were brought up under slogans of the type ''all real problems of the
mechanics of continuum medium are ill-posed'' that were repeatedly
reiterated without any explanations by ''greats'' at different conferences.
As a result, we have an implementation at a folklore level of the thesis
supported only by the practice of scientific research.

N. N. Yanenko, who, in contrast to some colleagues, was well aware of the
losses of numerical simulations from the breakup of ties of the techniques
of numerical realization with the basics of functional analysis, can be
called a flagship of this ideology. However, he considered to be of crucial
importance the principal difference between classical and computational
mathematics consisting in the fact that the former dealt with abstract
symbols without the loss of information, whereas the objects of the latter
were numerical arrays whose transformation was inevitably accompanied by
errors of different kinds (see \cite{3.3,3.31}).

The arguments of the methodologically oriented works of N. N. Yanenko allow
us to suggest that a certain role in the formation of his ideas was played
by ambitious motivations of being a co-participant of the emergence of
\textquotedblright new\textquotedblright\ mathematics that, while partly
employing the \textquotedblright old\textquotedblright\ one, was, in
general, substantially superior. A grotesque manifestation of this position
is contained in the materials of the monographs \cite{3.21,3.32}. Extracts
from these monographs are given in section 3.8.

It seems that we are facing a distortion of the essence of the problem,
because Banach's theorem on the inverse operator is an entity of a higher
level than numerical operations and, at the same time, is most important
exactly for them. Indeed, the boundedness of the inverse operator yields
practically a unique possibility to prevent both inadequate dependence of
the solution on the data of the problem and the accumulation of
computational errors.

\section{Ideas of the development of a constructive theory}

Thus, let us suppose that the kernel $k\left( x,\xi \right) $ of the
Fredholm integral equation of the first kind (\ref{3.1}) is symmetric and
closed, and the function $\psi \left( x\right) $ satisfying this equation in 
$L_{2}\left( 0,1\right) $ exists. Correspondingly, $f\left( x\right) \in
l_{2}^{\prime }$, i.e., the following condition \cite{3.13} is fulfilled:

\begin{equation}
\sum_{n=1}^{\infty }\alpha _{n}^{2}\lambda _{n}^{2}<\infty ,\quad \alpha
_{n}=\int\limits_{0}^{1}f\left( x\right) \bar{\psi}_{n}\left( x\right) dx,
\label{3.4}
\end{equation}%
where $\lambda _{n}$, $\bar{\psi}_{n}\left( x\right) $ are the
characteristic numbers and the eigenfunctions of the kernel $k\left( x,\xi
\right) $. Note also that the system of elements $\left\{ \bar{\psi}%
_{n}\right\} $ is complete in $R\left( A\right) $ or in the space $%
l_{2}^{\prime }$ (\cite{3.33}, p. 69).

In this case, the operator $A^{-1}$ that maps from the space $l_{2}^{\prime
} $ into $L_{2}\left( 0,1\right) $ is bounded (Banach's theorem). Does it
mean that the function $\psi \left( x\right) $ can be determined from (\ref%
{3.2}) without accumulation of errors?

From this point of view, the Inverse World of Banach is rather captivating.
However, it does not allow for any differentiation of the employed spaces
with respect to preference. They are determined by the content of the
problem, i.e., by the operator $A$. The dominant tendencies in the sphere of
computational mathematics are purely alternative. Therefore, both openly and
mainly implicitly, introduced is the thesis that Banach's theorem on the
inverse operator is useless.

At the first sight, there is a serious reason for this. Indeed, the
smallness of the perturbation of the data and of the error admitted in
computational operations is implied in $l_{2}^{\prime }$. However, a
practical possibility to satisfy this condition is absent. The space $%
l_{2}^{\prime }$ is, in a sense, illusive because it deals with an infinite
set of features of the data of the problem that, for large values of $n$ in (%
\ref{3.2}), in essence, cannot be identified.

One can also not that Eq. (\ref{3.1}) is, in a sense, nonlinear. Indeed, let
us represent the function, integrated according to (\ref{3.1}), in the form $%
\psi =\psi _{1}+\psi _{2}$. Correspondingly,%
\[
\int\limits_{0}^{1}k\left( x,\xi \right) \psi _{i}\left( \xi \right) d\xi
=f_{i}\left( x\right) ,\quad i=1,2, 
\]
and each of these two equations is solvable in the sense of the fulfillment
of a condition of the type (\ref{3.4}).

However, the function $f=f_{1}+f_{2}$ can be represented as a sum of an
infinite number of summands. If we assume that the equation%
\[
\int\limits_{0}^{1}k\left( x,\xi \right) \psi _{i}^{\prime }\left( \xi
\right) d\xi =f_{i}^{\prime }\left( x\right) ,\quad i=1,2, 
\]%
where $\psi _{1}^{\prime }+\psi _{2}^{\prime }=\psi $, is solvable for an
arbitrary subdivision of $f$ into $f_{1}^{\prime }$ and $f_{2}^{\prime }$,
we arrive at a contradiction. Indeed, the solution of Eq. (\ref{3.1}) is
unique, and a condition of the type (\ref{3.4}) is fulfilled only for $%
f_{i}\in R\left( A\right) $.

Thus, the principle of linear superposition does not apply to the free term
of Eq. (\ref{3.1}).\footnote{%
This point partly overlaps the material of section 7.5.} This situation
results from the fact that the range of the operator $A$ is not closed,
which was mentioned in section 2.3.

In general, the fact that the function $f\left( x\right) $, theoretically,
belongs to $l_{2}^{\prime }$, in reality, does not yield anything. However,
such a conclusion cannot serve as a basis for the neglect of the space $%
l_{2}^{\prime }$ in the consideration of the problem (\ref{3.1}). It seems
that constructiveness is possible here only in the context of the agreement
of, generally speaking, alternative aspirations:

- the function $f\left( x\right) $, employed in the calculations, belongs to 
$L_{2}$;

- the operator $A$ maps from $L_{2}$ into $l_{2}^{\prime }$.

The motivation is obvious: to preserve the potential of continuous inversion
of the operator $A$ for practical realization. At the same time, the
outlined contradiction is clear, and it cannot be overcome exclusively in
the framework of the Fredholm integral equation of the first kind (\ref{3.1}%
). In this situation, it is quite natural to turn, figuratively speaking, to
the origin of this equation, that is, to the issues related to the
formulation of the problem.

Consider a certain process described by the operator $A$. The direct problem
consists in the evaluation of the integral according to (\ref{3.1}) under
the substitution of the given function $\psi \left( x\right) $. This
procedure has a lot of interpretations and is mathematically correct.

A key element is the formulation of the inverse problem for the same
operator $A$, which is related to the restoration of the function $\psi
\left( x\right) $ from the realization of the above-mentioned integration,
that is, $f\left( x\right) $. Correspondingly, implied is the determination
of the cause from its consequence. Whereas the formulation of the direct
problem is transparent, the status of the inverse problem is diametrically
opposed. A priority of its solution is the actual algorithmic procedure (on
the basis of an adequate mathematical model) that is not an analog of the
process occurring in the regime of real time.\footnote{%
Indeed, the cause as an outcome of the consequence has no physical sense.}

In light of the above, is it possible not to turn to the statement of
Hadamard that all problems having practical interpretation admit a
mathematically correct formulation? From this point of view, since the
function $\psi \left( x\right) $ entering (\ref{3.1}) objectively exists,
the problem of its determination has to be only adequately posed. At the
same time, Hadamard did mot give corresponding recommendations of practical
character, and, as already mentioned, his methodology turned out to be, in
essence, completely rejected.

Let us try, however, to outline a formulation of the problem, inverse of the
evaluation of the integral (\ref{3.1}), that is carried out, in general,
with a certain error: 
\begin{equation}
A\psi =f+\delta f^{\prime },\quad x\in \left[ 0,1\right] .  \label{3.5}
\end{equation}

In the direct formulation, taking into account this error has no principal
importance. Nevertheless, solutions to the Fredholm integral equations of
the first kind (\ref{3.1}) and (\ref{3.5}) can be completely different. At
this point, it is senseless to pose the question about any quantitative
interpretation of $\delta f^{\prime }$. One can only assume that the error $%
\delta f^{\prime }$ is small compared to the values of the functions $\psi $
and $f$.

By general considerations, the presence of $\delta f^{\prime }$ in (\ref{3.5}%
) increases the potential of the formulation of the inverse problem, and the
question of a functional representation of the error arises alongside. In
this regard, one must take into account that the mechanism of its generation
is governed by the factor of smoothing of $\psi \left( x\right) $ by the
integration procedure; therefore, the structure of $\delta f^{\prime }$ must
reflect this situation

In light of the above, let us use an operator model of the error in the form 
\begin{equation}
\delta _{\bullet }=I_{\bullet }-\lambda B_{\bullet },  \label{3.6}
\end{equation}%
where $I$ is the identity operator; $B$ is a certain integral operator; $%
\delta =\mu \delta ^{\prime }$; $\mu $ and $\lambda $ are parameters.

Thus, instead of Eq. (\ref{3.1}), we propose to consider the following
problem: 
\begin{equation}
\mu A\psi =\mu f+\delta f;\quad \delta f=0,~x\in \left[ 0,1\right] .
\label{3.7}
\end{equation}%
The aim is to reduce this problem to the solution of the Fredholm integral
equation of the second kind. The parameter $\mu $, like $\lambda $, in the
inversion of the operator $I-\lambda B$ serves to prevent this equation from
positioning itself on the spectrum, which is equivalent to the existence and
uniqueness of its solution.

Note that we have just added a function representing "zero" to the free term
of (\ref{3.1}). At the same time, the transformation of the ill-posed
problem (\ref{3.1}) into the formulation (\ref{3.6}) creates conditions for
a radical change of the situation. We can demand, generally speaking, that $%
\delta f$ adaptively compensate for the errors of numerical operations that
take $f\left( x\right) $ out of the space $l_{2}^{\prime }$. As a result, a
prospects for a realization of the bounded operator $A^{-1}$ emerges. For $%
f+\delta f/\mu \in R\left( A\right) $, the negative factor of the
incorrectness of Eq. (\ref{3.1}) is fully neutralized.

Let us assume that the operator $B$ in (\ref{3.6}), for which $\delta f=0$
in the spaces $C$ or $L_{2}$, can be represented in the form 
\[
B_{\bullet }=\int\limits_{-1}^{1}h\left( x,\xi \right) _{\bullet }d\xi 
\]%
under certain conditions on the kernel $h\left( x,\xi \right) $. In this
case, the problem (\ref{3.7}) takes the form 
\begin{equation}
\psi \left( x\right) =\lambda \int\limits_{-1}^{1}h\left( x,\xi \right) \psi
\left( \xi \right) d\xi +\mu \int\limits_{0}^{1}k\left( x,\xi \right) \psi
\left( \xi \right) d\xi -\mu f\left( x\right) ,  \label{3.8}
\end{equation}%
\begin{equation}
\psi \left( x\right) =\lambda \int\limits_{-1}^{1}h\left( x,\xi \right) \psi
\left( \xi \right) d\xi ,\quad x\in \left[ 0,1\right] .  \label{3.9}
\end{equation}%
Thus, the condition that $\delta f$ be equal to zero, which equivalent to
Eq. (\ref{3.9}), is supposed to be satisfied with the help of $\psi \left(
x\right) $ on $x\in \left[ -1,0\right) $, i.e., a new unknown function.

There exists a well-known opinion that prospects of obtaining new
substantial results by simple transformation of mathematical relations are
not great. Indeed, by applying to Eqs. (\ref{3.8}), (\ref{3.9}) a
subtraction operation we again obtain the initial problem which is
ill-posed. However, first, we are not going to do this, and, second, behind
the integral equation with the sought function in an explicit form, we
intuitively feel a constructive potential.

From this point of view, a \textquotedblright refusal\textquotedblright\ of
the well-known example, given in a number of references, that illustrates
the smooting of information about the function $\psi $ by means of
integration of (\ref{3.1}) seems to be very significant. Indeed, assuming
that the function $\psi =\psi _{\ast }\left( x\right) $ satisfying the
system of equations (\ref{3.8}), (\ref{3.9}) is known, we give it a
perturbation of the type $\epsilon \sin \left( n\pi x\right) $. A
substitution into (\ref{3.7}) shows that this perturbation influences the
free term $f\left( x\right) $ both via a reduction coefficient (smoothing)
and without it, at the expense of an integral component and of explicit
presence of $\psi \left( x\right) $, respectively.

What is said does not apply to $\psi \left( x\right) $, $x\in \left[
-1,0\right) $. However, the determination of this function is beyond the
scope of the considered problem. We want to emphasize that the latter
arguments bear exclusively heuristic character.

From the position of a practical realization of the above, an interrelation
between the spaces $L_{2}$, $l_{2}$ and $l_{2}^{\prime }$ seems to be rather
significant. As is well-known, it is tightest in the pair of the spaces $%
L_{2}$ and $l_{2}$. The Riesz-Fischer theorem \cite{3.34} establishes a
one-to-one, continuos and linear relationship between functions from $L_{2}$
and numerical sequences $\left\{ c_{n}\right\} $ with a convergent sum of
the squares. In other words, there always exists a $L_{2}$-function for which%
\[
\sum_{n=1}^{\infty }c_{n}\varphi _{n}\left( x\right) 
\]%
is a Fourier series in terms of a system of orthonormal elements $\left\{
\varphi _{n}\left( x\right) \right\} $.

However, there is also a rather interesting relationship between the spaces $%
l_{2}$ and $l_{2}^{\prime }$, and, correspondingly, $L_{2}$. Indeed,
equation (\ref{3.2}) represents a Fourier series in terms of the orthonormal
elements $\bar{\psi}_{n}\left( x\right) $, whose convergence condition is
given by (\ref{3.4}). If we assume that $\lambda _{n}=r^{-n}$, where $0<r<1$%
, the space $l_{2}^{\prime }$ turns into $l_{2}$ under the condition $%
r\rightarrow 1$.

At the same time, the kernel $k\left( x,\xi \right) $ in (\ref{3.1})
possesses objectively inherent characteristic numbers and, consequently,
cannot be used for such transformation. However, there appeared the kernel $%
h\left( x,\xi \right) $, which is independent of the data of the problem:
hence a prospect of achieving what we set out to do. A considerable part of
our consideration below will be focused on this issue.

In conclusion of this section, we want to point out the inconsistency of the
wide-spread opinion that the formulation of problems of numerical simulation
should be left to specialists in applied sciences, whereas pure
mathematicians should be concerned exclusively with rigorous analytical
investigations, the development of computational methods and participation
in their realization.

It seems that specialists in applied sciences should be concerned with the
formulation of direct and, generally, well-posed problems. The factor of
incorrectness is directly related to the procedure of the numerical
realization. Therefore, the main concern of pure mathematicians should be a
reduction of formulations of problems describing the considered processes
and phenomena to the conditions of efficient implementation of Banach's
theorem on the inverse operator.

\chapter{A method of the reduction of problems, traditionally associated
with Fredholm integral equations of the first kind, to Fredholm integral
equations of the second kind}

\section{The formulation of the problem}

In light of the arguments of section 2.4 and 4.5, we proceed with the
consideration of the Fredholm integral equation of the first kind 
\begin{equation}
\left( A\psi \right) \left( x\right) \equiv \int\limits_{0}^{1}k\left( x,\xi
\right) \psi \left( \xi \right) d\xi =f\left( x\right) ,\quad x\in \left[ 0,1%
\right]  \label{4.1}
\end{equation}%
under the assumption that its solution exists and is unique, and the kernel $%
k\left( x,\xi \right) $ and the free term $f\left( x\right) $ belong to the
space $L_{2}$. In other words, using the terminology of \cite{4.1}, they are 
$L_{2}$ - functions:%
\[
\int\limits_{0}^{1}\int\limits_{0}^{1}k^{2}\left( x,\xi \right) dxd\xi
<\infty ;\quad \int\limits_{0}^{1}f^{2}\left( x\right) dx<\infty . 
\]

However, in reality, the determination of the function $\psi \left( x\right) 
$ from given $A$ and $f$ will carried out not by the use of the solution of
the Fredholm integral equation of the first kind (\ref{4.1}), but on the
basis of the following arguments. There is an operator $A$ describing a
certain phenomenon. This description is expressed in terms of the
integration of the function $\psi \left( x\right) \in L_{2}\left( 0,1\right) 
$ by (\ref{4.1}).

The evaluation of $f\left( x\right) $ is carried out with an error that we
denote as $\left( \delta f\right) \left( x\right) /\mu $, where $\mu $ is a
constant. In most cases this error, in virtue of its smallness, is
nonessential or can be reduced to a required level. Nevertheless, the
computational procedure can be interpreted as follows:%
\begin{equation}
\left( A\psi \right) \left( x\right) =f\left( x\right) +\left( \delta
f\right) \left( x\right) /\mu ,\quad x\in \left[ 0,1\right] .  \label{4.2}
\end{equation}

The situation changes cardinally if, on the contrary, we pose a problem of
the restoration of the function $\psi \left( x\right) $ from the information
contained in (\ref{4.1}), i.e., $A$ and $f$. Indeed, such a problem is, in
general, ill-posed, which, in fact, means that Eq. (\ref{4.1}) is insolvable.

From this point of view, Eq. (\ref{4.2}) is different because of the
presence of a potential of the reduction of the problem to a well-posed one.%
{\em \ }A necessary condition of this reduction consists in such a
representation of the error $\delta f$ that, irrespective of the data (\ref%
{4.1}) and of the function $\psi \left( x\right) $,%
\begin{equation}
f\left( x\right) +\left( \delta f\right) \left( x\right) /\mu \in R\left(
A\right) ,  \label{4.3}
\end{equation}%
where $R\left( A\right) $ is the range of the operator $A$. In other words,
the operator $\delta $ [see (\ref{3.6})] must endow the algorithm with
adaptive properties.

Thus, the following problem is posed: From given $A$ and $f$, determine
constructively the function $\psi \left( x\right) $ that, upon substitution
in (\ref{4.1}), would satisfy this equation. Here, constructiveness implies
a possibility to use a stable procedure of the numerical realization as a
result of the reduction of the problem to the solution of he Fredholm
integral equation of the second kind.\footnote{%
It is supposed that the kernel of this equation does not possess any
singularities incurred by the method of the realization of corresponding
transformations.}

The basis of further transformations will be formed by Eq. (\ref{4.2}),
where the central point is the establishment of adequate mutual dependence
of $\psi $ and $\delta f$. Equation (\ref{4.1}) is considered exclusively in
the context of the direct problem of the evaluation of the integral and as a
source of initial information.

\section{The model of the representation of the error}

Following the considerations of section 4.5, we present the error of the
evaluation of $f$ from (\ref{4.1})\ as a difference between the sought
function $\psi $ and the integral component 
\begin{equation}
\left( \delta f\right) \left( x\right) =\psi \left( x\right) -\lambda \left(
B\psi \right) \left( x\right) ,\quad x\in \left[ 0,1\right] ,  \label{4.4}
\end{equation}%
where $\lambda $ is a constant; the operator is given by 
\begin{equation}
B_{\bullet }=\int\limits_{-1}^{1}h\left( x,\xi \right) _{\bullet }d\xi ;
\label{4.5}
\end{equation}%
$\psi \left( x\right) \equiv \varphi \left( x\right) $, $x\in \left[
-1,0\right) $; the kernel $h\left( x,\xi \right) $ will be discussed later.

However, we intend to construct a stable algorithm of evaluation of the
function $\psi \left( x\right) $ satisfying (\ref{4.1}); hence small
variations of the data should not substantially influence the solution. In
this regard, consider a possibility of the fulfillment of the condition%
\begin{equation}
\left( \delta f\right) \left( x\right) =0,\quad x\in \left[ 0,1\right] ,
\label{4.6}
\end{equation}%
which means an assumption that the problem posed in section 5.1 can be
constructively solved (merely) by means of addition to the free term of Eq. (%
\ref{4.1}) of the "zero" from (\ref{4.4}) that has the following form:%
\footnote{%
Here, the error $\delta f$ or the function dependent on this error are
interpreted as a component of the free term of the Fredholm integral
equation of the second kind, employed for the determination of $\psi $.}%
\[
0=\psi \left( x\right) -\lambda \int\limits_{-1}^{1}h\left( x,\xi \right)
\psi \left( \xi \right) d\xi . 
\]%
This equation can be rewritten as%
\begin{equation}
\int\limits_{-1}^{0}h\left( x,\xi \right) \varphi \left( \xi \right) d\xi
=g\left( x\right) ,\quad x\in \left[ 0,1\right] ,  \label{4.7}
\end{equation}%
where 
\begin{equation}
g\left( x\right) =\frac{1}{\lambda }\psi \left( x\right)
-\int\limits_{0}^{1}h\left( x,\xi \right) \psi \left( \xi \right) d\xi .
\label{4.8}
\end{equation}%
Making the change of variables 
\begin{equation}
\zeta =2\pi x,\quad \theta =2\pi \left( 1+\xi \right) ,  \label{4.9}
\end{equation}%
we reduce it to the canonical form 
\begin{equation}
\int\limits_{0}^{2\pi }h\left( \zeta ,\theta \right) \varphi \left( \theta
\right) d\theta =g\left( \zeta \right) ,\quad \zeta \in \left[ 0,2\pi \right]
.  \label{4.10}
\end{equation}

As is obvious, the satisfaction of (\ref{4.6}) is equivalent to the
solvability of this equation. Let the kernel $h\left( \zeta ,\theta \right) $
belong to the space $L_{2}$ and be closed. In this case, Eq. (\ref{4.10}) is
a Fredholm integral equation of the first kind, whose the solution, if it
exists, is unique \cite{4.1}. By satisfying the above conditions, we
represent (\ref{4.10}) in the form of a Poisson integral (\cite{4.2}, pp.
202-205). Accordingly, the kernel is given by%
\begin{equation}
h\left( \zeta ,\theta \right) =\frac{1-r^{2}}{2\pi \left[ 1-2r\cos \left(
\zeta -\theta \right) +r^{2}\right] },\quad 0<r<1;  \label{4.11}
\end{equation}%
its characteristic numbers and orthonormal on $x\in \left[ 0,2\pi \right] $
eigenfunctions (\cite{4.1}, pp. 187-188) are%
\[
\lambda _{0}=1,\quad \lambda _{2n-1}=\lambda _{2n}=r^{-n},\quad n=1,2,\ldots
; 
\]%
\[
\bar{\varphi}_{0}\left( \zeta \right) =\frac{1}{\sqrt{2\pi }},\quad \bar{%
\varphi}_{2n-1}\left( \zeta \right) =\frac{1}{\sqrt{\pi }}\cos \left( n\zeta
\right) , 
\]

\begin{equation}
\bar{\varphi}_{2n}\left( \zeta \right) =\frac{1}{\sqrt{\pi }}\sin \left(
n\zeta \right) ,\quad n=1,2,\ldots ;  \label{4.12}
\end{equation}%
and, in (\ref{4.7}), 
\begin{equation}
h\left( x,\xi \right) =\frac{1-r^{2}}{1-2r\cos \left[ 2\pi \left( x-\xi
\right) \right] +r^{2}},\quad 0<r<1.  \label{4.13}
\end{equation}

If, in Eq. (\ref{4.10}), the function%
\begin{equation}
\varphi \left( \zeta \right) =\frac{1}{2}\alpha _{0}+\sum_{n=1}^{\infty
}\alpha _{n}\cos \left( n\zeta \right) +\alpha _{n}^{\prime }\sin \left(
n\zeta \right) ,  \label{4.14}
\end{equation}%
where $\alpha _{0}$, $\alpha _{n}$ and $\alpha _{n}^{\prime }$ are the
coefficients of its expansion into the Fourier series, is absolutely
integrable, i.e.,%
\[
\int\limits_{0}^{2\pi }\left\vert \varphi \left( \zeta \right) \right\vert
d\zeta <\infty , 
\]%
the function%
\begin{equation}
g\left( \zeta \right) =\frac{1}{2}\alpha _{0}+\sum_{n=1}^{\infty }r^{n}\left[
\alpha _{n}\cos \left( n\zeta \right) +\alpha _{n}^{\prime }\sin \left(
n\zeta \right) \right]  \label{4.15}
\end{equation}%
is the real part of an analytical inside a unity circle function and is
harmonic (\cite{4.3}, pp. 160-161; \cite{4.4}):%
\[
\partial _{X}^{2}g+\partial _{Y}^{2}g=0, 
\]%
where $X=r\cos \left( \zeta \right) $, $Y=r\sin \left( \zeta \right) $ are
Cartesian coordinates.\footnote{%
Here, the parameter $r$ is interpreted as a radial coordinate and $\zeta $
is, respectively, a polar angle.}

Since the above-mentioned property is independent of a linear change of
variables, it follows from (\ref{4.8}) with (\ref{4.9}) and (\ref{4.15})
that, under the condition (\ref{4.6}), the function $\psi \left( x\right) $
satisfying (\ref{4.1}) can only be harmonic. This means that it belongs to a
much narrower class of functions than it is supposed in the formulation of
the problem in section 5.1.

Nevertheless, one can conclude that the "zero" error of integration by (\ref%
{4.1}) of the harmonic function $\psi \left( x\right) $ is actually
representable in the form (\ref{4.4}) with the kernel $h\left( x,\xi \right) 
$ from (\ref{4.13}). This is an important point of our consideration.

The components (\ref{4.12}) satisfy the homogeneous equation%
\[
\varphi \left( \zeta \right) =\lambda \int\limits_{0}^{2\pi }h\left( \zeta
,\theta \right) \varphi \left( \theta \right) d\theta ,\quad \zeta \in \left[
0,2\pi \right] 
\]%
that, by the change of variables%
\[
\zeta =-2\pi x,\quad \theta =-2\pi \xi ;\quad \zeta =2\pi x,\quad \theta
=2\pi \xi 
\]%
is transformed to the following form:%
\[
\varphi \left( x\right) =\lambda \int\limits_{-1}^{0}h\left( x,\xi \right)
\varphi \left( \xi \right) d\xi ,\quad x\in \left[ -1,0\right) ; 
\]%
\begin{equation}
\varphi \left( x\right) =\lambda \int\limits_{0}^{1}h\left( x,\xi \right)
\varphi \left( \xi \right) d\xi ,\quad x\in \left[ 0,1\right] ,  \label{4.16}
\end{equation}%
which allows us, taking also account of (\ref{4.12}) and (\ref{4.9}), to
determine the characteristic numbers and the orthonormal on $x\in \left[
-1,0\right) $; $\left[ 0,1\right] $ eigenfunctions of the kernel (\ref{4.13}%
):%
\[
\lambda _{0}=1,\quad \lambda _{2n-1}=\lambda _{2n}=r^{-n},\quad n=1,2,\ldots
; 
\]%
\[
\bar{\varphi}_{0}\left( x\right) =1,\quad \bar{\varphi}_{2n-1}\left(
x\right) =\sqrt{2}\cos \left( 2\pi nx\right) , 
\]

\begin{equation}
\bar{\varphi}_{2n}\left( x\right) =\sqrt{2}\sin \left( 2\pi nx\right) ,\quad
n=1,2,\ldots .  \label{4.17}
\end{equation}

The solution of the problem (\ref{4.1}) is unique. Accordingly, by comparing
the homogeneous Fredholm integral equation of the second kind with respect
to $\psi \left( x\right) $ that corresponds to (\ref{4.8}) (i.e., for $%
g\equiv 0$) with (\ref{4.16}), we arrive at the condition%
\begin{equation}
\lambda \neq r^{-n},\quad n=0,1,\ldots .  \label{4.18}
\end{equation}

As the kernel in (\ref{4.16}) is symmetric, continuos, and all $\lambda
_{2n}>0$, by Mercer's theorem \cite{4.1},%
\[
h\left( x,\xi \right) =\frac{\bar{\varphi}_{0}\left( x\right) \bar{\varphi}%
_{0}\left( \xi \right) }{\lambda _{0}}+\sum_{n=1}^{\infty }\frac{\bar{\varphi%
}_{2n-1}\left( x\right) \bar{\varphi}_{2n-1}\left( \xi \right) +\bar{\varphi}%
_{2n}\left( x\right) \bar{\varphi}_{2n}\left( \xi \right) }{\lambda _{2n}} 
\]%
\begin{equation}
=1+2\sum_{n=1}^{\infty }r^{n}\left[ \cos \left( 2n\pi x\right) \cos \left(
2n\pi \xi \right) +\sin \left( 2n\pi x\right) \sin \left( 2n\pi \xi \right) %
\right] ,  \label{4.19}
\end{equation}%
where the series can be absolutely and uniformly convergent.

In what follows, we will need the resolvent of the operator $B$. From the
bilinear expansion (\ref{4.19}), by same Mercer's theorem, it follows that
the characteristic numbers and the orthonormal on $x\in \left[ -1,1\right] $
eigenfunctions of its kernel have the form%
\[
\lambda _{0}=\frac{1}{2},\quad \lambda _{2n-1}=\lambda _{2n}=\frac{1}{2}%
r^{-n},\quad n=1,2,\ldots ; 
\]%
\[
\bar{\psi}_{0}\left( x\right) =\frac{1}{\sqrt{2}},\quad \bar{\psi}%
_{2n-1}\left( x\right) =\cos \left( 2\pi nx\right) , 
\]

\[
\bar{\psi}_{2n}\left( x\right) =\sin \left( 2\pi nx\right) ,\quad
n=1,2,\ldots ; 
\]%
hence a necessity to impose one more condition:%
\begin{equation}
\lambda \neq \frac{1}{2}r^{-n},\quad n=0,1,\ldots .  \label{4.20}
\end{equation}

One should take into account that the use of Mercer's theorem is different
from the former representation of the kernel $h\left( x,\xi \right) $ be the
series (\ref{4.19}). Here, on the contrary, there exists an expansion of the
kernel $h\left( x,\xi \right) $ into a uniformly convergent bilinear series
in terms of an orthonormal on $-1\leq x\leq 1$ system of elements.
Accordingly, these elements, under a correction with respect to a
normalization factor and the value $1/2r^{n}$, are the eigenfunctions and
the characteristic numbers of the operator $B$.

We also note that the functions $\bar{\psi}_{2n-1}\left( x\right) $, $\bar{%
\psi}_{2n}\left( x\right) $ are orthogonal not only on $x\in \left[ -1,1%
\right] $, but on $x\in \left[ -1,0\right) $; $\left[ 0,1\right] $ as well.
This point will play a rather important role in the context of the
simplification of the procedure of the numerical realization.

The resolvent of the kernel (\ref{4.5}) is represented by the series \cite%
{4.1}%
\[
H\left( x,\xi ,\lambda \right) =\frac{\bar{\psi}_{0}\left( x\right) \bar{\psi%
}_{0}\left( \xi \right) }{\lambda _{0}-\lambda } 
\]%
\[
+\sum_{n=1}^{\infty }\frac{\bar{\psi}_{2n-1}\left( x\right) \bar{\psi}%
_{2n-1}\left( \xi \right) +\bar{\psi}_{2n}\left( x\right) \bar{\psi}%
_{2n}\left( \xi \right) }{\lambda _{2n}-\lambda }=\frac{1}{1-2\lambda } 
\]%
\begin{equation}
+2\sum_{n=1}^{\infty }\frac{r^{n}}{1-2\lambda r^{n}}\left[ \cos \left( 2n\pi
x\right) \cos \left( 2n\pi \xi \right) +\sin \left( 2n\pi x\right) \sin
\left( 2n\pi \xi \right) \right]  \label{4.21}
\end{equation}%
that, under the condition (\ref{4.20}), is also absolutely and uniformly
convergent.

From (\ref{4.8}) and (\ref{4.15}), taking into account (\ref{4.9}), we get:%
\[
\psi \left( x\right) =\frac{\alpha _{0}\lambda }{2\left( 1-\lambda \right) }%
+\sum_{n=1}^{\infty }\frac{\lambda r^{n}}{1-\lambda r^{n}}\left[ \alpha
_{n}\cos \left( 2n\pi x\right) +\alpha _{n}^{\prime }\sin \left( 2n\pi
x\right) \right] . 
\]%
Thus, under the condition (\ref{4.6}), Eq. (\ref{4.1}) can be satisfied only
in the case when%
\[
f\left( x\right) =\frac{\alpha _{0}\lambda }{2\left( 1-\lambda \right) }%
\int\limits_{0}^{1}k\left( x,\xi \right) d\xi 
\]%
\begin{equation}
+\sum_{n=1}^{\infty }\frac{\lambda r^{n}}{1-\lambda r^{n}}%
\int\limits_{0}^{1}k\left( x,\xi \right) \left[ \alpha _{n}\cos \left( 2n\pi
x\right) +\alpha _{n}^{\prime }\sin \left( 2n\pi x\right) \right] d\xi .
\label{4.22}
\end{equation}

In what follows, we assume that the function $\psi \left( r,x\right) $ is
harmonic and the free term of Eq. (\ref{4.1}) has the form (\ref{4.22}). As
already mentioned, this fact strongly narrows the sphere of practical
applications. As will be shown below (section 5.6), a solution, obtained for
this case, by means of the passage to the limit $r\rightarrow 1$ turns into
an $L_{2}$ - function $\psi \left( x\right) $ that satisfies Eq. (\ref{4.1}).%
\footnote{%
Simultaneously, Eq. (\ref{4.22}) takes the form (\ref{4.1}).}

\section{A transformed formulation of the problem}

Let us extend Eq. (\ref{4.4}), under the condition (\ref{4.6}), in the
following way:%
\[
\varphi \left( x\right) =\lambda \int\limits_{-1}^{0}h\left( x,\xi \right)
\varphi \left( \xi \right) d\xi +\lambda \int\limits_{0}^{1}h\left( x,\xi
\right) \psi \left( \xi \right) d\xi 
\]%
\begin{equation}
+\kappa \left( x\right) ,\quad x\in \left[ -1,0\right) ,  \label{4.23}
\end{equation}%
where $\kappa \left( x\right) \in L_{1}\left( -1,0\right) $, as a result of (%
\ref{4.14}), is a certain undefined function.

We represent the equation that unifies (\ref{4.7}) and (\ref{4.23}) in the
following form:%
\begin{equation}
\left. 
\begin{array}{c}
\psi \left( x\right) \\ 
\varphi \left( x\right)%
\end{array}%
\right\} =\lambda B\left( 
\begin{array}{c}
\psi \\ 
\varphi%
\end{array}%
\right) \left( x\right) +\left\{ 
\begin{array}{c}
0,\quad x\in \left[ 0,1\right] ; \\ 
\kappa \left( x\right) ,\quad x\in \left[ -1,0\right) ,%
\end{array}%
\right.  \label{4.24}
\end{equation}%
i.e.,%
\[
B\left( 
\begin{array}{c}
\psi \\ 
\varphi%
\end{array}%
\right) \left( x\right) =\int\limits_{-1}^{0}h\left( x,\xi \right) \varphi
\left( \xi \right) d\xi +\int\limits_{0}^{1}h\left( x,\xi \right) \psi
\left( \xi \right) d\xi . 
\]

Let us introduce also a related and close with respect to its structure
equation%
\begin{equation}
\left. 
\begin{array}{c}
\psi \left( x\right) \\ 
\varphi ^{\prime }\left( x\right)%
\end{array}%
\right\} =\lambda B\left( 
\begin{array}{c}
\psi \\ 
\varphi ^{\prime }%
\end{array}%
\right) \left( x\right) +\left\{ 
\begin{array}{c}
\chi \left( x\right) ,\quad x\in \left[ 0,1\right] ; \\ 
0,\quad x\in \left[ -1,0\right) ,%
\end{array}%
\right.  \label{4.25}
\end{equation}%
where $\varphi ^{\prime }\left( x\right) $ and $\chi \left( x\right) $ are
two more undefined functions (like $\psi $, they are harmonic). The
expediency of this step will be clear from what follows.

It is not difficult to represent the procedure of the construction of Eqs. (%
\ref{4.24}) and (\ref{4.25}) from the practical point of view. There is a
harmonic function $\psi \left( x\right) $ that is integrated according to (%
\ref{4.1}). As is shown above, there exist the kernel $h\left( x,\xi \right) 
$ and an absolutely integrable function $\varphi \left( x\right) $ for which
Eq. (\ref{4.24}) is satisfied on $x\in \left[ 0,1\right] $. One can assume
that the function $\varphi \left( x\right) $ is specified in a certain way.
Now both $\psi \left( x\right) $ and $\varphi \left( x\right) $ are given
functions. Equation (\ref{4.24}) is satisfied by means of the function $%
\kappa \left( x\right) $ on $x\in \left[ -1,0\right) $ and on the whole.

The function $\psi \left( x\right) $ is again given. The function $\varphi
^{\prime }\left( x\right) $ is determined from Eq. (\ref{4.25}) on $x\in %
\left[ -1,0\right) $:%
\[
\varphi ^{\prime }\left( x\right) =\lambda \int\limits_{-1}^{0}h\left( x,\xi
\right) \varphi ^{\prime }\left( \xi \right) d\xi +g^{\prime }\left(
x\right) , 
\]%
\begin{equation}
g^{\prime }\left( x\right) =\lambda \int\limits_{0}^{1}h\left( x,\xi \right)
\psi \left( \xi \right) d\xi .  \label{4.26}
\end{equation}

This is a Fredholm integral equation of the second kind with respect to $%
\varphi ^{\prime }\left( x\right) $. According to the foundations of the
general theory \cite{4.1}, under the condition (\ref{4.18}), the solution of
(\ref{4.26}) exists and is unique. The functions $\psi \left( x\right) $ and 
$\varphi ^{\prime }\left( x\right) $ are given, and Eq. (\ref{4.25}) is
satisfied by means $\chi \left( x\right) $ on $x\in \left[ 0,1\right] $ and
on the whole.

In terms of the notation%
\[
\Psi \left( x\right) =\left\{ 
\begin{array}{c}
\psi \left( x\right) ,\quad x\in \left[ 0,1\right] ;~ \\ 
\varphi \left( x\right) ,\quad x\in \left[ -1,0\right) ,%
\end{array}%
\right. 
\]%
\begin{equation}
\Psi ^{\prime }\left( x\right) =\left\{ 
\begin{array}{c}
\psi \left( x\right) ,\quad x\in \left[ 0,1\right] ;~ \\ 
\varphi ^{\prime }\left( x\right) ,\quad x\in \left[ -1,0\right) ,%
\end{array}%
\right.  \label{4.27}
\end{equation}%
Eqs. (\ref{4.24}), (\ref{4.25}) are Fredholm integral equations of the
second kind with respect to $\Psi $ and $\Psi ^{\prime }$, with the free
terms%
\[
P\left( x\right) =\left\{ 
\begin{array}{c}
0,\quad x\in \left[ 0,1\right] ;~ \\ 
\kappa \left( x\right) ,\quad x\in \left[ -1,0\right) ,%
\end{array}%
\right. 
\]%
\[
P^{\prime }\left( x\right) =\left\{ 
\begin{array}{c}
\chi \left( x\right) ,\quad x\in \left[ 0,1\right] ;~ \\ 
0,\quad x\in \left[ -1,0\right) ,%
\end{array}%
\right. 
\]%
respectively.

Under the condition (\ref{4.20}), the solutions of these equations are given
by%
\begin{equation}
\psi \left( x\right) =\lambda \int\limits_{-1}^{0}H\left( x,\xi ,\lambda
\right) \kappa \left( \xi \right) d\xi ,\quad x\in \left[ 0,1\right] ;
\label{4.28}
\end{equation}%
\begin{equation}
\varphi \left( x\right) =\kappa \left( x\right) +\lambda
\int\limits_{-1}^{0}H\left( x,\xi ,\lambda \right) \kappa \left( \xi \right)
d\xi ,\quad x\in \left[ -1,0\right)  \label{4.29}
\end{equation}%
and%
\begin{equation}
\psi \left( x\right) =\chi \left( x\right) +\lambda
\int\limits_{0}^{1}H\left( x,\xi ,\lambda \right) \chi \left( \xi \right)
d\xi ,\quad x\in \left[ 0,1\right] ;  \label{4.30}
\end{equation}%
\begin{equation}
\varphi ^{\prime }\left( x\right) =\lambda \int\limits_{0}^{1}H\left( x,\xi
,\lambda \right) \chi \left( \xi \right) d\xi ,\quad x\in \left[ -1,0\right)
,  \label{4.31}
\end{equation}%
where $H\left( x,\xi ,\lambda \right) $ is the resolvent of the operator $B$
that has the form (\ref{4.21}).

By subtracting (\ref{4.25}) from Eq. (\ref{4.24}), we get%
\begin{equation}
\chi \left( x\right) =\lambda \int\limits_{-1}^{0}h\left( x,\xi \right) 
\left[ \varphi \left( \xi \right) -\varphi ^{\prime }\left( \xi \right) %
\right] d\xi ,\quad x\in \left[ 0,1\right] ;  \label{4.32}
\end{equation}%
\begin{equation}
\varphi \left( x\right) -\varphi ^{\prime }\left( x\right) =\lambda
\int\limits_{-1}^{0}h\left( x,\xi \right) \left[ \varphi \left( \xi \right)
-\varphi ^{\prime }\left( \xi \right) \right] d\xi +\kappa \left( x\right)
\quad x\in \left[ -1,0\right) .  \label{4.33}
\end{equation}

From these relations, it follows that the function $\chi $ can be
constructively expressed via $\kappa $, i.e., by means of the solution of
the Fredholm integral equation of the second kind. Indeed, under the
condition (\ref{4.18}), $\varphi -\varphi ^{\prime }$ is determined via the
resolvent of the kernel $h\left( x,\xi \right) $ in (\ref{4.33}). However,
the inverse procedure, i.e., a representation of the function $\kappa $ via $%
\chi $, would be related to the solution of the Fredholm integral equation
of the first kind.

Let us add to Eqs. (\ref{4.24}), (\ref{4.25}) the "zero" from (\ref{4.1}),
i.e., $\mu A\psi -\mu f$ with the free term of the form (\ref{4.22}). As a
result, we obtain, respectively,%
\begin{equation}
\left. 
\begin{array}{c}
\psi \left( x\right) \\ 
\varphi \left( x\right)%
\end{array}%
\right\} =\lambda B\left( 
\begin{array}{c}
\psi \\ 
\varphi%
\end{array}%
\right) \left( x\right) +\left\{ 
\begin{array}{c}
\mu \left( A\psi \right) \left( x\right) -\mu f\left( x\right) ,\quad x\in 
\left[ 0,1\right] ; \\ 
\kappa \left( x\right) ,\quad x\in \left[ -1,0\right) ,%
\end{array}%
\right.  \label{4.34}
\end{equation}%
\begin{equation}
\left. 
\begin{array}{c}
\psi \left( x\right) \\ 
\varphi ^{\prime }\left( x\right)%
\end{array}%
\right\} =\lambda B\left( 
\begin{array}{c}
\psi \\ 
\varphi ^{\prime }%
\end{array}%
\right) \left( x\right) +\left\{ 
\begin{array}{c}
\mu \left( A\psi \right) \left( x\right) -\mu f\left( x\right) +\chi \left(
x\right) ,\quad x\in \left[ 0,1\right] ; \\ 
0,\quad x\in \left[ -1,0\right) .%
\end{array}%
\right.  \label{4.35}
\end{equation}

Thus, instead of the ill-posed problem (\ref{4.1}), in what follows we will
consider the two systems of integral equations (\ref{4.24}), (\ref{4.34})
and (\ref{4.25}), (\ref{4.35}).\footnote{%
Note that (\ref{4.34}), (\ref{4.35}) do not constitute Fredholm integral
equations of the second kind with respect to the functions (\ref{4.27}).}

\section{A constructive algorithm of practical realization}

A further orientation of transformations is, in a sense, opposed to the
previous one. Indeed, above, in a fact, we have done our best [beginning
with the model of the error (\ref{4.4})] to ensure that the sought function $%
\psi \left( x\right) $, as well as $\varphi \left( x\right) $ and $\varphi
^{\prime }\left( x\right) $, appear in specially constructed equations not
only under the sign of integration but also in an explicit form. As a
consequence, we have obtained (\ref{4.30}), a representation of the solution 
$\psi \left( x\right) $ with the function $\chi \left( x\right) $ also in an
explicit form.

It would be highly desirable to derive a different representation of $\psi
\left( x\right) $ that would apparently contain the data of the problem (\ref%
{4.1}) and where the function $\chi \left( x\right) $ would appear only
under sign of integration. Upon elimination of the function $\psi \left(
x\right) $ both from this representation and from (\ref{4.30}), we could
obtain a Fredholm integral equation of the second kind with respect to $\chi
\left( x\right) $.

Another way of achieving the same goal consists in the determination of the
integrand (\ref{4.32}) via $\chi \left( x\right) $. Since the function $%
\varphi ^{\prime }\left( x\right) $ is, in this sense, known [see (\ref{4.31}%
)], it is necessary to establish a relationship between $\varphi $, $\chi $
and the data of the problem.

The realization of each of the two outlined versions can be represented in
the context of the reduction of (\ref{4.35}) to the form (\ref{4.34}). The
grounds for this reduction lie in the fact that the function $\psi \left(
x\right) $ enters both the equations and that their structure is analogous.
These are heuristic arguments.

In order to eliminate the function $\chi \left( x\right) $ from (\ref{4.35}%
), we use the equation%
\begin{equation}
\left. 
\begin{array}{c}
\psi _{0}\left( x\right) \\ 
\varphi _{0}^{\prime }\left( x\right)%
\end{array}%
\right\} =\lambda B\left( 
\begin{array}{c}
\psi _{0} \\ 
\varphi _{0}^{\prime }%
\end{array}%
\right) \left( x\right) +\left\{ 
\begin{array}{c}
\mu \left( A\psi _{0}\right) \left( x\right) +\chi \left( x\right) ,\quad
x\in \left[ 0,1\right] ; \\ 
0,\quad x\in \left[ -1,0\right) .%
\end{array}%
\right.  \label{4.36}
\end{equation}%
By subtracting this equation, we get%
\[
\left. 
\begin{array}{c}
\psi \left( x\right) -\psi _{0}\left( x\right) \\ 
\varphi ^{\prime }\left( x\right) -\varphi _{0}^{\prime }\left( x\right)%
\end{array}%
\right\} =\lambda B\left( 
\begin{array}{c}
\psi -\psi _{0} \\ 
\varphi ^{\prime }-\varphi _{0}^{\prime }%
\end{array}%
\right) \left( x\right) 
\]%
\begin{equation}
+\left\{ 
\begin{array}{c}
\mu A\left( \psi -\psi _{0}\right) \left( x\right) -\mu f\left( x\right)
,\quad x\in \left[ 0,1\right] ; \\ 
0,\quad x\in \left[ -1,0\right) ,%
\end{array}%
\right.  \label{4.37}
\end{equation}%
or%
\begin{equation}
\left. 
\begin{array}{c}
\psi _{1}\left( x\right) \\ 
\varphi _{1}^{\prime }\left( x\right)%
\end{array}%
\right\} =\lambda B\left( 
\begin{array}{c}
\psi _{1} \\ 
\varphi _{1}^{\prime }%
\end{array}%
\right) \left( x\right) +\left\{ 
\begin{array}{c}
\mu \left( A\psi _{1}\right) \left( x\right) -\mu f\left( x\right) ,\quad
x\in \left[ 0,1\right] ; \\ 
0,\quad x\in \left[ -1,0\right) ,%
\end{array}%
\right.  \label{4.38}
\end{equation}%
where%
\begin{equation}
\psi _{1}\left( x\right) =\psi \left( x\right) -\psi _{0}\left( x\right)
;\quad \varphi _{1}^{\prime }\left( x\right) =\varphi ^{\prime }\left(
x\right) -\varphi _{0}^{\prime }\left( x\right) .  \label{4.39}
\end{equation}

If introduce the notation%
\[
\Psi _{1}\left( x\right) =\left\{ 
\begin{array}{c}
\psi _{1}\left( x\right) ,\quad x\in \left[ 0,1\right] ; \\ 
\varphi _{1}^{\prime }\left( x\right) ,\quad x\in \left[ -1,0\right) ,%
\end{array}%
\right. 
\]%
equation (\ref{4.38}) takes the form%
\[
\Psi _{1}\left( x\right) =\lambda \left( B\Psi _{1}\right) \left( x\right)
+\left\{ 
\begin{array}{c}
\mu \left( A\psi _{1}\right) \left( x\right) -\mu f\left( x\right) ,\quad
x\in \left[ 0,1\right] ; \\ 
0,\quad x\in \left[ -1,0\right) .%
\end{array}%
\right. 
\]

This is a Fredholm integral equation of the second kind with respect to $%
\Psi _{1}$. The inversion of the operator $I-\lambda B$ under the condition (%
\ref{4.20}), taking into account (\ref{4.1}), yields:%
\begin{equation}
\psi _{1}\left( x\right) =\lambda \int\limits_{0}^{1}K\left( x,\xi \right)
\psi _{1}\left( \xi \right) d\xi +f_{1}\left( x\right) ,\quad x\in \left[ 0,1%
\right] ;  \label{4.40}
\end{equation}%
\[
\varphi _{1}^{\prime }\left( x\right) =\mu \lambda \int\limits_{0}^{1}\left[
\psi _{1}\left( \xi \right) \int\limits_{0}^{1}H\left( x,\zeta ,\lambda
\right) k\left( \zeta ,\xi \right) d\zeta \right. 
\]%
\begin{equation}
\left. -H\left( x,\xi ,\lambda \right) f\left( \xi \right) \right] d\xi
,\quad x\in \left[ -1,0\right) ,  \label{4.41}
\end{equation}%
where%
\[
K\left( x,\xi \right) =k\left( x,\xi \right) +\lambda
\int\limits_{0}^{1}H\left( x,\zeta ,\lambda \right) k\left( \zeta ,\xi
\right) d\zeta ; 
\]%
\[
f_{1}\left( x\right) =-\mu \left[ f\left( x\right) +\lambda
\int\limits_{0}^{1}H\left( x,\xi ,\lambda \right) f\left( \xi \right) d\xi %
\right] . 
\]

Thus, the function $\psi _{1}\left( x\right) $ is determined from the
Fredholm integral equation of the second kind (\ref{4.40}) and depends only
on the data (\ref{4.1}) and on the chosen kernel $h\left( x,\xi \right) $.%
\footnote{%
As a matter of fact, $\psi _{1}$ represents the part of the solution (\ref%
{4.35}) on $x\in \left[ 0,1\right] $ that is stipulated by the component of
the free term $-\mu f$.} Here, we assume that%
\[
\mu \neq \mu _{n},\quad n=1,2,\ldots , 
\]%
where $\mu _{n}$ are the characteristic numbers of the homogeneous equation
obtained from (\ref{4.40}) in the case $f_{1}\equiv 0$. The values of $\mu
_{n}$, as well as the solution of (\ref{4.40}), should be found by means of
approximate methods \cite{4.5}. After that, the function $\varphi
_{1}^{\prime }\left( x\right) $ is evaluated from the formula (\ref{4.41}).

However, Eq. (\ref{4.37}) can be regarded as Eq. (\ref{4.34}). Indeed, the
elimination of $\chi \left( x\right) $ from Eq. (\ref{4.35}) is,
figuratively, equivalent to a flow of this function to $\kappa \left(
x\right) $ with the appearance of Eq. (\ref{4.34}). Consequently, what is
needed is an identification of the functions $\varphi \left( x\right) $ and $%
\kappa \left( x\right) $ on the basis of (\ref{4.34}) in the structure of
Eq. (\ref{4.37}).

To this end, we use Eq. (\ref{4.37}) on $x\in \left[ -1,0\right) $,%
\[
\varphi ^{\prime }\left( x\right) -\varphi _{0}^{\prime }\left( x\right)
=\lambda \int\limits_{-1}^{0}h\left( x,\xi \right) \left[ \varphi ^{\prime
}\left( \xi \right) -\varphi _{0}^{\prime }\left( \xi \right) \right] d\xi 
\]%
\begin{equation}
+\lambda \int\limits_{0}^{1}h\left( x,\xi \right) \left[ \psi \left( \xi
\right) -\psi _{0}\left( \xi \right) \right] d\xi ,  \label{4.42}
\end{equation}%
paying attention to the method of its derivation. It consists in the
elimination from Eq. (\ref{4.35}) of the part of the solution that depends
on the component of the free term $\chi \left( x\right) $. However, in this
procedure the functions satisfying this equation both on $x\in \left[
-1,0\right) $ and on $x\in \left[ 0,1\right] $ have changed. In other words,
both the functions $\varphi ^{\prime }$ and $\psi $ have undergone change.

At the same time, the structure of Eqs. (\ref{4.34}), (\ref{4.35}) implies a
transformation of one of these equations into the other by means of a change
of the contained functions only on $x\in \left[ -1,0\right) $, that is, of $%
\varphi ^{\prime }$ and $\varphi $.\footnote{%
What was in position of $\psi $ in (\ref{4.35}) must remain unchanged.}
Therefore, we will correct $\varphi ^{\prime }\left( x\right) $ in Eq. (\ref%
{4.42}) in order to eliminate the term with the function $\psi _{0}\left(
x\right) $. Accordingly, we must include in $\kappa \left( x\right) $ the
terms of Eq. (\ref{4.42}) that contain the function $\varphi _{0}^{\prime
}\left( x\right) $.

As a result, there appear the relations%
\begin{equation}
\kappa \left( x\right) =\varphi _{0}^{\prime }\left( x\right) -\lambda
\int\limits_{-1}^{0}h\left( x,\xi \right) \varphi _{0}^{\prime }\left( \xi
\right) d\xi ,\quad x\in \left[ -1,0\right) ;  \label{4.43}
\end{equation}%
\begin{equation}
\varphi \left( x\right) =\varphi ^{\prime }\left( x\right) -\varphi
_{0}\left( x\right) ,\quad x\in \left[ -1,0\right) .  \label{4.44}
\end{equation}

Here, $\varphi _{0}\left( x\right) $ is the solution of the Fredholm
integral equation of the second kind%
\begin{equation}
\varphi _{0}\left( x\right) =\lambda \int\limits_{-1}^{0}h\left( x,\xi
\right) \varphi _{0}\left( \xi \right) d\xi +f_{0}\left( x\right) ,\quad
x\in \left[ -1,0\right) ,  \label{4.45}
\end{equation}%
where%
\[
f_{0}\left( x\right) =-\lambda \int\limits_{0}^{1}h\left( x,\xi \right) \psi
_{0}\left( \xi \right) d\xi , 
\]%
under the condition (\ref{4.18}).

Subtracting (\ref{4.45}) from Eq. (\ref{4.42}), we get%
\[
\varphi ^{\prime }\left( x\right) -\varphi _{0}^{\prime }\left( x\right)
-\varphi _{0}\left( x\right) =\lambda \int\limits_{-1}^{0}h\left( x,\xi
\right) \left[ \varphi ^{\prime }\left( \xi \right) -\varphi _{0}^{\prime
}\left( \xi \right) -\varphi _{0}\left( \xi \right) \right] d\xi 
\]%
\begin{equation}
+\lambda \int\limits_{-1}^{0}h\left( x,\xi \right) \psi \left( \xi \right)
d\xi ,\quad x\in \left[ -1,0\right) .  \label{4.46}
\end{equation}

Equation (\ref{4.35}) on $x\in \left[ -1,0\right) $ has the form%
\begin{equation}
\varphi ^{\prime }\left( x\right) =\lambda \int\limits_{-1}^{0}h\left( x,\xi
\right) \varphi ^{\prime }\left( \xi \right) d\xi +\lambda
\int\limits_{0}^{1}h\left( x,\xi \right) \psi \left( \xi \right) d\xi .
\label{4.47}
\end{equation}%
Its comparison with (\ref{4.46}) yields:%
\begin{equation}
\varphi _{0}\left( x\right) =-\varphi _{0}^{\prime }\left( x\right) ,
\label{4.48}
\end{equation}%
that is, we have, in fact, returned from (\ref{4.42}) to Eq. (\ref{4.35}) on 
$x\in \left[ -1,0\right) $ in such a way that allows us to establish this
relation.

It should be noted that relations (\ref{4.43}), (\ref{4.44}) transform (\ref%
{4.46}) into (\ref{4.23}). Now we will show that relations (\ref{4.43}), (%
\ref{4.44}) and (\ref{4.48}) indeed reduce Eq. (\ref{4.35}) to the form (\ref%
{4.34}). To this end, we turn to Eq. (\ref{4.35}) on $x\in \left[ 0,1\right] 
$:%
\[
\psi \left( x\right) =\lambda \int\limits_{-1}^{0}h\left( x,\xi \right)
\varphi ^{\prime }\left( \xi \right) d\xi +\lambda
\int\limits_{0}^{1}h\left( x,\xi \right) \psi \left( \xi \right) d\xi 
\]%
\begin{equation}
+\mu \left( A\psi \right) \left( x\right) -\mu f\left( x\right) +\chi \left(
x\right) .  \label{4/49}
\end{equation}

Using (\ref{4.44}) and (\ref{4.32}), we get:%
\[
\lambda \int\limits_{-1}^{0}h\left( x,\xi \right) \varphi ^{\prime }\left(
\xi \right) d\xi =\lambda \int\limits_{-1}^{0}h\left( x,\xi \right) \varphi
\left( \xi \right) d\xi +\lambda \int\limits_{-1}^{0}h\left( x,\xi \right)
\varphi _{0}\left( \xi \right) d\xi , 
\]%
where 
\[
\lambda \int\limits_{-1}^{0}h\left( x,\xi \right) \varphi _{0}\left( \xi
\right) d\xi =-\lambda \int\limits_{-1}^{0}h\left( x,\xi \right) \left[
\varphi \left( \xi \right) -\varphi ^{\prime }\left( \xi \right) \right]
d\xi =-\chi \left( x\right) , 
\]%
which, by means of substitution of the above expressions (\ref{4/49}), is
transformed into Eq. (\ref{4.34}) on $x\in \left[ 0,1\right] $.

The substitution of the function $\varphi ^{\prime }\left( x\right) $ from (%
\ref{4.44}) into (\ref{4.47}), with the use of (\ref{4.48}) and (\ref{4.43}%
), leads to Eq. (\ref{4.34}) on $x\in \left[ -1,0\right) $.\footnote{%
What was in position of $\psi $ in (\ref{4.23}) must remain unchanged.}
Thus, by means of the established relations, Eq. (\ref{4.35}), both on $x\in %
\left[ -1,0\right) $ and on $x\in \left[ 0,1\right] $, is transformed into
Eq. (\ref{4.34}).

By (\ref{4.44}), (\ref{4.48}) and (\ref{4.39}),%
\begin{equation}
\varphi \left( x\right) -\varphi ^{\prime }\left( x\right) =-\varphi
_{0}\left( x\right) =\varphi _{0}^{\prime }\left( x\right) =\varphi ^{\prime
}\left( x\right) -\varphi _{1}^{\prime }\left( x\right) ,  \label{4.50}
\end{equation}%
and, as a result, expression (\ref{4.32}) takes the form%
\begin{equation}
\chi \left( x\right) =\lambda \int\limits_{-1}^{0}h\left( x,\xi \right) 
\left[ \varphi ^{\prime }\left( \xi \right) -\varphi _{1}^{\prime }\left(
\xi \right) \right] d\xi .  \label{4.51}
\end{equation}

The derivation of relations (\ref{4.43}), (\ref{4.44}) as well as (\ref{4.48}%
) and, finally, (\ref{4.50}) is the main link in the construction of the
algorithm.

The substitution of expression (\ref{4.51}) into (\ref{4.31}) leads to a
Fredholm integral equation of the second kind:%
\begin{equation}
\chi \left( x\right) =\Lambda \int\limits_{0}^{1}l\left( x,\xi \right) \chi
\left( \xi \right) d\xi +q\left( x\right) ,\quad x\in \left[ 0,1\right] ,
\label{4.52}
\end{equation}%
where $\Lambda =\lambda ^{2}$;%
\begin{equation}
l\left( x,\xi \right) =\int\limits_{-1}^{0}h\left( x,\zeta \right) H\left(
\zeta ,\xi ,\lambda \right) d\zeta ;  \label{4.53}
\end{equation}%
\begin{equation}
q\left( x\right) =-\lambda \int\limits_{-1}^{0}h\left( x,\xi \right) \varphi
_{1}^{\prime }\left( \xi \right) d\xi .  \label{4.54}
\end{equation}

Expression (\ref{4.53}), after the substitution of (\ref{4.19}) and (\ref%
{4.21}), takes the form%
\[
l\left( x,\xi \right) =\frac{1}{1-2\lambda }+2\sum_{n=1}^{\infty }\frac{%
r^{2n}}{1-2\lambda r^{n}}\left[ \cos \left( 2n\pi x\right) \cos \left( 2n\pi
\xi \right) \right. 
\]%
\[
\left. +\sin \left( 2n\pi x\right) \sin \left( 2n\pi \xi \right) \right] =%
\frac{\bar{\varphi}_{0}\left( x\right) \bar{\varphi}_{0}\left( \xi \right) }{%
1-2\lambda } 
\]%
\begin{equation}
+\sum_{n=1}^{\infty }\left( \frac{1-2\lambda r^{n}}{r^{2n}}\right) ^{-1}%
\left[ \bar{\varphi}_{2n-1}\left( x\right) \bar{\varphi}_{2n-1}\left( \xi
\right) +\bar{\varphi}_{2n}\left( x\right) \bar{\varphi}_{2n}\left( \xi
\right) \right] ,\quad x\in \left[ 0,1\right] ,  \label{4.55}
\end{equation}%
where $\bar{\varphi}_{2n-1}\left( x\right) $ and $\bar{\varphi}_{2n}\left(
x\right) $ are the eigenfunctions of (\ref{4.17}), orthonormal on $x\in %
\left[ 0,1\right] $. This fact allows us to determine the resolvent of the
kernel $l\left( x,\xi \right) $. Indeed, its characteristic numbers are%
\[
\Lambda _{0}=1-2\lambda ;\quad \Lambda _{2n-1}=\Lambda _{2n}=\frac{%
1-2\lambda r^{n}}{r^{2n}},\quad n=1,2,\ldots , 
\]%
and, because of the property $0<r<1$ for a bounded $\lambda $, which is
assumed, only a limited number of these can take on negative values. By
Mercer's theorem \cite{4.1}, expression (\ref{4.55}) is a bilinear expansion
of the symmetric continuos kernel $l\left( x,\xi \right) $, $0\leq x,\xi
\leq 1$. Under the condition%
\[
\Lambda \neq \Lambda _{0};\quad \Lambda \neq \Lambda _{2n},\quad
n=1,2,\ldots , 
\]%
which is equivalent to (\ref{4.20}), its resolvent takes the form%
\[
L\left( x,\xi ,\Lambda \right) =\frac{1}{1-2\lambda -\lambda ^{2}} 
\]%
\begin{equation}
+2\sum_{n=1}^{\infty }\frac{r^{2n}}{1-2\lambda r^{n}-\lambda ^{2}r^{2n}}%
\left[ \cos \left( 2n\pi x\right) \cos \left( 2n\pi \xi \right) +\sin \left(
2n\pi x\right) \sin \left( 2n\pi \xi \right) \right] .  \label{4.56}
\end{equation}%
As a result, the solution of (\ref{4.52}) can be represented as follows:%
\begin{equation}
\chi \left( x\right) =q\left( x\right) +\Lambda \int\limits_{0}^{1}L\left(
x,\xi ,\Lambda \right) q\left( \xi \right) d\xi .  \label{4.57}
\end{equation}

Obviously, for the convergence of the series (\ref{4.56}), in addition to (%
\ref{4.18}) and (\ref{4.20}), it is necessary that the following condition
be fulfilled:%
\begin{equation}
\lambda \neq \left( -1\pm \sqrt{2}\right) r^{-n},\quad n=0,1,\ldots .
\label{4.58}
\end{equation}

The substitution of expression (\ref{4.57}) into (\ref{4.30}), by use of (%
\ref{4.21}), allows us to evaluate the function $\psi \left( x\right) $,
which is the solution of the considered problem.

The procedure of the numerical realization includes the following stages:

- concretization of the parameter $0<r<1;$

- determination of the parameter $\lambda $ from the conditions (\ref{4.18}%
), (\ref{4.20}) and (\ref{4.58}), taking also account of (\ref{4.4}), that
is, 
\begin{equation}
\lambda \neq 0,\quad \lambda \neq r^{-n},\quad \lambda \neq \frac{1}{2}%
r^{-n},\quad \lambda \neq \left( -1\pm \sqrt{2}\right) r^{-n},\quad
n=1,2,\ldots ;  \label{4.59}
\end{equation}

- determination of the parameter $\mu $ in (\ref{4.40}), so that the equation%
\begin{equation}
\psi \left( x\right) =\mu \int\limits_{0}^{1}K\left( x,\xi \right) \psi
\left( \xi \right) d\xi ,\quad x\in \left[ 0,1\right]  \label{4.60}
\end{equation}%
possess only the trivial solution;

- determination of the function $\psi _{1}$ from Eq. (\ref{4.40});

- evaluation of the function $\varphi _{1}^{\prime }$ by formula (\ref{4.41}%
);

- evaluation of the function $q$ by formula (\ref{4.54});

- evaluation of the function $\chi $ by formula (\ref{4.57});

- evaluation of the sought function $\psi $ by formula (\ref{4.30}).

Note that the realization of the algorithm is related with the use of
quadrature and cubature formulas on a two-dimensional domain \cite{4.6}.
Simultaneously, one can apply the technique of the improvement of
convergence of trigonometric series (\cite{4.7}, pp. 187-193) and the
methods of integration of oscillating functions (\cite{4.8}, pp. 112-115).

\section{The reliability of the obtained results}

Thus, the function $\psi \left( x\right) $, i.e., the solution of the
problem (\ref{4.1}) in its restricted formulation (see section 5.2), is
determined by formula (\ref{4.30}). At the same time, expressions (\ref{4.28}%
)-(\ref{4.31}) that represent the solution of Eqs. (\ref{4.24}), (\ref{4.25}%
) satisfy these equations identically, irrespective of the form of $\kappa
\left( x\right) $ and $\chi \left( x\right) $.

Therefore, one cannot argue on the basis of simple subtraction of Eq. (\ref%
{4.25}) from (\ref{4.35}) that the solution of the former equation also
satisfies Eq. (\ref{4.1}). In the general case, solutions of these equations
can be completely different.

Accordingly, one has to show that the function $\psi \left( x\right) $,
determined by expression (\ref{4.30}), that together with $\varphi ^{\prime
}\left( x\right) $ satisfies Eq. (\ref{4.25}) is also the solution of Eq. (%
\ref{4.35}) on $x\in \left[ 0,1\right] $. To this end, it is reasonable to
introduce new notation for the functions $\psi \left( x\right) $, $\varphi
\left( x\right) $, $\varphi ^{\prime }\left( x\right) $ entering Eqs. (\ref%
{4.24}), (\ref{4.25}) and (\ref{4.34}), (\ref{4.35}), namely, $\tilde{\psi}%
\left( x\right) $, $\tilde{\varphi}\left( x\right) $, $\tilde{\varphi}%
^{\prime }\left( x\right) $ and $\check{\psi}\left( x\right) $, $\check{%
\varphi}\left( x\right) $, $\check{\varphi}^{\prime }\left( x\right) $,
respectively.

By use of the above-mentioned pairs of equations, respectively, the
following relations have been obtained in section 5.4:%
\begin{equation}
\chi \left( x\right) =\lambda \int\limits_{-1}^{0}h\left( x,\xi \right) 
\left[ \tilde{\varphi}\left( \xi \right) -\tilde{\varphi}^{\prime }\left(
\xi \right) \right] d\xi  \label{4.61}
\end{equation}%
and%
\[
\check{\varphi}_{1}^{\prime }\left( x\right) =\check{\varphi}^{\prime
}\left( x\right) -\check{\varphi}_{0}^{\prime }\left( x\right) ;\quad \check{%
\varphi}\left( x\right) =\check{\varphi}^{\prime }\left( x\right) -\check{%
\varphi}_{0}\left( x\right) ;\quad \check{\varphi}_{0}\left( x\right) =-%
\check{\varphi}_{0}^{\prime }\left( x\right) 
\]%
[see (\ref{4.32}) and (\ref{4.39}), (\ref{4.44}), (\ref{4.48})] or%
\[
\check{\varphi}\left( x\right) -\check{\varphi}^{\prime }\left( x\right) =-%
\check{\varphi}_{0}\left( x\right) =\check{\varphi}_{0}^{\prime }\left(
x\right) =\check{\varphi}^{\prime }\left( x\right) -\check{\varphi}%
_{1}^{\prime }\left( x\right) 
\]%
[see (\ref{4.50})]. Finally, in a short form,%
\begin{equation}
\check{\varphi}\left( x\right) -\check{\varphi}^{\prime }\left( x\right) =%
\check{\varphi}^{\prime }\left( x\right) -\check{\varphi}_{1}^{\prime
}\left( x\right) .  \label{4.62}
\end{equation}

However, only the relation%
\begin{equation}
\tilde{\varphi}\left( x\right) -\tilde{\varphi}^{\prime }\left( x\right) =%
\tilde{\varphi}^{\prime }\left( x\right) -\check{\varphi}_{1}^{\prime
}\left( x\right) ,  \label{4.63}
\end{equation}%
where%
\[
\tilde{\varphi}^{\prime }\left( x\right) =\lambda \int\limits_{0}^{1}H\left(
x,\xi ,\lambda \right) \chi \left( \xi \right) d\xi , 
\]%
has been used [see (\ref{4.31})]. Upon substitution into (\ref{4.61}), this
leads to the relation%
\[
\chi \left( x\right) =\lambda \int\limits_{-1}^{0}h\left( x,\xi \right) 
\left[ \tilde{\varphi}^{\prime }\left( \xi \right) -\check{\varphi}%
_{1}^{\prime }\left( \xi \right) \right] d\xi 
\]%
[see (\ref{4.51})], which has, as a result, Eq. (\ref{4.52}).

In other words, we have substituted the function $\tilde{\varphi}^{\prime
}\left( x\right) $ into the right-hand side of (\ref{4.62}), in place of $%
\check{\varphi}^{\prime }\left( x\right) $. In general, the derivation of (%
\ref{4.63}) has been as follows:%
\[
\tilde{\varphi}\left( x\right) -\tilde{\varphi}^{\prime }\left( x\right)
\equiv \check{\varphi}\left( x\right) -\check{\varphi}^{\prime }\left(
x\right) =\check{\varphi}^{\prime }\left( x\right) -\check{\varphi}%
_{1}^{\prime }\left( x\right) =\tilde{\varphi}^{\prime }\left( x\right) -%
\check{\varphi}_{1}^{\prime }\left( x\right) , 
\]%
i.e., the two premises%
\[
\tilde{\varphi}\left( x\right) -\tilde{\varphi}^{\prime }\left( x\right)
\equiv \check{\varphi}\left( x\right) -\check{\varphi}^{\prime }\left(
x\right) ,\quad \tilde{\varphi}^{\prime }\left( x\right) \equiv \check{%
\varphi}^{\prime }\left( x\right) 
\]%
and relation (\ref{4.62}) have been used.

Indeed, when these identities are satisfied, relation (\ref{4.63}) turns
into (\ref{4.62}). Thus, one can conclude that the above-mentioned premises,
i.e., the identities%
\begin{equation}
\tilde{\varphi}\left( x\right) \equiv \check{\varphi}\left( x\right) ;\quad 
\tilde{\varphi}^{\prime }\left( x\right) \equiv \check{\varphi}^{\prime
}\left( x\right) ,  \label{4.64}
\end{equation}%
constitute sufficient conditions for the reduction of the problem (\ref{4.24}%
), (\ref{4.34}) and (\ref{4.25}), (\ref{4.35}) to the solution of Eq. (\ref%
{4.52}).

At the same time, they are also necessary. Indeed, by (\ref{4.62}), (\ref%
{4.63}),%
\begin{equation}
\tilde{\varphi}\left( x\right) -\check{\varphi}\left( x\right) =2\left[ 
\tilde{\varphi}^{\prime }\left( x\right) -\check{\varphi}^{\prime }\left(
x\right) \right] .  \label{4.65}
\end{equation}%
Analogously, that is, by subtraction of Eqs. (\ref{4.34}), (\ref{4.35}) from
(\ref{4.24}), (\ref{4.25}), respectively, we get%
\[
\tilde{\varphi}\left( x\right) -\check{\varphi}\left( x\right) =\lambda
\int\limits_{-1}^{0}h\left( x,\xi \right) \left[ \tilde{\varphi}\left( \xi
\right) -\check{\varphi}\left( \xi \right) \right] d\xi 
\]%
\[
+\lambda \int\limits_{0}^{1}h\left( x,\xi \right) \left[ \tilde{\psi}\left(
\xi \right) -\check{\psi}\left( \xi \right) \right] d\xi ; 
\]%
\[
\tilde{\varphi}^{\prime }\left( x\right) -\check{\varphi}^{\prime }\left(
x\right) =\lambda \int\limits_{-1}^{0}h\left( x,\xi \right) \left[ \tilde{%
\varphi}^{\prime }\left( \xi \right) -\check{\varphi}^{\prime }\left( \xi
\right) \right] d\xi 
\]%
\[
+\lambda \int\limits_{0}^{1}h\left( x,\xi \right) \left[ \tilde{\psi}\left(
\xi \right) -\check{\psi}\left( \xi \right) \right] d\xi . 
\]

As a result of the subtraction, with the use of (\ref{4.65}), there arise
homogeneous equations:%
\[
\tilde{\varphi}\left( x\right) -\check{\varphi}\left( x\right) =\lambda
\int\limits_{-1}^{0}h\left( x,\xi \right) \left[ \tilde{\varphi}\left( \xi
\right) -\check{\varphi}\left( \xi \right) \right] d\xi ; 
\]%
\[
\tilde{\varphi}^{\prime }\left( x\right) -\check{\varphi}^{\prime }\left(
x\right) =\lambda \int\limits_{-1}^{0}h\left( x,\xi \right) \left[ \tilde{%
\varphi}^{\prime }\left( \xi \right) -\check{\varphi}^{\prime }\left( \xi
\right) \right] d\xi ,\quad x\in \left[ -1,0\right) , 
\]%
whose solution under the condition (\ref{4.18}) is trivial. Thus, relations (%
\ref{4.62}), (\ref{4.63}) automatically result in the identities (\ref{4.64}%
). In other words, the existence of the above-mentioned relations imply that
the functions $\varphi \left( x\right) $, $\varphi ^{\prime }\left( x\right) 
$ in Eqs. (\ref{4.24}), (\ref{4.25}) and (\ref{4.34}), (\ref{4.35}),
respectively, are the same.

The procedure of subtraction in each of the pairs of the equations yields:%
\[
\int\limits_{0}^{1}h\left( x,\xi \right) \left[ \tilde{\psi}\left( \xi
\right) -\check{\psi}\left( \xi \right) \right] d\xi =0,\quad x\in \left[
-1,0\right) , 
\]%
and, by the change of variables%
\[
\zeta =2\pi \left( 1+x\right) ;\quad \theta =2\pi \xi , 
\]%
we get%
\[
\int\limits_{0}^{2\pi }h\left( \zeta ,\theta \right) \left[ \tilde{\psi}%
\left( \theta \right) -\check{\psi}\left( \theta \right) \right] d\theta
=0,\quad \theta \in \left[ 0,2\pi \right] . 
\]

This is a homogeneous Fredholm integral equation of the first kind with the
kernel (\ref{4.11}). As it is closed, we can conclude that%
\[
\tilde{\psi}\left( x\right) \equiv \check{\psi}\left( x\right) \equiv \check{%
\psi}_{0}\left( x\right) +\check{\psi}_{1}\left( x\right) \equiv \psi
_{0}\left( x\right) +\psi _{1}\left( x\right) \equiv \psi \left( x\right) . 
\]

Consequently, in order that the functions $\psi \left( x\right) $, $\varphi
^{\prime }\left( x\right) $, determined by formulas (\ref{4.30}), (\ref{4.31}%
), satisfy both Eq. (\ref{4.25}) and Eq. (\ref{4.35}) as well as their
difference, Eq. (\ref{4.1}), the function $\chi \left( x\right) $ must
represent the solution of the Fredholm integral equation of the second kind (%
\ref{4.52}). This is a very important point of the whole consideration.

Note that, instead of (\ref{4.64}), one could employ a single identity%
\[
\tilde{\varphi}\left( x\right) -\tilde{\varphi}^{\prime }\left( x\right)
\equiv \check{\varphi}\left( x\right) -\check{\varphi}^{\prime }\left(
x\right) . 
\]%
However, in this case, a Fredholm integral equation of the second kind,
obtained by the substitution of expression (\ref{4.63}) into (\ref{4.61}),
would be more cumbersome.

\section{An arbitrary function from $L_{2}$ as the solution}

Beginning from section 5.2 and up to the present point, we have assumed that
the function $\psi \left( x\right) $ satisfying Eq. (\ref{4.1}) can be only
harmonic. Accordingly, its free term $f\left( x\right) $ is determined by
expression (\ref{4.22}). Here, we present a generalization of the algorithm
of section 5.4. To this end, we will employ an approach which is analogous
to Abel-Poisson's method of the summation of Fourier series \cite{4.2,4.3}:

- execution of the transformation in an analytical form with a harmonic
function $\psi \left( r,x\right) $ that is represented by a well convergent
series for $0<r<1$ in (\ref{4.19});

- a passage to the limit $r\rightarrow 1$ in the expression for $\psi \left(
r,x\right) $ via the data of the problem that is represented by a series
whose terms explicitly depend on the parameter $r$.

In this way, we will obtained the solution of the problem posed in section
5.1: namely, the restoration of the $L_{2}$ - function $\psi \left( x\right) 
$ from the results of integration according to formula (\ref{4.1}) or from a
given related expression for $f\left( x\right) $.

In Eq. (\ref{4.40}), we use the following representations:%
\begin{equation}
\psi _{1}\left( x\right) =\frac{1}{2}s_{0}+\sum_{n=1}^{\infty }s_{n}\cos
\left( 2n\pi x\right) +s_{n}^{\prime }\sin \left( 2n\pi x\right) ,
\label{4.66}
\end{equation}%
where $s_{0}$, $s_{n}$ and $s_{n}^{\prime }$ are undefined coefficients;%
\begin{equation}
k\left( x,\xi \right) =\frac{1}{2}k_{0}\left( \xi \right)
+\sum_{n=1}^{\infty }k_{n}\left( \xi \right) \cos \left( 2n\pi x\right)
+k_{n}^{\prime }\left( \xi \right) \sin \left( 2n\pi x\right) ,  \label{4.67}
\end{equation}%
\begin{equation}
f\left( x\right) =\frac{1}{2}c_{0}+\sum_{n=1}^{\infty }c_{n}\cos \left(
2n\pi x\right) +c_{n}^{\prime }\sin \left( 2n\pi x\right) ,  \label{4.68}
\end{equation}%
where the Fourier coefficients are given by%
\[
k_{0}\left( \xi \right) =2\int\limits_{0}^{1}k\left( x,\xi \right) dx;\quad
k_{n}\left( \xi \right) =2\int\limits_{0}^{1}k\left( x,\xi \right) \cos
\left( 2n\pi x\right) dx; 
\]%
\begin{equation}
k_{n}^{\prime }\left( \xi \right) =2\int\limits_{0}^{1}k\left( x,\xi \right)
\sin \left( 2n\pi x\right) dx,\quad n=1,2,\ldots  \label{4.69}
\end{equation}%
(note that explicit evaluation of these functions is unnecessary);%
\[
c_{0}=2\int\limits_{0}^{1}f\left( \xi \right) d\xi ;\quad
c_{n}=2\int\limits_{0}^{1}f\left( \xi \right) \cos \left( 2n\pi \xi \right)
d\xi ; 
\]%
\begin{equation}
c_{n}^{\prime }=2\int\limits_{0}^{1}f\left( \xi \right) \sin \left( 2n\pi
\xi \right) d\xi ,\quad n=1,2,\ldots .  \label{4.70}
\end{equation}

Accordingly,%
\[
K\left( x,\xi \right) =\frac{1-\lambda }{2\left( 1-2\lambda \right) }%
k_{0}\left( \xi \right) 
\]%
\begin{equation}
+\sum_{n=1}^{\infty }\frac{1-\lambda r^{n}}{1-2\lambda r^{n}}\left[
k_{n}\left( \xi \right) \cos \left( 2n\pi x\right) +k_{n}^{\prime }\left(
\xi \right) \sin \left( 2n\pi x\right) \right] ;  \label{4.71}
\end{equation}%
\[
f_{1}\left( x\right) =\frac{\mu \left( 1-\lambda \right) }{2\left(
1-2\lambda \right) }c_{0} 
\]%
\begin{equation}
-\sum_{n=1}^{\infty }\frac{\mu \left( 1-\lambda r^{n}\right) }{1-2\lambda
r^{n}}\left[ c_{n}\cos \left( 2n\pi x\right) +c_{n}^{\prime }\sin \left(
2n\pi x\right) \right] .  \label{4.72}
\end{equation}

On substitution of expressions (\ref{4.66})-(\ref{4.68}) into Eq. (\ref{4.40}%
) and reduction of the factors multiplying $\cos \left( 2n\pi x\right) $, $%
\sin \left( 2n\pi x\right) $, the evaluation of the coefficients $s_{0}$, $%
s_{n}$, $s_{n}^{\prime }$ reduce to the solution of the following linear
algebraic equations:%
\[
\left[ 2\left( 1-2\lambda \right) -\mu \left( 1-\lambda \right) p_{00}\right]
s_{0}=2\mu \left( 1-\lambda \right) \sum_{m=1}^{\infty
}p_{0m}s_{m}+p_{0m}^{\prime }s_{m}^{\prime }-2\mu \left( 1-\lambda \right)
c_{0}; 
\]%
\[
2\left[ 1-2\lambda r^{n}-\mu \left( 1-\lambda r^{n}\right) p_{nn}\right]
s_{n}=\mu \left( 1-\lambda r^{n}\right) p_{n0}s_{0}+2\mu \left( 1-\lambda
r^{n}\right) \sum_{m=1,\,m\neq n}^{\infty }p_{nm}s_{m} 
\]%
\[
+2\mu \left( 1-\lambda r^{n}\right) \sum_{m=1}^{\infty }p_{nm}^{\prime
}s_{m}^{\prime }-2\mu \left( 1-\lambda r^{n}\right) c_{n}; 
\]%
\[
2\left[ 1-2\lambda r^{n}-\mu \left( 1-\lambda r^{n}\right) p_{nn}^{\prime
\prime \prime }\right] s_{n}^{\prime }=\mu \left( 1-\lambda r^{n}\right)
p_{n0}^{\prime }s_{0}^{\prime }+2\mu \left( 1-\lambda r^{n}\right)
\sum_{m=1}^{\infty }p_{nm}^{\prime \prime }s_{m} 
\]%
\begin{equation}
+2\mu \left( 1-\lambda r^{n}\right) \sum_{m=1,\,m\neq n}^{\infty
}p_{nm}^{\prime \prime \prime }s_{m}^{\prime }-2\mu \left( 1-\lambda
r^{n}\right) c_{n}^{\prime };\quad n=1,2,\ldots ,  \label{4.73}
\end{equation}%
where, by (\ref{4.69}),%
\[
p_{00}=2\int\limits_{0}^{1}\int\limits_{0}^{1}k\left( x,\xi \right) dxd\xi ; 
\]%
\[
p_{0m}=2\int\limits_{0}^{1}\int\limits_{0}^{1}k\left( x,\xi \right) \cos
\left( 2m\pi x\right) dxd\xi ; 
\]%
\[
p_{0m}^{\prime }=2\int\limits_{0}^{1}\int\limits_{0}^{1}k\left( x,\xi
\right) \sin \left( 2m\pi x\right) dxd\xi ; 
\]%
\[
p_{0n}=2\int\limits_{0}^{1}\int\limits_{0}^{1}k\left( x,\xi \right) \cos
\left( 2n\pi x\right) dxd\xi ; 
\]%
\[
p_{0n}^{\prime }=2\int\limits_{0}^{1}\int\limits_{0}^{1}k\left( x,\xi
\right) \sin \left( 2n\pi x\right) dxd\xi ; 
\]%
\[
p_{nm}=2\int\limits_{0}^{1}\int\limits_{0}^{1}k\left( x,\xi \right) \cos
\left( 2n\pi x\right) \cos \left( 2m\pi \xi \right) dxd\xi ; 
\]%
\[
p_{nm}^{\prime }=2\int\limits_{0}^{1}\int\limits_{0}^{1}k\left( x,\xi
\right) \cos \left( 2n\pi x\right) \sin \left( 2m\pi \xi \right) dxd\xi ; 
\]%
\[
p_{nm}^{\prime \prime }=2\int\limits_{0}^{1}\int\limits_{0}^{1}k\left( x,\xi
\right) \sin \left( 2n\pi x\right) \cos \left( 2m\pi \xi \right) dxd\xi ; 
\]%
\begin{equation}
p_{nm}^{\prime \prime \prime \prime
}=2\int\limits_{0}^{1}\int\limits_{0}^{1}k\left( x,\xi \right) \sin \left(
2n\pi x\right) \sin \left( 2m\pi \xi \right) dxd\xi ,\quad n,m=1,2,\ldots .
\label{4.74}
\end{equation}

Obviously, to ensure the solvability of the system of equations (\ref{4.73}%
), the parameter $\mu $ must be such that, as in section 5.4, Eq. (\ref{4.60}%
) would admit only of the trivial solution. Note that for $\lambda =r^{-n}$, 
$n=0,1,\ldots $, that is, in the case when the condition (\ref{4.18}) is not
fulfilled, the elements of the column of the free terms (\ref{4.73}) tend to
zero.

In expression (\ref{4.41}),%
\[
\int\limits_{0}^{1}H\left( x,\zeta ,\lambda \right) k\left( \zeta ,\xi
\right) d\zeta =\frac{1}{2\left( 1-2\lambda \right) }k_{0}\left( \xi \right) 
\]%
\[
+\sum_{n=1}^{\infty }\frac{r^{n}}{1-2\lambda r^{n}}\left[ k_{n}\left( \xi
\right) \cos \left( 2n\pi x\right) +k_{n}^{\prime }\left( \xi \right) \sin
\left( 2n\pi x\right) \right] , 
\]%
and, accordingly,%
\[
\varphi _{1}^{\prime }\left( x\right) =\frac{1}{2}a_{0}+\sum_{n=1}^{\infty }%
\frac{r^{n}}{1-2\lambda r^{n}}\left[ a_{n}\cos \left( 2n\pi x\right)
+a_{n}^{\prime }\sin \left( 2n\pi x\right) \right] , 
\]%
where%
\[
a_{0}=\frac{\mu \lambda }{4\left( 1-2\lambda \right) }\left(
p_{00}s_{0}+2\sum_{m=1}^{\infty }p_{0m}s_{m}+p_{0m}^{\prime }s_{0m}^{\prime
}-2c_{0}\right) ; 
\]%
\[
a_{n}\left( r\right) =\frac{\mu \lambda r^{n}}{1-2\lambda r^{n}}\left( \frac{%
1}{2}s_{0}p_{0n}+2\sum_{m=1}^{\infty }p_{nm}s_{m}+p_{nm}^{\prime
}s_{nm}^{\prime }-c_{n}\right) ; 
\]%
\begin{equation}
a_{n}^{\prime }\left( r\right) =\frac{\mu \lambda r^{n}}{1-2\lambda r^{n}}%
\left( \frac{1}{2}s_{0}^{\prime }p_{0n}^{\prime }+2\sum_{m=1}^{\infty
}p_{nm}^{\prime \prime }s_{m}+p_{nm}^{\prime \prime \prime }s_{nm}^{\prime
}-c_{n}^{\prime }\right) ,\quad n=1,2,\ldots .  \label{4.75}
\end{equation}

The substitution of this function into (\ref{4.54}) and subsequent
substitution of $q\left( x\right) $ into (\ref{4.57}) lead to the expression%
\[
\chi \left( x\right) =-\frac{\left( 1-2\lambda \right) \lambda }{2\left(
1-2\lambda -\lambda ^{2}\right) }a_{0} 
\]%
\[
-\sum_{n=1}^{\infty }\frac{\left( 1-2\lambda r^{n}\right) \lambda }{%
1-2\lambda r^{n}-\lambda ^{2}r^{2n}}\left[ a_{n}\left( r\right) \cos \left(
2n\pi x\right) +a_{n}^{\prime }\left( r\right) \sin \left( 2n\pi x\right) %
\right] . 
\]%
As a result, by formula (\ref{4.30}), we obtain%
\begin{equation}
\psi \left( x\right) =\frac{1}{2}t_{0}+\sum_{n=1}^{\infty }t_{n}\cos \left(
2n\pi x\right) +t_{n}^{\prime }\sin \left( 2n\pi x\right) ,  \label{4.76}
\end{equation}%
where%
\[
t_{0}=-\frac{\left( 1-\lambda \right) \lambda }{1-2\lambda -\lambda ^{2}}%
a_{0};\quad t_{n}=-\frac{\left( 1-\lambda r^{n}\right) \lambda }{1-2\lambda
r^{n}-\lambda ^{2}r^{2n}}a_{n}\left( r\right) ; 
\]%
\[
t_{n}^{\prime }=-\frac{\left( 1-\lambda r^{n}\right) \lambda }{1-2\lambda
r^{n}-\lambda ^{2}r^{2n}}a_{n}^{\prime }\left( r\right) ,\quad n=1,2,\ldots
. 
\]

The passage to the limit $r\rightarrow 1$ yields the following coefficients
of the series (\ref{4.76}):%
\begin{equation}
t_{0}=\sigma b_{0};\quad t_{n}=\sigma b_{n};\quad t_{n}^{\prime }=\sigma
b_{n}^{\prime },\quad n=1,2,\ldots ,  \label{4.77}
\end{equation}%
where, by (\ref{4.75}),%
\[
b_{0}=\frac{1}{4}\left( p_{00}s_{0}+2\sum_{m=1}^{\infty
}p_{0m}s_{m}+p_{0m}^{\prime }s_{0m}^{\prime }-2c_{0}\right) ; 
\]%
\[
b_{n}=\frac{1}{2}s_{0}p_{0n}+\sum_{m=1}^{\infty }p_{nm}s_{m}+p_{nm}^{\prime
}s_{nm}^{\prime }-c_{n}; 
\]%
\[
b_{n}^{\prime }=\frac{1}{2}s_{0}^{\prime }p_{0n}^{\prime
}+\sum_{m=1}^{\infty }p_{nm}^{\prime \prime }s_{m}+p_{nm}^{\prime \prime
\prime }s_{nm}^{\prime }-c_{n}^{\prime };. 
\]%
\[
\sigma =-\frac{\mu \lambda ^{2}\left( 1-\lambda \right) }{\left( 1-2\lambda
\right) \left( 1-2\lambda -\lambda ^{2}\right) } 
\]%
is a constant. Moreover,%
\[
b_{0}=\frac{1-2\lambda }{\mu \lambda }a_{0};\quad b_{n}=\frac{1-2\lambda }{%
\mu \lambda }a_{n}\left( 1\right) ;\quad b_{n}^{\prime }=\frac{1-2\lambda }{%
\mu \lambda }a_{n}^{\prime }\left( 1\right) .\quad 
\]

As an example clarifying the mechanism of the performed transformations, let
us consider the determination of the functions (\ref{4.54}) and (\ref{4.30}):%
\[
q\left( x\right) =-\lambda \int\limits_{-1}^{0}h\left( x,\xi \right) \varphi
_{1}^{\prime }\left( \xi \right) d\xi 
\]%
\[
=-\lambda \int\limits_{-1}^{0}\left\{ 1+2\sum_{n=1}^{\infty }r^{n}\left[
\cos \left( 2n\pi x\right) \cos \left( 2n\pi \xi \right) +\sin \left( 2n\pi
x\right) \sin \left( 2n\pi \xi \right) \right] \right\} \varphi _{1}^{\prime
}\left( \xi \right) d\xi 
\]%
\[
=-\lambda \left\{ \int\limits_{-1}^{0}\varphi _{1}^{\prime }\left( \xi
\right) d\xi +2\sum_{n=1}^{\infty }r^{n}\left[ \cos \left( 2n\pi x\right)
\int\limits_{-1}^{0}\varphi _{1}^{\prime }\left( \xi \right) \cos \left(
2n\pi \xi \right) d\xi \right. \right. 
\]%
\[
\left. \left. +\sin \left( 2n\pi x\right) \int\limits_{-1}^{0}\varphi
_{1}^{\prime }\left( \xi \right) \sin \left( 2n\pi \xi \right) d\xi \right]
\right\} ; 
\]%
\[
\psi \left( x\right) =\chi \left( x\right) +\lambda
\int\limits_{0}^{1}H\left( x,\xi ,\lambda \right) \chi \left( \xi \right)
d\xi 
\]%
\[
=\frac{1-\lambda }{1-2\lambda }\int\limits_{0}^{1}\chi \left( \xi \right)
d\xi +2\sum_{n=1}^{\infty }\frac{1-\lambda r^{n}}{1-2\lambda r^{n}}\left[
\cos \left( 2n\pi x\right) \int\limits_{0}^{1}\chi \left( \xi \right) \cos
\left( 2n\pi \xi \right) d\xi \right. 
\]%
\[
\left. +\sin \left( 2n\pi x\right) \int\limits_{0}^{1}\chi \left( \xi
\right) \sin \left( 2n\pi \xi \right) d\xi \right] . 
\]

Here,%
\[
-2\lambda \int\limits_{-1}^{0}\varphi _{1}^{\prime }\left( \xi \right) d\xi
;\quad -2\lambda r^{n}\int\limits_{-1}^{0}\varphi _{1}^{\prime }\left( \xi
\right) \cos \left( 2n\pi \xi \right) d\xi ; 
\]%
\[
-2\lambda r^{n}\int\limits_{-1}^{0}\varphi _{1}^{\prime }\left( \xi \right)
\sin \left( 2n\pi \xi \right) d\xi 
\]%
and%
\[
\frac{2\left( 1-\lambda \right) }{1-2\lambda }\int\limits_{0}^{1}\chi \left(
\xi \right) d\xi ;\quad \frac{2\left( 1-\lambda r^{n}\right) }{1-2\lambda
r^{n}}\int\limits_{0}^{1}\chi \left( \xi \right) \cos \left( 2n\pi \xi
\right) d\xi ; 
\]%
\[
\frac{2\left( 1-\lambda r^{n}\right) }{1-2\lambda r^{n}}\int\limits_{0}^{1}%
\chi \left( \xi \right) \sin \left( 2n\pi \xi \right) d\xi 
\]%
are the Fourier coefficients of the functions $q\left( x\right) $ and $\psi
\left( x\right) $, respectively.

In other words, in the limit $r\rightarrow 1$, in the first case there
occurs a redefinition of the Fourier coefficients by the factor $-\lambda $ (%
$\varphi _{1}^{\prime }$ and $q$ are determined on $x\in \left[ -1,0\right) $
and on $x\in \left[ 0,1\right] $, respectively), whereas in the second case
the function $\psi \left( x\right) $ is expressed via $\chi \left( x\right) $
by simple multiplication by the factor $\left( 1-\lambda \right) /\left(
1-2\lambda \right) $.

The system of the algebraic equations (\ref{4.73}) in the limit $%
r\rightarrow 1$ takes the form%
\[
\left[ 2\left( 1-2\lambda \right) -\mu \left( 1-\lambda \right) p_{00}\right]
s_{0}=2\mu \left( 1-\lambda \right) \sum_{m=1}^{\infty
}p_{0m}s_{m}+p_{0m}^{\prime }s_{m}^{\prime }-2\mu \left( 1-\lambda \right)
c_{0};
\]%
\[
2\left[ 1-2\lambda -\mu \left( 1-\lambda \right) p_{nn}\right] s_{n}=\mu
\left( 1-\lambda \right) p_{n0}s_{0}+2\mu \left( 1-\lambda \right)
\sum_{m=1,\,m\neq n}^{\infty }p_{nm}s_{m}
\]%
\[
+2\mu \left( 1-\lambda \right) \sum_{m=1}^{\infty }p_{nm}^{\prime
}s_{m}^{\prime }-2\mu \left( 1-\lambda \right) c_{n};
\]%
\[
2\left[ 1-2\lambda -\mu \left( 1-\lambda \right) p_{nn}^{\prime \prime
\prime }\right] s_{n}^{\prime }=\mu \left( 1-\lambda \right) p_{n0}^{\prime
}s_{0}^{\prime }+2\mu \left( 1-\lambda \right) \sum_{m=1}^{\infty
}p_{nm}^{\prime \prime }s_{m}
\]%
\begin{equation}
+2\mu \left( 1-\lambda \right) \sum_{m=1,\,m\neq n}^{\infty }p_{nm}^{\prime
\prime \prime }s_{m}^{\prime }-2\mu \left( 1-\lambda \right) c_{n}^{\prime
},\quad n=1,2,\ldots ,  \label{4.78}
\end{equation}

The elements of its matrix predominate on the diagonal owing to the
component $1-2\lambda $ that does not depend on $n$ and $m$. As a result,
for the determination of the coefficients $s_{0}$, $s_{n}$, $s_{n}^{\prime }$%
, contained in (\ref{4.66}), various methods prove to be efficient \cite{4.9}%
.

If $k\left( x,\xi \right) $ and $f\left( x\right) $ are $L_{2}$ - functions,
the corresponding Fourier series (\ref{4.67}), (\ref{4.68}) converge in the
mean. From (\ref{4.71}) and (\ref{4.72}), for $r=1$, we get%
\begin{equation}
K\left( x,\xi \right) =\frac{1-\lambda }{1-2\lambda }k\left( x,\xi \right)
,\quad f_{1}\left( x\right) =-\frac{\mu \left( 1-\lambda \right) }{%
1-2\lambda }f\left( x\right) ,  \label{4.79}
\end{equation}%
and the factors contained herein are bounded. Hence, the series obtained by
the substitution of expressions (\ref{4.67}), (\ref{4.68}) are analogously
convergent.

The solvability condition for the system of equations (\ref{4.78}) is
equivalent to the absence of nontrivial solutions to Eq. (\ref{4.60}) with
the kernel and the free term (\ref{4.79}). At the same time, the class of
functions that may contain the solution of the Fredholm integral equation of
the second kind (\ref{4.40}) is extended to the space $L_{2}$.

Accordingly, the series (\ref{4.66}) approximating the function $\psi
_{1}\left( x\right) $ converges in the mean, and by Parseval's relation%
\[
s_{0}^{2}+\sum_{n=1}^{\infty }s_{n}^{2}+s_{n}^{\prime 2}<\infty . 
\]%
From (\ref{4.77}), it follows that the series (\ref{4.76}) is analogously
convergent:%
\[
t_{0}^{2}+\sum_{n=1}^{\infty }t_{n}^{2}+t_{n}^{\prime 2}<\infty . 
\]

By the Riesz-Fischer theorem \cite{4.1} and on the basis of the previous
consideration, we can conclude that it represents an expansion of the $L_{2}$
- function $\psi \left( x\right) $ satisfying Eq. (\ref{4.1}) into a Fourier
series in terms of the elements $\left\{ \cos \left( 2n\pi x\right) ,\sin
\left( 2n\pi x\right) \right\} $.

It should be noted that the parameter $0<r<1$ plays here an exclusively
important role, because in its absence it would be impossible:

- to construct the algorithm that lead to Eqs. (\ref{4.40}) and (\ref{4.52});

- to perform transformations of integrals whose kernels have the form of the
series (\ref{4.19}), (\ref{4.21}), (\ref{4.55}) and (\ref{4.56}) that
diverge for $r\rightarrow 1$.

Thus, the values of $t_{0}$, $t_{n}$ and $t_{n}^{\prime }$ in (\ref{4.76})
are determined by the Fourier coefficients of the data of the problem by
means of a stable procedure of the numerical realization that include the
following stages:

- determination of the parameter $\lambda $ from the condition (\ref{4.59})
with $r=1$, i.e.,%
\[
\lambda \neq 0,\quad \lambda \neq 1,\quad \lambda \neq 1/2,\quad \lambda
\neq -1\pm \sqrt{2}; 
\]

- determination of the parameter $\mu $ from the condition that Eq. (\ref%
{4.60}) with the data (\ref{4.79}) admit only of the trivial solution;

- evaluation of the coefficients $c_{0}$, $c_{n}$, $c_{n}^{\prime }$ and $%
p_{00}$, $p_{nm}$, \ldots , $p_{nm}^{\prime \prime \prime }$ using,
respectively, formulas (\ref{4.70}) and (\ref{4.74});

- determination of the coefficients $s_{0}$, $s_{n}$ and $s_{n}^{\prime }$
from the system of linear algebraic equations (\ref{4.78});

- evaluation of the coefficients $t_{0}$, $t_{n}$ and $t_{n}^{\prime }$
using formulas (\ref{4.77}).

The fulfillment of the condition (\ref{4.6}), after substitution into (\ref%
{4.8}) of expressions (\ref{4.76}) with the coefficients (\ref{4.77}) and%
\[
\varphi \left( x\right) =\frac{1}{2}\alpha _{0}+\sum_{n=1}^{\infty }\alpha
_{n}\cos \left( 2n\pi x\right) +\alpha _{n}^{\prime }\sin \left( 2n\pi
x\right) , 
\]%
reduces to redefinition of the Fourier coefficients:%
\begin{equation}
\alpha _{0}=\frac{1-\lambda }{\lambda }t_{0};\quad \alpha _{n}=\frac{%
1-\lambda }{\lambda }t_{n};\quad \alpha _{n}^{\prime }=\frac{1-\lambda }{%
\lambda }t_{n}^{\prime },\quad n=1,2,\ldots .  \label{4.80}
\end{equation}%
Accordingly, a limit procedure with respect to $r$ transforms also $\varphi
\left( x\right) $ into a $L_{2}$-function. The condition (\ref{4.6}) is now
understood in the sense that%
\begin{equation}
\left\Vert \delta f\right\Vert _{L_{2}\left( 0,1\right) }=\left\Vert \psi
-\lambda B\psi \right\Vert _{L_{2}\left( 0,1\right) }=0.  \label{4.81}
\end{equation}

Thus, for $\psi \left( x\right) \in L_{2}\left( 0,1\right) $, one can find
the Fourier coefficients of the function $\varphi \left( x\right) $ that
allow for the fulfillment of the condition (\ref{4.81}). However, this
discretization done at the very beginning, i.e., without the transformation
with the parameter $0<r<1$, as already mentioned, would completely exclude
any possibility of the construction of the algorithm permitting the
determination of the function $\psi \left( x\right) $ satisfying (\ref{4.1}).

In the limit $r\rightarrow 1$, expression (\ref{4.22}) turns into it by (\ref%
{4.80}) and (\ref{4.76}). Accordingly, the restriction on the form of the
free term $f\left( x\right) $, imposed by the "harmonic" case of the
solution of the problem, is no longer in force.

It is important to note that the conclusion of section 5.5 that the function 
$\psi \left( x\right) $ actually satisfies Eq. (\ref{4.1}) still holds for $%
r\rightarrow 1$. Relation (\ref{4.61}) is fulfilled in this case by analogy
with (\ref{4.81}), that is, in virtue of mutual dependence between the
Fourier coefficients of the functions $\chi \left( x\right) $ and $\varphi
^{\prime }\left( x\right) $, $\varphi _{1}^{\prime }\left( x\right) $.

In section 6.3, we present a method of the solution of the problem (\ref{4.1}%
) without proceeding to the limit with respect to the parameter $r$. This is
achieved at the expense of satisfaction of the condition (\ref{4.6}) in the
sense of generalized functions. The general orientation of transformations
remain unchanged and the results of section 5.4 will be used to a full
extent.

\chapter{An analysis of the material of the previous section and some
additions}

\section{Comments on the material of the sections}

In section 5.1, we have developed the previous arguments that the
restoration of the function $\psi \left( x\right) $ from the results of
integration of $f\left( x\right) $ cannot be considered in terms of the
solution of the Fredholm integral equation of the first kind (\ref{4.1}). We
have formulated the problem of the determination of $\psi $ from the data $A$
and $f$ taking account of an inevitable error of the calculations. To this
end, we have proposed to use a functional relationship between the error of
integration, $\left( \delta f\right) \left( x\right) $, and $\psi \left(
x\right) $ in order to compensate adaptively for a small mismatch between $%
R\left( A\right) $ and $\left( A\psi \right) \left( x\right) $ that are
actually known [see (\ref{4.2}),(\ref{4.3})].

Further, it is shown in section 5.2 that a functional model of the error of
evaluation of the integral (\ref{4.1}), see section 4.5, can indeed be
represented by expression (\ref{4.4}). The latter is a difference between
the sought function $\psi \left( x\right) $ and an integral over this
function as well as one more unknown function $\varphi \left( x\right) $,
with the kernel $h\left( x,\xi \right) $ that has the form (\ref{4.13}). In
this case, the fulfillment of (\ref{4.6}), the condition that reflects the
smallness of $\left( \delta f\right) \left( x\right) $, requires that the
function $\psi \left( x\right) $ be harmonic.

Such an assumption is, apparently, applicable to the type of problems that
are concerned with the determination of the heat transfer (described by the
Laplace equation) from the result $f\left( r,x\right) $ of its action on a
system characterized by $k\left( x,\xi \right) $. At the same time, it is
desirable that the function $\psi \left( x\right) $ satisfying Eq. (\ref{4.1}%
) be more or less arbitrary and, ideally, belong to the space $L_{2}$.

Obviously, the above-mentioned harmonicity is stipulated by the presence in
the expression $h\left( x,\xi \right) $ of the parameter $0<r<1$. Moreover,
the use, instead of (\ref{4.13}), of a different, also bounded, kernel, in
practice, does not yield anything new, because the range of a completely
continuous operator is not closed.\footnote{%
Any closed subspace of $R\left( A\right) $ is finite-dimensional (\cite{0.1}%
, p. 96).}

A rather important point is the extension of (\ref{4.4}) to $x\in \left[
-1,0\right) $ under the condition (\ref{4.6}), carried out in section 5.3,
which led to Eq. (\ref{4.24}). In contrast to this equation, Eq. (\ref{4.25}%
) is more abstractly related to the problem (\ref{4.1}). This equation
arises as a result of the suggestion that the efficiency of the
transformations will be facilitated by the use, together with (\ref{4.24}),
of an analogous equation that is distinguished by its free term going to
zero on the other part of the interval of definition, $x\in \left[
-1,0\right) $. By means of simple transformations, it proved to be possible
to obtain the key, in this case, relations, i.e., (\ref{4.32}), further, (%
\ref{4.43}), (\ref{4.44}), (\ref{4.48}), and, finally, (\ref{4.50}).

Equations (\ref{4.24}) and (\ref{4.25}) are rather specific. Obviously, on
the subtraction of%
\begin{equation}
\varphi _{0}^{\prime \prime }\left( x\right) =\lambda
\int\limits_{-1}^{0}h\left( x,\xi \right) \varphi _{0}^{\prime \prime
}\left( \xi \right) d\xi +\kappa \left( x\right) ,\quad x\in \left[
-1,0\right)  \label{0.1}
\end{equation}%
from Eq. (\ref{4.23}), on this part of the interval of definition appears
Eq. (\ref{4.25}), and, accordingly, taking into account also (\ref{4.44})
and (\ref{4.48}),%
\[
\varphi ^{\prime }\left( x\right) =\varphi \left( x\right) -\varphi
_{0}^{\prime \prime }\left( x\right) =\varphi ^{\prime }\left( x\right)
-\varphi _{0}\left( x\right) -\varphi _{0}^{\prime \prime }\left( x\right) 
\]%
\[
=\varphi ^{\prime }\left( x\right) +\varphi _{0}^{\prime }\left( x\right)
-\varphi _{0}^{\prime \prime }\left( x\right) . 
\]%
Hence, $\varphi _{0}^{\prime }\equiv \varphi _{0}^{\prime \prime }$, i.e.,
Eqs. (\ref{0.1}) and (\ref{4.36}) are identical on $x\in \left[ -1,0\right) $%
.

Simultaneously, the free term of Eq. (\ref{4.24}), $\kappa \left( x\right) $%
, "flows" (it is difficult to characterize this procedure otherwise) to the
free term of Eq. (\ref{4.25}), $\chi \left( x\right) $. Indeed, by (\ref%
{4.32}), equation (\ref{4.24}) on $x\in \left[ 0,1\right] $ undergoes the
transformation%
\[
\lambda \int\limits_{-1}^{0}h\left( x,\xi \right) \varphi \left( \xi \right)
d\xi =\lambda \int\limits_{-1}^{0}h\left( x,\xi \right) \left[ \varphi
^{\prime }\left( \xi \right) +\varphi \left( \xi \right) -\varphi ^{\prime
}\left( \xi \right) \right] d\xi 
\]%
\[
=\lambda \int\limits_{-1}^{0}h\left( x,\xi \right) \varphi ^{\prime }\left(
\xi \right) d\xi +\chi \left( x\right) . 
\]

However, whatever one might say about the premises of the construction of (%
\ref{4.24}), (\ref{4.25}), these equations are, both formally and actually,
Fredholm integral equations of the second kind, whose the solution has the
form (\ref{4.28})-(\ref{4.31}). On one part of the interval of definition
their free terms are contained in an explicit form, whereas on the other
part they enter under the sign of integration. This issue, being absolutely
nonessential from the point of view of both general theory of this type of
equations as well as methods of their numerical realization, is a very
important factor of the realization of further transformations.

In section 5.3, we have presented a scheme of the construction of Eqs. (\ref%
{4.24}) and (\ref{4.25}) starting from a hypothetically given function $\psi
\left( x\right) $. In other words, the structure of these equations does not
contain contradictions.

A trivial, at the first sight, addition of (\ref{4.1}) to (\ref{4.24}) and (%
\ref{4.25}), which led to Eqs. (\ref{4.34}), (\ref{4.35}), has rather
substantial meaning of embedding the model of the error in the procedure of
the determination of the function $\psi \left( x\right) $.

Turning to section 5.4, we note that, with the help of (\ref{4.45}) and (\ref%
{4.48}), relations (\ref{4.43}), (\ref{4.44}) reduce to the following:%
\begin{equation}
\kappa \left( x\right) =\lambda \int\limits_{0}^{1}h\left( x,\xi \right)
\psi _{0}\left( \xi \right) d\xi ,\quad x\in \left[ -1,0\right) ;
\label{0.2}
\end{equation}%
\begin{equation}
\varphi \left( x\right) =\varphi ^{\prime }\left( x\right) +\varphi
_{0}^{\prime }\left( x\right) ,\quad x\in \left[ -1,0\right) .  \label{0.3}
\end{equation}%
(It seems that, irrespective of the above reduction, this result is by no
means obvious.)

Let us demonstrate the reduction of (\ref{4.34}) to Eq. (\ref{4.35}).%
\footnote{%
This procedure is inverse to that of section 5.4.} The substitution of $%
\varphi \left( x\right) $ from (\ref{0.3}) into (\ref{4.34}) leads to the
equations%
\[
\psi \left( x\right) =\lambda \int\limits_{-1}^{0}h\left( x,\xi \right) 
\left[ \varphi ^{\prime }\left( \xi \right) +\varphi _{0}^{\prime }\left(
\xi \right) \right] d\xi +\lambda \int\limits_{0}^{1}h\left( x,\xi \right)
\psi \left( \xi \right) d\xi 
\]%
\begin{equation}
+\mu \left( A\psi \right) \left( x\right) -\mu f\left( x\right) ,\quad x\in 
\left[ 0,1\right] ;  \label{0.4}
\end{equation}%
\[
\varphi ^{\prime }\left( x\right) +\varphi _{0}^{\prime }\left( x\right)
=\lambda \int\limits_{-1}^{0}h\left( x,\xi \right) \left[ \varphi ^{\prime
}\left( \xi \right) +\varphi _{0}^{\prime }\left( \xi \right) \right] d\xi 
\]%
\begin{equation}
+\lambda \int\limits_{0}^{1}h\left( x,\xi \right) \psi \left( \xi \right)
d\xi +\kappa \left( x\right) ,\quad x\in \left[ -1,0\right) .  \label{0.5}
\end{equation}

From (\ref{0.3}), (\ref{4.32}) and (\ref{4.43}), it follows that in (\ref%
{0.4}) and (\ref{0.5}) we have, respectively,%
\[
\lambda \int\limits_{-1}^{0}h\left( x,\xi \right) \varphi _{0}^{\prime
}\left( \xi \right) d\xi =\lambda \int\limits_{-1}^{0}h\left( x,\xi \right) 
\left[ \varphi \left( \xi \right) -\varphi ^{\prime }\left( \xi \right) %
\right] d\xi 
\]%
\begin{equation}
=\chi \left( x\right) ,\quad x\in \left[ 0,1\right] ;  \label{0.6}
\end{equation}%
\[
\varphi ^{\prime }\left( x\right) +\kappa \left( x\right) =\lambda
\int\limits_{-1}^{0}h\left( x,\xi \right) \varphi ^{\prime }\left( \xi
\right) d\xi 
\]%
\begin{equation}
+\lambda \int\limits_{0}^{1}h\left( x,\xi \right) \psi \left( \xi \right)
d\xi +\kappa \left( x\right) ,\quad x\in \left[ -1,0\right) .  \label{0.7}
\end{equation}

The fact that (\ref{0.4}), (\ref{0.5}) are identical to Eq. (\ref{4.35}) is
obvious [the function $\kappa $ is eliminated from (\ref{0.7})].
Analogously, vice versa, equation (\ref{4.35}), by the use of relations (\ref%
{0.2}), (\ref{0.3}) and (\ref{0.6}), is reduced to Eq. (\ref{4.34}).

From (\ref{4.39}), it follows that%
\begin{equation}
\psi \left( x\right) =\psi _{0}\left( x\right) +\psi _{1}\left( x\right) ;
\label{0.8}
\end{equation}%
that is, the function satisfying (\ref{4.1}) is a sum of the solutions of
the Fredholm integral equations of the second kind (\ref{4.36}) and (\ref%
{4.38}) that are stipulated by the components of the free term of (\ref{4.35}%
), i.e., $\chi $ and $-\mu f$, respectively.

The function $\psi _{1}\left( x\right) $ depends on the data of the problem
and, as such, represents the solution of the modified Eq. (\ref{4.1}),
artificially "shifted" into the plane of the stability of the procedures of
numerical realization.

This is the solution of a problem that is completely different from the
considered one, and, quite naturally, the function $\psi _{1}\left( x\right) 
$ does not satisfy Eq. (\ref{4.1}).

In its turn, the function $\psi _{0}\left( x\right) $ depends on $\psi
_{1}\left( x\right) \,$, which follows from Eqs. (\ref{4.36}), (\ref{4.52})
and expressions (\ref{4.54}), (\ref{4.41}). The addition of $\psi _{0}$ and $%
\psi _{1}$ in (\ref{0.8}) compensates adaptively for the effect of the
above-mentioned "shift", which makes the function $\psi \left( x\right) $
satisfy Eq. (\ref{4.1}).

Here, it should be emphasized that, at every stage of the solution, the
transformations, i.e., the "shift" and "compensation for the shift" are
carried out in association with a well-posed problem. The procedure (\ref%
{0.8}) can be interpreted as discarding a part of the function $\psi
_{1}\left( x\right) $ that prevents satisfaction of Eq. (\ref{4.1}).

Let us employ the relation $\varphi =2\varphi _{0}^{\prime }+\varphi
_{1}^{\prime }$ that follows from (\ref{0.3}) and (\ref{4.39}). Accordingly, 
$\varphi -\varphi _{0}^{\prime }=\varphi _{0}^{\prime }+\varphi _{1}^{\prime
}$ and, in virtue of $\varphi _{0}^{\prime }+\varphi _{1}^{\prime }=\varphi
^{\prime }$, we get $\varphi =\varphi ^{\prime }+\varphi _{0}^{\prime }$,
that is, we return to (\ref{0.3}).\footnote{%
In other words, the solution of Eq. (\ref{4.34}) on $x\in \left[ -1,0\right) 
$ is the sum of the solutions of Eqs. (\ref{4.35}) and (\ref{4.36}) on this
interval.} This situation is completely in line with the logic of the "flow"
of the functions $\kappa $ and $\chi $ from one to another. Indeed, by
"giving away" $\varphi _{0}^{\prime }$, the function $\varphi $ turns into $%
\varphi ^{\prime }$, and, instead of $\kappa $, there appears $\chi $.
Equation (\ref{4.34}) takes the form (\ref{4.35}). The inverse procedure,
i.e., a transformation of (\ref{4.35}) into (\ref{4.34}), is, naturally,
related to the "return" of $\varphi _{0}^{\prime }$.

Thus, under the assumption that the function $\psi \left( x\right) $ is
harmonic, the problem has been reduced to the solution of Eq. (\ref{4.52}).
Its free term depends on the function $\varphi _{1}^{\prime }\left( x\right) 
$ that, in turn, is also determined by the solution of the Fredholm integral
equation of the second kind (\ref{4.40}) and by expression (\ref{4.41}).

The above-mentioned results of the transformations (they can be
characterized as equivalent) should be interpreted in the following way.
There is a harmonic function $\psi \left( x\right) $. After integration
according to (\ref{4.1}), it is determined by expression (\ref{4.30}). The
latter, in virtue of $0<r<1$, is a Fredholm integral equation of the second
kind with respect to the function $\chi \left( x\right) $. Under the
condition (\ref{4.18}), its solution, $\chi _{\ast }\left( x\right) $, can
be determined in a certain way. From this point of view, the substitution of 
$\chi =\chi _{\ast }\left( x\right) $ into Eq. (\ref{4.52}), irrespective of
the form of the kernel $l\left( x,\xi \right) $, allows us to evaluate the
free term $q=q_{\ast }\left( x\right) $. Hence, equation (\ref{4.52}) has
every right to exist.

In other words, for any given kernel $l\left( x,\xi \right) $ that,
specifically, has the form (\ref{4.55}) and for a corresponding value of the
parameter $\lambda $, there exists a free term $q\left( x\right) $ such that
the solution of Fredholm integral equation of the second kind (\ref{4.52}), $%
\chi _{\ast }\left( x\right) $, after substitution into expression (\ref%
{4.30}), allows us to determine the function $\psi =\psi _{\ast }\left(
x\right) $ that satisfies Eq. (\ref{4.1}).

The above transformations consisted, in essence, both in the determination
of Eq. (\ref{4.52}) itself and in effective determination its free term $q$.
Here, the kernel $l\left( x,\xi \right) $ does not depend on the data of the
problem and is stipulated exclusively by the interests of a constructive
side of the transformations. Implied is a possibility to make use of the
techniques of the theory of Fredholm integral equations of the second kind
with symmetric kernels resulting from the model of the error (\ref{4.4}),
condition (\ref{4.6}), the kernel (\ref{4.13}) and the way of further
extension of the problem to $x\in \left[ -1,0\right) $.

Carrying out the transformations in an analytical form, including finding
the resolvent (\ref{4.56}), was substantially facilitated by the properties
of the kernel $h\left( x,\xi \right) $.\footnote{%
A list of these properties is given in the next section.} At the same time,
for this purpose, instead of (\ref{4.19}), we could use other convergent
series in terms of the elements $\bar{\varphi}_{2n-1}\left( x\right) $, $%
\bar{\varphi}_{2n}\left( x\right) $ from (\ref{4.17}).

However, the kernel (\ref{4.13}) has an inherent special property that
consists in the fact that, for $r\rightarrow 1$, the integral 
\[
\int\limits_{0}^{1}h\left( x,\xi \right) \psi \left( \xi \right) d\xi =\frac{%
1}{2}t_{0}+\sum_{n=1}^{\infty }t_{n}\cos \left( 2n\pi x\right)
+t_{n}^{\prime }\sin \left( 2n\pi x\right) ,\quad x\in \left[ 0,1\right] 
\]%
[see (\ref{4.19}),(\ref{4.76})], where%
\[
t_{0}=2\int\limits_{0}^{1}\psi \left( \xi \right) d\xi ;\quad
t_{n}=2\int\limits_{0}^{1}\psi \left( \xi \right) \cos \left( 2n\pi \xi
\right) d\xi ; 
\]%
\[
t_{n}^{\prime }=2\int\limits_{0}^{1}\psi \left( \xi \right) \sin \left(
2n\pi \xi \right) d\xi ,\quad n=1,2,\ldots , 
\]%
is a Fourier series of the function $\psi \left( x\right) $ in terms of the
elements (\ref{4.17}). As is known (see, e.g., \cite{0.2}, pp. 110-116),
using such a series, one can approach in the mean an arbitrary function from
the space $L_{2}$.\footnote{%
In this sense, an alternative is given by the kernel (\ref{4.19}) for $r=1$,
which is the series $h\left( x,\xi \right) =1+2\sum_{n=1}^{\infty }\cos %
\left[ 2n\pi \left( x-\xi \right) \right] $ whose sum is not bounded.}

Here, a one-to-one, continuous and linear correspondence between the spaces $%
l_{2}$ and $L_{2}$, resulting from the Riesz-Fischer theorem ( \cite{0.3},
pp. 116-119), manifests itself to a full extent. At the same time, a passage
to the limit $r\rightarrow 1$ can be regarded as a realization of the
objective to transform $l_{2}^{\prime }$ into the space $l_{2}$ (see section
4.5).

It should be noted that one can draw a conclusion about the stability of the
computational procedure of section 5.6 using the passage to the limit $%
r\rightarrow 1$ from the linear dependence of the Fourier coefficients $%
t_{0} $, $t_{n}$, $t_{n}^{\prime }$; $s_{0}$, $s_{n}$, $s_{n}^{\prime }$ and 
$c_{0} $, $c_{n}$, $c_{n}^{\prime }$ of the sought function $\psi \left(
x\right) $, the function $\psi _{1}\left( x\right) $ satisfying Eq. (\ref%
{4.40}) and of the free term $f\left( x\right) $ from (\ref{4.1}),
respectively [see (\ref{4.76}), (\ref{4.77}) and (\ref{4.78})].

The following point seems to be characteristic. Upon the substitution of
expressions (\ref{4.71}) and (\ref{4.72}) with $r=1$, that is, (\ref{4.79}),
equation (\ref{4.40}) does not change its status as a Fredholm integral
equation of the second kind. In this regard, it should be noted that the
expansion of $\psi _{1}\left( x\right) $ into the series (\ref{4.66}) is
merely one of possible ways of its solution. If one carries out the same
substitution into (\ref{4.41}), evaluates numerically $\psi _{1}\left(
x\right) $ from Eq. (\ref{4.40}) and, after that, the function $\varphi
_{1}^{\prime }\left( x\right) $, the function $\psi \left( x\right) $ is
determined by means of multiplication by the coefficient%
\[
-\frac{\lambda \left( 1-\lambda \right) }{1-2\lambda -\lambda ^{2}} 
\]%
[see (\ref{4.77})]. At the same time, this fact became clear only as a
result of the transformations with the parameter $r$ and letting it go to $1$%
.

The proof that $\psi \left( x\right) $ satisfies (\ref{4.1}) (see section
5.5) is a very important point whose meaning lies in the following. In the
derivation of Eq. (\ref{4.52}), the condition concerning the identity of the
solutions of Eqs. (\ref{4.25}) and (\ref{4.35}), although in an implicit
form, has been employed. An analysis of the actual transformations has
allowed us to draw a conclusion that this condition is indeed fulfilled and
that the function $\psi \left( x\right) $, determined by means of the
solution of (\ref{4.52}), satisfies Eq. (\ref{4.1}).

In this way, we have essentially confirmed a possibility to realize in (\ref%
{4.52}) the free term $q\left( x\right) $ that is adequate to the
substitution for $\psi \left( x\right) $ of a function whose integration by (%
\ref{4.1}) yields, as a result, $f\left( x\right) $.

\section{Additional arguments}

There exist a number of works concerned with the issue of the perturbation
of linear operators (\cite{0.4}, \cite{0.5} section 7, and others). Therein,
mostly completely continuous perturbations as well as perturbations of the
spectrum are studied. The zero error (\ref{4.4}) is an incompletely
continuos perturbation. As shown in section 5.4, such a perturbation (in
contrast to a completely continuous one) can qualitatively change the
formulation of the problem and introduce principally new possibilities of
its numerical realization.

In this regard, condition (\ref{4.6}) that subsequently terns into (\ref%
{4.81}) is necessary. Indeed, there arises (\ref{4.8}), a Fredholm integral
equation of the second kind with respect to the sought function $\psi \left(
x\right) $, that creates the premises of far-reaching transformations. Taken
together, equations (\ref{4.4}) and (\ref{4.6}) can be characterized as the
main factor of the construction of a stable algorithm of numerical
realization of the problem (\ref{4.1}).

Nonetheless, the above does not suffice to carry out the transformations of
Chapter 5. Let us set in (\ref{4.4}) $\lambda =1$ and, instead of (\ref{4.5}%
), let the operator be%
\begin{equation}
B_{\bullet }=\int\limits_{-1}^{x}h\left( x,\xi \right) _{\bullet }d\xi .
\label{0.9}
\end{equation}

For unique solvability of Eq. (\ref{4.7}), it is necessary here to have a
kernel $h\left( x,\xi \right) $ that possesses the property of being closed.
Therefore, it can be taken in the form (\ref{4.13}). Instead of (\ref{4.8}),
we now have%
\[
g\left( x\right) =\psi \left( x\right) -\int\limits_{-1}^{x}h\left( x,\xi
\right) \psi \left( \xi \right) d\xi . 
\]

Taking into account this point, by extending Eq. (\ref{4.7}) to $x\in \left[
-1,0\right) $, analogously to section 5.3, we obtain%
\begin{equation}
\psi \left( x\right) =\int\limits_{-1}^{0}h\left( x,\xi \right) \varphi
\left( \xi \right) d\xi +\int\limits_{0}^{x}h\left( x,\xi \right) \psi
\left( \xi \right) d\xi ,\quad x\in \left[ 0,1\right] ;  \label{0.10}
\end{equation}%
\begin{equation}
\varphi \left( x\right) =\int\limits_{-1}^{x}h\left( x,\xi \right) \varphi
\left( \xi \right) d\xi +\kappa \left( x\right) ,\quad x\in \left[
-1,0\right) ,  \label{0.11}
\end{equation}%
where $\kappa \left( x\right) $ is an undefined function.

The solution of Eq. (\ref{0.11}) is expressed via the resolvent of the
kernel $h\left( x,\xi \right) $. Its substitution into (\ref{0.10}) leads to
an equation of the form%
\[
\psi \left( x\right) =\int\limits_{0}^{x}h\left( x,\xi \right) \psi \left(
\xi \right) d\xi +\chi \left( x\right) ,\quad x\in \left[ 0,1\right] , 
\]%
where the function $\chi $ depends on $\kappa $.

However, this equation cannot be related to Eq. (\ref{0.11}), that is, the
procedure of extension to $x\in \left[ -1,0\right) $ does not yield anything
in reality. The reason lies in the absence of the function $\psi $ in Eq. (%
\ref{0.11}). If the extension of (\ref{0.10}) to $x\in \left[ -1,0\right) $
is done with the use of a definite integral over $\psi $, we get the
algorithm of section 5.4 in a complicated form.

At the same time, the actual reason for the invalidity of the operator (\ref%
{0.9}) for application in (\ref{4.5}) is rooted deeper. The essence lies in
a qualitative mismatch between the ranges of the Fredholm and Volterra
integral operators of the first kind. Whereas in the first case the solution
of the corresponding equation exists only under the conditions of Picard's
theorem, in the second case, it is sufficient for its definition that the
kernel and the free term be continuous.\footnote{%
Implied is a reduction to the Volterra integral equation of the second kind
by differentiation.}

In light of the above, the second factor of the achieved efficiency should
be noted. It is related essentially with the extension of Eq. (\ref{4.4}),
where the operator $B$ has the form (\ref{4.5}), under the condition (\ref%
{4.6}), to $x\in \left[ -1,0\right) $ by (\ref{4.23}). In this case, the
solution of Eq. (\ref{4.24}) on $x\in \left[ -1,0\right) $; $x\in \left[ 0,1%
\right] $ contains the function $\kappa \left( x\right) $ only in an
explicit form and under the sign of integration, respectively. This point
constitutes an important prerequisite of obtaining a Fredholm integral
equation of the second kind for the function $\kappa \left( x\right) $.

The third factor consists in the use of Eqs. (\ref{4.25}) and (\ref{4.35})
along with (\ref{4.24}), (\ref{4.34}). With the help of these equations, the
construction of the algorithm moves into the plane of practical realization.
In the process of the reduction of (\ref{4.35}) to Eq. (\ref{4.34}) that has
the same form of the solution on $x\in \left[ 0,1\right] $, we have obtained
the basic computational relations.

And, finally, the fourth factor is related, in fact, to the choice of the
kernel $h\left( x,\xi \right) $ that allowed us to do the following:

- carry out the transformations in an analytical form up to their final
stage;

- determine the function $\psi \left( x\right) $ for the data of (\ref{4.1})
from the space $L_{2}$ by means of a passage to the limit in the solution
obtained for the case when $0<r<1$.

In addition, the kernel (\ref{4.13}) has a whole spectrum of positive
properties: namely, it is closed, symmetric and positive definite; it
depends on the difference of the arguments, and the eigenfunctions of the
operator $B$ are orthogonal both on the interval $x\in \left[ -1,1\right] $
and on $x\in \left[ -1,0\right) $; $\left[ 0,1\right] $.

Let us turn to the question that is related to Eq. (\ref{4.40}). For $r=1$
in (\ref{4.71}), (\ref{4.72}), we get (\ref{4.79}). Accordingly, equation (%
\ref{4.40}) takes the form%
\begin{equation}
\psi _{1}\left( x\right) =\mu \frac{1-\lambda }{1-2\lambda }%
\int\limits_{0}^{1}k\left( x,\xi \right) \psi _{1}\left( \xi \right) d\xi -%
\frac{\mu \left( 1-\lambda \right) }{1-2\lambda }f\left( x\right) ,\quad
x\in \left[ 0,1\right] ,  \label{0.12}
\end{equation}%
or%
\begin{equation}
-\frac{1-2\lambda }{\mu \left( 1-\lambda \right) }\psi _{1}\left( x\right)
+\int\limits_{0}^{1}k\left( x,\xi \right) \psi _{1}\left( \xi \right) d\xi
=f\left( x\right) ,\quad x\in \left[ 0,1\right] ,  \label{0.13}
\end{equation}%
which makes it rather interesting. As a matter of fact, instead of an
ill-posed problem, as a basic object of investigation, there actually arises
a Fredholm integral equation of the second kind obtained just by adding to (%
\ref{4.1}) the sought function with a coefficient whose set of admissible
values is practically unlimited.

Indeed, for $\lambda \neq :0,1,1/2,-1\pm \sqrt{2}$, it is not difficult to
choose the parameter $\mu $ in such a way that the solution of the
homogeneous equation (\ref{5.12}), i.e.,%
\[
\psi \left( x\right) =\mu _{\lambda }\int\limits_{0}^{1}k\left( x,\xi
\right) \psi \left( \xi \right) d\xi ,\quad x\in \left[ 0,1\right] , 
\]%
where%
\begin{equation}
\mu _{\lambda }=\frac{\mu \left( 1-\lambda \right) }{1-2\lambda },
\label{0.14}
\end{equation}%
be trivial.

Note that\thinspace\ in the process of the evaluation of the function $%
\varphi _{1}^{\prime }\left( x\right) $, information about the data of the
problem contained in $\psi _{1}\left( x\right) $ undergoes substantial
changes that involve the kernel and the free term of Eq. (\ref{4.1}).
Simultaneously, the next stage of the calculations concerned with the
determination of $q\left( x\right) $ is transferred from $x\in \left[
-1,0\right) $ to $x\in \left[ 0,1\right] $.

After that, i.e., in the process of the evaluation of the Fourier
coefficients of the functions $q\left( x\right) $, $\chi \left( x\right) $
and $\psi \left( x\right) $, no new information about the data of the
problem is introduced. As a matter of fact, the Fourier coefficients of the
function $\varphi _{1}^{\prime }\left( x\right) $ are triply multiplied by
the corresponding constants. At the same time, by turning to the system of
equations (\ref{4.78}), we can notice that a relationship between the
Fourier coefficients of the functions $\psi \left( x\right) $ and $\psi
_{1}\left( x\right) $, i.e., $t_{0}$, $t_{n}$, $t_{n}^{\prime }$ and $s_{0}$%
, $s_{n}$, $s_{n}^{\prime }$, respectively, has, by (\ref{4.70}) and (\ref%
{4.74}), rather substantial meaning.

Upon the substitution of the function $\psi _{1}\left( x\right) $ from (\ref%
{4.39}) into (\ref{0.12}), taking account of (\ref{4.1}) and (\ref{0.14}),
we get%
\begin{equation}
\psi _{0}\left( x\right) =\mu _{\lambda }\int\limits_{0}^{1}k\left( x,\xi
\right) \psi _{0}\left( \xi \right) d\xi +f_{0}^{\prime }\left( x\right) ,
\label{0.15}
\end{equation}%
where%
\[
f_{0}^{\prime }\left( x\right) =\psi \left( x\right) -\mu _{\lambda
}\int\limits_{0}^{1}k\left( x,\xi \right) \psi \left( \xi \right) d\xi +\mu
_{\lambda }f\left( x\right) ,\quad x\in \left[ 0,1\right] , 
\]%
which leads to rather interesting, as it seems, conclusions:

1) There exists a Fredholm integral equation of the second kind with the
kernel $k\left( x,\xi \right) $ from (\ref{4.1}) and the parameter (\ref%
{0.14}) whose free term, on the subtraction of $\mu _{\lambda }f\left(
x\right) $, is the same as in a completely identical equation for the sought
function $\psi \left( x\right) $. Moreover,%
\[
\psi _{0}\left( x\right) =\psi \left( x\right) -\psi _{1}\left( x\right) , 
\]%
where $\psi _{1}\left( x\right) $ is the solution of (\ref{0.12}) that is
also a Fredholm integral equation of the second kind.

2) And, vice versa, the function $\psi \left( x\right) $ satisfying (\ref%
{4.1}) is expressed from (\ref{0.15}) via the solution of the Fredholm
integral equation of the second kind (\ref{4.36}) and the data of the
problem. Note that, as a result of the subtraction of Eqs. (\ref{4.36}) and (%
\ref{0.15}), the function $\psi \left( x\right) $ can also be represented in
terms of integral dependence on $\psi _{0}\left( x\right) $, $\varphi
_{0}^{\prime }\left( x\right) $.

3) For $\lambda =1/2$, equation (\ref{0.13}) turns into (\ref{4.1}). At the
same time, for a different value of the parameter $\lambda $, the solution
of this equation is a well-posed problem, and, as shown above, it serves for
the determination of $\psi \left( x\right) $ by means of a stable procedure
of numerical realization. In other words, equation (\ref{4.1}) corresponds
to a set of well-posed problems for the critical values of the parameters
contained therein.

Thus, the functions $\psi \left( x\right) $, $\psi _{0}\left( x\right) $ and 
$\psi _{1}\left( x\right) $ in (\ref{4.39}) are mutually related by means of
the Fredholm integral operator (\ref{4.5}). It seems that we have outlined
an important point that deserves further interpretation.

The next issue is related to a possibility of the realization of other
methods of the determination of the function $\chi \left( x\right) $ or $%
\kappa \left( x\right) $ that allow us to find the solution of the problem
from (\ref{4.28}), (\ref{4.30}). We outline schematically one of these
methods: substitution into (\ref{0.2}) from (\ref{4.30}) and (\ref{4.40}) of%
\[
\psi _{0}\left( x\right) =\psi \left( x\right) -\psi _{1}\left( x\right) 
\]%
\begin{equation}
=\chi \left( x\right) +\lambda \int\limits_{0}^{1}H\left( x,\xi ,\lambda
\right) \chi \left( \xi \right) d\xi -\psi _{1}\left( x\right) ;
\label{0.16}
\end{equation}%
substitution of $\kappa \left( x\right) $ into (\ref{4.28}); elimination of $%
\psi \left( x\right) $ from expressions (\ref{4.28}) and (\ref{4.30}).

However, in this situation, the kernel of the integral equation that serves
for the determination of $\chi \left( x\right) $ depends analytically on the
parameter $\lambda $. As a result, there appear unnecessary complications.
As a matter of fact, generally speaking, such equations may prove to be
insolvable irrespective of the value of the parameter $\lambda $ (\cite{0.6}%
, pp. 130-132; \cite{0.7}). In general, an approach to the solution of the
problem based on the use of (\ref{0.16}) seems to be less efficient

An analogue of Eq. (\ref{4.52}) can be constructed also for the function $%
\kappa \left( x\right) $. To this end, we substitute $\psi _{0}\left(
x\right) $ from (\ref{4.39}) into (\ref{0.2}), where the function $\psi
\left( x\right) $ has the form (\ref{4.28}), i.e.,%
\[
\psi _{0}\left( x\right) =\lambda \int\limits_{-1}^{0}H\left( x,\xi ,\lambda
\right) \kappa \left( \xi \right) d\xi -\psi _{1}\left( x\right) . 
\]

As a result,%
\begin{equation}
\kappa \left( x\right) =\Lambda \int\limits_{-1}^{0}l^{\prime }\left( x,\xi
\right) \kappa \left( \xi \right) d\xi +q^{\prime }\left( x\right) ,\quad
x\in \left[ -1,0\right) ,  \label{0.17}
\end{equation}%
where $\Lambda =\lambda ^{2}$;%
\[
l^{\prime }\left( x,\xi \right) =\int\limits_{0}^{1}h\left( x,\zeta \right)
H\left( \zeta ,\xi ,\lambda \right) d\zeta . 
\]%
A comparison with (\ref{4.53}), by (\ref{4.19}), (\ref{4.21}), shows that $%
l^{\prime }\left( x,\xi \right) \equiv l\left( x,\xi \right) $;%
\begin{equation}
q^{\prime }\left( x\right) =-\lambda \int\limits_{0}^{1}h\left( x,\xi
\right) \psi _{1}\left( \xi \right) d\xi .  \label{0.18}
\end{equation}

Compared to the algorithm of section 5.4, in this case, the necessity of
intermediate determination of the function $\varphi _{0}^{\prime }\left(
x\right) $ drops out, which, by the way, does not simplify substantially the
transformations.

Note that if the free term $f\left( x\right) $ has discontinuities or other
singularities stipulated by the kernel $k\left( x,\xi \right) $, its
explicit presence in the solution may prove to be desirable. To this end,
the function $\psi \left( x\right) $ should be expressed via $\kappa \left(
x\right) $ with the help of Eq. (\ref{4.34}). As regards Eq. (\ref{4.24}),
it is satisfied as a result of the use of (\ref{4.31}) in the process of the
construction of Eq. (\ref{4.52}).\footnote{%
As a matter of fact, this issue is discussed in section 5.5.}

We want to conclude this section by returning to the condition (\ref{4.3}).
Let us draw attention to an adaptive connection of $f\left( x\right) $ with
the space $l_{2}^{\prime }$ and the logic of its practical realization.
Preliminarily, we note main facts concerning the mapping (\ref{4.1}) from
one of the spaces $L_{2}$, $l_{2}^{\prime }$ into another, irrespective of
supposed use of $\delta f$.

Thus, the range $R\left( A\right) =l_{2}^{\prime }$ is not closed. As a
result, the operator $A$ is incompletely continuos, and its inverse $A^{-1}$
acting from $l_{2}^{\prime }$ into $L_{2}\left( 0,1\right) $ is bounded.
However, in the framework of the traditional object of investigation, (\ref%
{4.1}), there is no possibility to use somehow this fact for the
construction of the operator $A^{-1}$.

At the same time, the following sequence of arguments arises:

- objectively, a bounded operator $A^{-1}$ from $l_{2}^{\prime }$ does
exist, that is, the solution of Eq. (\ref{4.1}) in the pair of spaces $%
\left( L_{2},l_{2}^{\prime }\right) $ is a well-posed problem;

- the property of limited inversion is directly associated with the Fredholm
integral operator of the second kind, which in our case is $I-\lambda B$;

- insertion in the scheme of transformation of the identity operator $I$
organically combines with the modelling of the error of integration induced
[together with the prime cause, i.e., non-closed character of $R\left(
A\right) $] by the incorrectness of the problem (\ref{4.1});

- the adaptation of $f\left( x\right) $ to the space $l_{2}^{\prime }$ and a
functional representation of the error are, thus, closely related;

- the fact that $R\left( A\right) $ is non-closed does not prevent the
determination of the function $\psi _{1}\left( x\right) $ from Eq. (\ref%
{0.12}). As shown above, this function serves for finding the solution of
the problem, the function $\psi \left( x\right) $;

- from this point of view, equation (\ref{4.7}) that follows from (\ref{4.4}%
), (\ref{4.6}) realizes a connection of the Fredholm integral equations of
the first and the second kind via their common range, which creates a
serious prerequisite of the solution of the problem of the determination of
the function $\psi \left( x\right) $ satisfying (\ref{4.1}) in the correct
formulation.

The outlined orientation will be developed in the next section.

\section{The second version of the solution of the problem}

Here, the transformations of sections 5.2-5.4 are extended to the case when
the data of Eq. (\ref{4.1}), i.e., $k\left( x,\xi \right) $ and $f\left(
x\right) $ as well as the function $\psi \left( x\right) $ satisfying this
equation, from the very beginning belong to the space $L_{2}$. The kernel~$%
h\left( x,\xi \right) $ has the previous form (\ref{4.13}); the parameter $%
0<r<1$ is fixed.\footnote{%
Below, we point out that, in the framework of the present version of the
solution, expression (\ref{4.13}) is, generally speaking, has alternatives.}

Let us return to Eq. (\ref{4.24}):%
\begin{equation}
\left. 
\begin{array}{c}
\psi \left( x\right) \\ 
\varphi \left( x\right)%
\end{array}%
\right\} =\lambda B\left( 
\begin{array}{c}
\psi \\ 
\varphi%
\end{array}%
\right) \left( x\right) +\left\{ 
\begin{array}{c}
0,\quad x\in \left[ 0,1\right] ; \\ 
\kappa \left( x\right) ,\quad x\in \left[ -1,0\right) ,%
\end{array}%
\right.  \label{0.19}
\end{equation}%
or%
\begin{equation}
\psi \left( x\right) =\lambda \int\limits_{-1}^{0}h\left( x,\xi \right)
\varphi \left( \xi \right) d\xi +\lambda \int\limits_{0}^{1}h\left( x,\xi
\right) \psi \left( \xi \right) d\xi ,\quad x\in \left[ 0,1\right] ;
\label{0.20}
\end{equation}%
\begin{equation}
\varphi \left( x\right) =\lambda \int\limits_{-1}^{0}h\left( x,\xi \right)
\varphi \left( \xi \right) d\xi +\lambda \int\limits_{0}^{1}h\left( x,\xi
\right) \psi \left( \xi \right) d\xi +\kappa \left( x\right) ,\quad x\in 
\left[ -1,0\right) ,  \label{0.21}
\end{equation}%
which follows from the representation of the error (\ref{4.4}), under the
condition (\ref{4.6}) and the extension of (\ref{0.20}) to $x\in \left[
-1,0\right) $.

A possibility to satisfy this equation on the interval $x\in \left[ 0,1%
\right] $ by $\varphi \left( x\right) $ was considered in section 5.2 [the
function $\psi \left( x\right) $ was given]. In order to satisfy the
condition (\ref{4.6}) that implied fitting in the space $C\left[ 0,1\right] $%
, we had to restrict the class of admissible functions for $\psi \left(
x\right) $ by harmonic functions.

In this case, the issue essentially reduced to an investigation into the
solvability of the Fredholm integral equation of the first kind (\ref{4.10})
stipulated by the conditions of Picard's theorem \cite{0.2}:%
\begin{equation}
\sum_{n=1}^{\infty }\alpha _{n}^{2}\lambda _{n}^{2}<\infty ,\quad \alpha
_{n}=\int\limits_{0}^{2\pi }g\left( \zeta \right) \bar{\varphi}_{n}\left(
\zeta \right) d\zeta ,  \label{0.22}
\end{equation}%
where $\lambda _{n}$ and $\bar{\varphi}_{n}\left( \zeta \right) $ are,
respectively, the characteristic numbers and the eigenfunctions of the
kernel (\ref{4.11}) numbered in order of their sequence order [see (\ref%
{4.12})]. Moreover, the system of the eigenfunctions $\bar{\varphi}%
_{n}\left( \zeta \right) $ must be complete on the interval $\zeta \in \left[
0,2\pi \right] $, and the kernel must be real and symmetric, which, in this
case, is certainly fulfilled.

And, nevertheless, as shown by E. Goursat (\cite{0.8}, pp. 141-143), even if
the condition (\ref{0.22}) is not fulfilled, one can always find such a
function $\varphi \left( \zeta \right) $ that the difference between the
integral%
\[
\int\limits_{0}^{2\pi }h\left( \zeta ,\theta \right) \varphi \left( \theta
\right) d\theta 
\]%
and the function $g\left( \zeta \right) $ from (\ref{4.10}) is arbitrarily
small. It should be noted that $h\left( \zeta ,\theta \right) $, in this
case, is assumed to be a much more general kernel than (\ref{4.11}).

A proof is based on the fact that%
\[
g_{n}\left( \zeta \right) =\int\limits_{0}^{2\pi }h\left( \zeta ,\theta
\right) \varphi ^{\left( n\right) }\left( \theta \right) d\theta , 
\]%
where%
\begin{equation}
\varphi ^{\left( n\right) }\left( \theta \right) =\sum_{i=1}^{n}a_{i}\lambda
_{i}\bar{\varphi}_{i}\left( \theta \right) ,  \label{0.23}
\end{equation}%
coincides with the sum of the first terms of the Fourier series of the
functions $g\left( \zeta \right) $ in terms of the elements $\bar{\varphi}%
_{n}\left( \theta \right) $. Therefore, one can establish the number $n$ for
which the integral%
\[
\int\limits_{0}^{2\pi }\left[ g\left( \zeta \right) -g_{n}\left( \zeta
\right) \right] ^{2}d\zeta 
\]%
is smaller than a certain $\varepsilon >0$.

However, the series (\ref{0.23}) that satisfies in this way Eq. (\ref{4.10})
diverges in the space $L_{2}$ with the growth of $n$. Accordingly, it is not
possible to regard (\ref{4.19}) as a Fredholm integral equation of the
second kind\ for the function%
\begin{equation}
\Psi \left( x\right) =\left\{ 
\begin{array}{c}
\psi \left( x\right) ,\quad x\in \left[ 0,1\right] ;~ \\ 
\varphi \left( x\right) ,\quad x\in \left[ -1,0\right) ,%
\end{array}%
\right.  \label{0.24}
\end{equation}%
and to invert the operator $I-\lambda B$. In terms of the terminology that
became predominant later, $\varphi \left( x\right) $ is interpreted as a
generalized function (distribution) (\cite{0.9}, section 2.1.5; \cite{0.10},
section 12).

Thus, there exists a function $\varphi \left( x\right) $ satisfying Eq. (\ref%
{0.20}), or (\ref{4.7}), in the sense that%
\[
\int\limits_{0}^{1}dx\left\{ \int\limits_{0}^{1}h\left( x,\xi \right) \left[
\psi \left( \xi \right) -\lambda \left( B\psi \right) \left( \xi \right) %
\right] d\xi \right\} ^{2}=0, 
\]%
where the operator $B$ has the form (\ref{4.5}); $\psi \left( x\right)
\equiv \varphi \left( x\right) $, $x\in \left[ -1,0\right) $ and the change
of variables (\ref{4.9}) is made.

At the same time, the kernel (\ref{4.13}) that is actually used is
infinitely differentiable and depends periodically on $\xi $, which allows
us to interpret $\varphi \left( x\right) $ as a generalized function in a
less restrictive sense of the convergence of the series (\ref{0.23}) (\cite%
{0.11}, pp. 17-18):%
\begin{equation}
\lim_{n\rightarrow \infty }\int\limits_{-1}^{0}h\left( x,\xi \right) \varphi
^{\left( n\right) }\left( \xi \right) d\xi =\int\limits_{-1}^{0}h\left(
x,\xi \right) \varphi \left( \xi \right) d\xi ,\quad x\in \left[ 0,1\right]
\label{0.25}
\end{equation}%
(the variable $x$ plays the role of a parameter).

Indeed, the substitution of expressions (\ref{0.23}) and (\ref{4.19}) into (%
\ref{0.25}), with the use of (\ref{4.17}) and under a redefinition of the
coefficients, yields the following:%
\[
\lim_{n\rightarrow \infty }\int\limits_{-1}^{0}\left\{ 1+2\sum_{i=1}^{\infty
}r^{i}\left[ \cos \left( 2i\pi x\right) \cos \left( 2i\pi \xi \right) +\sin
\left( 2i\pi x\right) \sin \left( 2i\pi \xi \right) \right] \right\} 
\]%
\[
\times \left\{ \frac{1}{2}a_{0}^{\prime }+\sum_{j=1}^{n}r^{-j}\left[
a_{2j-1}^{\prime }\cos \left( 2j\pi \xi \right) +a_{2j}^{\prime \prime }\sin
\left( 2j\pi \xi \right) \right] \right\} d\xi 
\]%
\[
=\frac{1}{2}a_{0}^{\prime }+\lim_{n\rightarrow \infty
}\sum_{i=1}^{n}a_{2i-1}^{\prime }\cos \left( 2i\pi x\right) +a_{2i}^{\prime
\prime }\sin \left( 2i\pi x\right) 
\]%
\[
=\frac{1}{2}a_{0}+\sum_{n=1}^{\infty }a_{2n-1}\cos \left( 2n\pi x\right)
+a_{2n}\sin \left( 2n\pi x\right) ,\quad x\in \left[ 0,1\right] . 
\]

Further, the function $\kappa \left( x\right) $ will be determined from an
equation constructed on the basis of relation (\ref{0.2}).\footnote{%
This is in contrast to section 5.4, where for an analogous purpose relation (%
\ref{4.32}) was used, which led to an equation for the function $\chi \left(
x\right) $.} By (\ref{4.39}), it takes the form%
\begin{equation}
\kappa \left( x\right) =\lambda \int\limits_{0}^{1}h\left( x,\xi \right) 
\left[ \psi \left( \xi \right) -\psi _{1}\left( \xi \right) \right] d\xi
,\quad x\in \left[ -1,0\right) ,  \label{0.26}
\end{equation}%
where $\psi _{1}\left( x\right) $ is the solution of the Fredholm integral
equation of the second kind (\ref{4.40}).

In this case, it is by no means necessary to substitute the function $\psi
\left( x\right) $ from (\ref{4.28}) into (\ref{0.26}).\footnote{%
In this way, we would arrive at Eq. (\ref{0.17}) obtained under the
assumption that the function $\psi \left( x\right) $ is harmonic.} It is
sufficient to express the integral%
\[
\int\limits_{0}^{1}h\left( x,\xi \right) \psi \left( \xi \right) d\xi 
\]%
via the function $\kappa \left( x\right) $, which is equivalent to
interpreting also $\psi \left( x\right) $ as a generalized function.

Realizing objectively in this sense $\varphi \left( x\right) $, we integrate
with the kernel $h\left( x,\xi \right) $ of Eqs. (\ref{0.20}), (\ref{0.21})
in the limits $0,1$ and $-1,0$, respectively. We get:%
\begin{equation}
\hat{\psi}\left( x\right) =\lambda \int\limits_{0}^{1}h\left( x,\xi \right) 
\hat{\varphi}\left( \xi \right) d\xi +\lambda \int\limits_{0}^{1}h\left(
x,\xi \right) \hat{\psi}\left( \xi \right) d\xi ,\quad x\in \left[ 0,1\right]
;  \label{0.27}
\end{equation}%
\begin{equation}
\hat{\varphi}\left( x\right) =\lambda \int\limits_{-1}^{0}h\left( x,\xi
\right) \hat{\varphi}\left( \xi \right) d\xi +\lambda
\int\limits_{-1}^{0}h\left( x,\xi \right) \hat{\psi}\left( \xi \right) d\xi +%
\hat{\kappa}\left( x\right) ,\quad x\in \left[ -1,0\right) ,  \label{0.28}
\end{equation}%
where%
\[
\hat{\psi}\left( x\right) =\int\limits_{0}^{1}h\left( x,\xi \right) \psi
\left( \xi \right) d\xi ;\quad \hat{\varphi}\left( x\right)
=\int\limits_{-1}^{0}h\left( x,\xi \right) \varphi \left( \xi \right) d\xi ; 
\]%
\begin{equation}
\hat{\kappa}\left( x\right) =\int\limits_{-1}^{0}h\left( x,\xi \right)
\kappa \left( \xi \right) d\xi .  \label{0.29}
\end{equation}

Given that%
\[
\int\limits_{0}^{1}h\left( x,\xi \right) \hat{\varphi}\left( \xi \right)
d\xi =\int\limits_{-1}^{0}h\left( x,\xi \right) \hat{\varphi}\left( \xi
\right) d\xi ; 
\]%
\[
\int\limits_{-1}^{0}h\left( x,\xi \right) \hat{\psi}\left( \xi \right) d\xi
=\int\limits_{0}^{1}h\left( x,\xi \right) \hat{\psi}\left( \xi \right) d\xi
,\quad x\in \left[ -1,0\right) ;\left[ 0,1\right] , 
\]%
by%
\[
\int\limits_{-1}^{0}h\left( x,\zeta \right) h\left( \zeta ,\xi \right)
d\zeta =\int\limits_{0}^{1}h\left( x,\zeta \right) h\left( \zeta ,\xi
\right) d\zeta =h\left( x,\xi \right) 
\]%
[see (\ref{4.19})], equations (\ref{0.27}) and (\ref{0.28}) are equivalent
to the following ones:%
\[
\hat{\psi}\left( x\right) =\lambda \int\limits_{-1}^{0}h\left( x,\xi \right) 
\hat{\varphi}\left( \xi \right) d\xi +\lambda \int\limits_{0}^{1}h\left(
x,\xi \right) \hat{\psi}\left( \xi \right) d\xi ,\quad x\in \left[ 0,1\right]
; 
\]%
\[
\hat{\varphi}\left( x\right) =\lambda \int\limits_{-1}^{0}h\left( x,\xi
\right) \hat{\varphi}\left( \xi \right) d\xi +\lambda
\int\limits_{0}^{1}h\left( x,\xi \right) \hat{\psi}\left( \xi \right) d\xi +%
\hat{\kappa}\left( x\right) ,\quad x\in \left[ -1,0\right) , 
\]%
or%
\begin{equation}
\hat{\Psi}\left( x\right) =\lambda \left( B\hat{\Psi}\right) \equiv \lambda
\int\limits_{-1}^{1}h\left( x,\xi \right) \hat{\Psi}\left( \xi \right) d\xi
+\left\{ 
\begin{array}{c}
0,\quad x\in \left[ 0,1\right] ;~ \\ 
\hat{\kappa}\left( x\right) ,\quad x\in \left[ -1,0\right) ,%
\end{array}%
\right.  \label{0.30}
\end{equation}%
where%
\begin{equation}
\hat{\Psi}\left( x\right) =\left\{ 
\begin{array}{c}
\hat{\psi}\left( x\right) ,\quad x\in \left[ 0,1\right] ;~ \\ 
\hat{\varphi}\left( x\right) ,\quad x\in \left[ -1,0\right) .%
\end{array}%
\right.  \label{0.31}
\end{equation}%
[Note that the choice of the kernel $h\left( x,\xi \right) $ again played a
positive role.]

Thus we have obtained an exact analogue of Eq. (\ref{0.19}), where the
generalized function (\ref{0.31}) substitutes for the function (\ref{0.24}).
The inversion of the operator $I-\lambda B$ in (\ref{0.30}) yields%
\[
\hat{\psi}\left( x\right) =\lambda \int\limits_{-1}^{0}H\left( x,\xi
,\lambda \right) \hat{\kappa}\left( \xi \right) d\xi 
\]%
[an analogue of (\ref{4.28})]. As a result, expression (\ref{0.26}), by (\ref%
{0.29}) and (\ref{0.18}), takes the form%
\[
\kappa \left( x\right) =\lambda ^{2}\int\limits_{-1}^{0}H\left( x,\xi
,\lambda \right) \hat{\kappa}\left( \xi \right) d\xi +q^{\prime }\left(
x\right) ,\quad x\in \left[ -1,0\right) . 
\]

The substitution of $\hat{\kappa}\left( x\right) $ from (\ref{0.29}), in
virtue of%
\[
\int\limits_{-1}^{0}H\left( x,\zeta ,\lambda \right) h\left( \zeta ,\xi
\right) d\zeta =\int\limits_{-1}^{0}h\left( x,\zeta \right) H\left( \zeta
,\xi ,\lambda \right) d\zeta 
\]%
[see (\ref{4.53}), (\ref{4.19}) and (\ref{4.21})], leads to the same
equation (\ref{0.17}).

Thus, the use of generalized functions, in this case, has ensured the
legitimacy of the transformations and has not affected the final result.

Further, let us turn to Eqs. (\ref{4.34}) and (\ref{4.35}):%
\begin{equation}
\left. 
\begin{array}{c}
\psi \left( x\right) \\ 
\varphi \left( x\right)%
\end{array}%
\right\} =\lambda B\left( 
\begin{array}{c}
\psi \\ 
\varphi%
\end{array}%
\right) \left( x\right) +\left\{ 
\begin{array}{c}
\mu \left( A\psi \right) \left( x\right) -\mu f\left( x\right) ,\quad x\in 
\left[ 0,1\right] ; \\ 
\kappa \left( x\right) ,\quad x\in \left[ -1,0\right) ,%
\end{array}%
\right.  \label{0.32}
\end{equation}%
\begin{equation}
\left. 
\begin{array}{c}
\psi \left( x\right) \\ 
\varphi ^{\prime }\left( x\right)%
\end{array}%
\right\} =\lambda B\left( 
\begin{array}{c}
\psi \\ 
\varphi ^{\prime }%
\end{array}%
\right) \left( x\right) +\left\{ 
\begin{array}{c}
\mu \left( A\psi \right) \left( x\right) -\mu f\left( x\right) +\chi \left(
x\right) ,\quad x\in \left[ 0,1\right] ; \\ 
0,\quad x\in \left[ -1,0\right) .%
\end{array}%
\right.  \label{0.33}
\end{equation}%
where $B$ is the operator (\ref{4.5}).

As shown above, the transformation of (\ref{0.33}) into Eq. (\ref{0.32}), or
vice versa, with the help of relations (\ref{4.32}), (\ref{0.2}), (\ref{0.3}%
) and others allows us to obtain the Fredholm integral equations of the
second kind (\ref{4.52}), (\ref{0.17}) whose solutions, for corresponding
values of the parameters $\mu $, $\lambda $ exist and are unique. For a
specific choice of the kernel $h\left( x,\xi \right) $, either in the form (%
\ref{4.13}) or in any other one, they depend exclusively on the data of the
problem.

In this sense, the solutions of Eqs. (\ref{0.32}), (\ref{0.33}) exist and
are unique under the assumption that their free terms including the
functions $\kappa \left( \xi \right) $, $\chi \left( x\right) $ are given.
Indeed, the inversion of the operator $I-\lambda B$ in (\ref{0.32}) leads to
a Fredholm integral equation of the second kind:\footnote{%
They are mentioned in the Introduction.}%
\begin{equation}
\psi \left( x\right) =\mu \int\limits_{0}^{1}K\left( x,\xi \right) \psi
\left( \xi \right) d\xi +F\left( x\right) ,\quad x\in \left[ 0,1\right] .
\label{0.34}
\end{equation}

Here,%
\begin{equation}
K\left( x,\xi \right) =k\left( x,\xi \right) +\lambda
\int\limits_{0}^{1}H\left( x,\zeta ,\lambda \right) k\left( \zeta ,\xi
\right) d\zeta ;  \label{0.35}
\end{equation}%
\[
F\left( x\right) =\lambda \int\limits_{-1}^{0}H\left( x,\xi ,\lambda \right)
\kappa \left( \xi \right) d\xi - 
\]%
\begin{equation}
-\mu \left[ f\left( x\right) +\int\limits_{0}^{1}H\left( x,\xi ,\lambda
\right) f\left( \xi \right) d\xi \right] ,  \label{0.36}
\end{equation}%
where $H\left( x,\xi ,\lambda \right) $ is the resolvent (\ref{4.21}). This
equation differs from (\ref{4.40}) only by a component of the free term, and
its solution can be found analogously. If $\psi \left( x\right) $ is known,
we can obtain%
\[
\varphi \left( x\right) =\kappa \left( x\right) +\lambda
\int\limits_{-1}^{0}H\left( x,\xi ,\lambda \right) \kappa \left( \xi \right)
d\xi 
\]%
\[
+\mu \lambda \int\limits_{0}^{1}H\left( x,\xi ,\lambda \right) \left[ \left(
A\psi \right) \left( \xi \right) -f\left( \xi \right) \right] d\xi 
\]%
[the same can be easily done with the use of Eq. (\ref{0.33})].

The function $\kappa \left( x\right) $ entering expression (\ref{0.36}) can
be found from Eq. (\ref{0.17}), whose solution is%
\begin{equation}
\kappa \left( x\right) =q^{\prime }\left( x\right) +\Lambda
\int\limits_{-1}^{0}L\left( x,\xi ,\Lambda \right) q^{\prime }\left( \xi
\right) d\xi ,\quad x\in \left[ -1,0\right) ,  \label{0.37}
\end{equation}%
where $\Lambda =\lambda ^{2}$; $L\left( x,\xi ,\Lambda \right) $ is the
resolvent (\ref{4.56}); the function $q^{\prime }\left( x\right) $ is
determined by expression (\ref{0.18}).

The latter depends on the function $\psi _{1}\left( x\right) $ that, in its
turn, represents the solution of the Fredholm integral equation of the
second kind (\ref{4.40}). In this case, it is assumed that the conditions on
the parameter $\lambda $, (\ref{4.59}), are fulfilled. The value of $\mu $
is chosen from the condition that Eq. (\ref{4.60}) possess no nontrivial
solutions.

Provided the function $\psi _{1}\left( x\right) $ is found, a general
sequence of computational procedures consists in the determination of the
following:

- the function $q^{\prime }$ by (\ref{0.18});

- the function $\kappa $ by (\ref{0.37});\footnote{%
Note that, irrespective of the data of (\ref{4.1}), the function $\kappa $
is infinitely differentiable.}

- the kernel and the free term of Eq. (\ref{0.34}) by (\ref{0.35}) and (\ref%
{0.36}) ;

- the sought function $\psi $ by (\ref{0.34}).

A proof that the so obtained solution satisfies (\ref{4.1}) is analogous to
that of section 5.5.

Let us touch on numerical realization of Eqs. (\ref{4.40}) and (\ref{0.34}).
As is known, there are a number of stable algorithms for the solution of the
Fredholm integral equation of the second kind, including both the evaluation
of spectral characteristics and of quadratures. Along with the handbook
referred to in Chapter 5 as \cite{4.5}, they are presented in \cite{0.12}, 
\cite{0.13} and in a number of other books. Generally speaking, analogous
approaches are discussed.

We can point out S. G. Mikhlin's algorithm of numerical realization of the
resolvent (\cite{0.14}, section 12) based on a finite-element approximation
of the kernel of the integral equation. Of special interest is the method of
G. N. Polozhij \cite{0.15} that transforms the Fredholm integral equation of
the second kind in such a way that its solution, irrespective of the value
of the parameter, is achieved by means of simple iterations. In this regard,
the initial approximation is established and the second iterated kernel is
used.

For convergence in the mean of simple iterations to the solution of Eq. (\ref%
{0.34}), it is necessary that%
\begin{equation}
\left\vert \mu \right\vert c_{1}<1,  \label{0.38}
\end{equation}%
where%
\[
c_{1}^{2}=\int\limits_{0}^{1}\int\limits_{0}^{1}\left\vert K\left( x,\xi
\right) \right\vert ^{2}dxd\xi . 
\]

If 
\[
\int\limits_{0}^{1}\left\vert K\left( x,\xi \right) \right\vert ^{2}d\xi
\leq c_{2}, 
\]%
where $c_{2}$ is a constant and condition (\ref{0.38}) is fulfilled,
corresponding Neumann's series is absolutely and uniformly convergent \cite%
{0.16}.

Note that a possibility of varying the parameter $0<r<1$ creates an
additional reserve of increasing the efficiency of the procedure of
numerical realization.

Instead of (\ref{4.13}), a different kernel $h\left( x,\xi \right) $ could
be used in the transformations of the present section. At the same time, the
above-mentioned numerous advantages of Poisson's kernel make such a change
absolutely unnecessary. Indeed, if $h\left( x,\xi \right) $ satisfies the
conditions of Mercer's theorem, which would be unreasonable to give up, then
(\cite{0.2}, p. 166) 
\[
\sum_{n=1}^{\infty }\lambda _{n}^{-1}<\infty , 
\]%
and, as characteristic numbers of an alternative kernel, one could accept,
for example, the inverse of the terms of an arithmetic (or geometric)
progression. Such a preference is rather problematic, whereas some of the
available advantages could be lost.

Let us now discuss the issue of the adaptation of Eq. (\ref{4.1}) to the
space $l_{2}^{\prime }$ [defined by the condition (\ref{0.22})], which was
more than once mentioned above. Thus, $R\left( A\right) $, the range of the
operator $A$, is not closed, which is the origin of the difficulties of the
determination of the function $\psi \left( x\right) $ satisfying Eq. (\ref%
{4.1}) in terms of the solution of the Fredholm integral equation of the
first kind.

As regards (\ref{0.32}), this issue is unimportant, because, by inverting
the operator $I-\lambda B$, we get a Fredholm integral equation of the
second kind, namely, (\ref{0.34}). At the same time, equation (\ref{0.32})
is constructed by a very interesting subject, namely, (\ref{4.7}). This is a
Fredholm integral equation of both the first and the second kind for the
functions $\varphi \left( x\right) $ and $\psi \left( x\right) $
respectively.

However, as regards the first of the above-mentioned properties, this
equation is principally different from (\ref{4.1}) in the following respect:
the free term $g\left( x\right) $, determined by expression (\ref{4.8}),
depends on the sought function $\psi \left( x\right) $ while not taking on
any concrete values. Therefore, one can assume that%
\[
g\left( x\right) =\frac{1}{\lambda }\psi \left( x\right)
-\int\limits_{0}^{1}h\left( x,\xi \right) \psi \left( \xi \right) d\xi \in
R\left( B^{\prime }\right) , 
\]%
where the operator is given by%
\[
B_{\bullet }^{\prime }=\int\limits_{-1}^{0}h\left( x,\xi \right) _{\bullet
}d\xi ,\quad x\in \left[ 0,1\right] ; 
\]%
and it seems that an element of the above-mentioned adaptivity is obvious.

After that, as a result of the extension of (\ref{4.7}) to $x\in \left[
-1,0\right) $, there appeared the Fredholm integral equation of the second
kind (\ref{4.24}) for $\Psi \left( x\right) $ [see (\ref{0.24})] with the
function $\kappa \left( x\right) $ representing its free term. Indeed, for a
given $\kappa \left( x\right) $, by inverting the operator $I-\lambda B$,
one can determine the function $\psi \left( x\right) $ as well as $\varphi
\left( x\right) $. Note, however: the problem of finding the function $\psi
\left( x\right) $, contained under the sigh of integration in (\ref{4.1}),
has transformed into the problem of the determination of the function $%
\kappa \left( x\right) $ that enters explicitly!

Is it possible not to return to the arguments of J. Hadamard for the
existence of correct formulation of physically substantial problems as well
as to our repeated suggestion to consider (\ref{4.1}) as a rule for carrying
out the integration of the function $\psi \left( x\right) $? As a matter of
fact, being substituted into a certain Fredholm integral equation of the
second kind, this function generates a corresponding free term.
Consequently, the problem reduces to a procedure of its determination, which
has a powerful resource, namely, a possibility of an arbitrary choice of the
kernel of the integral equation (in reality, constructed in a certain way).

We may assume that here we realize only one of a number of existing
approaches of the outlined orientation. Given such a favorable object as the
Fredholm integral equation of the second kind and the fact that a free term
that allows the function $\psi \left( x\right) $ to satisfy this equation
exists objectively, there is no alternative to correct formulation of
problems of mathematical physics!

Returning to the algorithm, let us follow the way this formulation is
actually realized. Thus, we have found the function $\psi _{1}\left(
x\right) $, part of the solution (\ref{0.34}), stipulated by $-\mu f\left(
x\right) $. From a computational point of view, the properties of the
employed Eq. (\ref{4.40}) are excellent.

Elimination from (\ref{0.33}) of the component of the solution depending on $%
\chi \left( x\right) $ has led to relation (\ref{0.2}). With the help of
this relation, by means of the resolvent, the function $\kappa \left(
x\right) $ has been determined. This function is exactly the free term to
whose determination the problem is reduced. In this regard, also taking
account of (\ref{0.37}), let us again draw attention to (\ref{4.28}), (\ref%
{4.30}), a representation of $\psi \left( x\right) $, which is,
respectively, "integral" and with the function $\chi \left( x\right) $
entering explicitly. Their variation in the process of transformations
ensuring the realization of a stable procedure of the evaluation of the
function $\kappa \left( x\right) $ has been one of the decisive factors.

Thus, the problem (\ref{4.1}) has turned into the one of the solution of the
Fredholm integral equation of the second kind (\ref{0.34}). By the
subtraction of (\ref{4.40}) from this equation, we get%
\begin{equation}
\psi _{0}\left( x\right) =\mu \int\limits_{0}^{1}K\left( x,\xi \right) \psi
_{0}\left( \xi \right) d\xi +F_{0}\left( x\right) ,\quad x\in \left[ 0,1%
\right] ,  \label{0.39}
\end{equation}%
where%
\[
F_{0}\left( x\right) =\lambda \int\limits_{-1}^{0}H\left( x,\xi ,\lambda
\right) \kappa \left( \xi \right) d\xi ; 
\]%
the function $\kappa \left( x\right) $ is determined by expressions (\ref%
{0.37}), (\ref{0.18}). Indeed, $\psi _{0}\left( x\right) $ is the same
function that enters (\ref{4.36}). This follows from Eqs. (\ref{4.34}), (\ref%
{4.35}) and the representations of the solution (\ref{4.28}), (\ref{4.31}),
i.e.,%
\[
\lambda \int\limits_{-1}^{0}H\left( x,\xi ,\lambda \right) \kappa \left( \xi
\right) d\xi =\chi \left( x\right) +\lambda \int\limits_{0}^{1}H\left( x,\xi
,\lambda \right) \chi \left( \xi \right) d\xi . 
\]

Thus the function $\psi \left( x\right) $ can be determined by adding the
solutions of (\ref{4.40}) and (\ref{0.39}):%
\[
\psi \left( x\right) =\psi _{0}\left( x\right) +\psi _{1}\left( x\right)
,\quad x\in \left[ 0,1\right] . 
\]

After the determination of the function $\psi _{1}\left( x\right) $, the
problem can be solved also by using the substitution of expression (\ref%
{0.26}) into (\ref{4.28}). In this case, the function $\psi \left( x\right) $
satisfies the Fredholm integral equation of the second kind%
\begin{equation}
\psi \left( x\right) =\lambda ^{2}\int\limits_{0}^{1}l\left( x,\xi \right)
\psi \left( \xi \right) d\xi +f^{\prime }\left( x\right) ,  \label{0.40}
\end{equation}%
where%
\[
f^{\prime }\left( x\right) =-\lambda \int\limits_{0}^{1}l\left( x,\xi
\right) \psi \left( \xi \right) d\xi 
\]%
[see also (\ref{4.55})].

Accordingly, the solution of (\ref{0.40}) is expressed via the resolvent (%
\ref{4.56}):%
\[
\psi \left( x\right) =f^{\prime }\left( x\right) +\Lambda
\int\limits_{0}^{1}L\left( x,\xi ,\Lambda \right) f^{\prime }\left( \xi
\right) d\xi ,\quad \Lambda =\lambda ^{2}; 
\]%
and, as can be noticed, in this case, only one Fredholm integral equation of
the second kind is solved numerically, namely, (\ref{4.40}). After that,
only procedures of integration are carried out.

Equations (\ref{4.40}) and (\ref{0.39}), or (\ref{4.40}) with the use of (%
\ref{0.40}) whose data are stipulated above, embody the problem (\ref{4.1})
in its correct formulation!

By comparing the algorithms of sections 5.4 and 5.6, we find that the latter
one is, obviously, more formalized, which may prove to be, in a sense, more
advantageous.

\section{A summary of computational relations (to section 6.3)}

As the necessary formulas are scattered throughout the text, we find it
reasonable to present them in a consecrative and the most convenient for
computation form.

Under the assumption that $f\left( x\right) \in R\left( A\right) $ and the
kernel $k\left( x,\xi \right) $ is closed, the function satisfying the
equation%
\[
\left( A\psi \right) \left( x\right) \equiv \int\limits_{0}^{1}k\left( x,\xi
,\lambda \right) \psi \left( \xi \right) d\xi +f\left( x\right) ,\quad x\in 
\left[ 0,1\right] 
\]%
is defined as%
\begin{equation}
\psi \left( x\right) =\psi _{0}\left( x\right) +\psi _{1}\left( x\right) ,
\label{0.41}
\end{equation}%
where $\psi _{0}\left( x\right) $ and $\psi _{1}\left( x\right) $ are the
solutions of the Fredholm integral equation of the second kind%
\begin{equation}
\psi \left( x\right) =\mu \int\limits_{0}^{1}K\left( x,\xi \right) \psi
\left( \xi \right) d\xi +F\left( x\right) ,\quad x\in \left[ 0,1\right] ,
\label{0.42}
\end{equation}%
with the free term, respectively%
\begin{equation}
F\left( x\right) \equiv F_{0}\left( x\right) =\lambda
\int\limits_{-1}^{0}H\left( x,\xi ,\lambda \right) \kappa \left( \xi \right)
d\xi ;  \label{0.43}
\end{equation}%
\begin{equation}
F\left( x\right) \equiv F_{1}\left( x\right) =-\mu \left[ f\left( x\right)
+\lambda \int\limits_{-1}^{0}H\left( x,\xi ,\lambda \right) f\left( \xi
\right) d\xi \right] ;  \label{0.44}
\end{equation}%
[in relation to Eq. (\ref{4.40}), $F_{1}\equiv f_{1}$].

The kernel of Eq. (\ref{0.42}) is given by%
\begin{equation}
K\left( x,\xi \right) =k\left( x,\xi \right) +\lambda
\int\limits_{0}^{1}H\left( x,\zeta ,\lambda \right) k\left( \zeta ,\xi
\right) d\zeta .  \label{0.45}
\end{equation}

Here and above,%
\begin{equation}
H\left( x,\xi ,\lambda \right) =\frac{1}{1-2\lambda }+2\sum_{n=1}^{\infty }%
\frac{r^{n}}{1-2\lambda r^{n}}\cos \left[ 2n\pi \left( x-\xi \right) \right]
,  \label{0.46}
\end{equation}%
with the parameter $0<r<1$;%
\begin{equation}
\kappa \left( x\right) =\rho \left( x\right) +\Lambda
\int\limits_{-1}^{0}L\left( x,\xi ,\Lambda \right) \rho \left( \xi \right)
d\xi  \label{0.47}
\end{equation}%
[in relation to Eq. (\ref{0.18}), $\rho =q^{\prime }$].

In this expression, $\Lambda =\lambda ^{2}$;%
\begin{equation}
\rho \left( x\right) =-\lambda \int\limits_{0}^{1}h\left( x,\xi \right) \psi
_{1}\left( \xi \right) d\xi ,  \label{0.48}
\end{equation}%
where%
\[
h\left( x,\xi \right) =\frac{1-r^{2}}{1-2r\cos \left[ 2n\pi \left( x-\xi
\right) \right] +r^{2}}= 
\]%
\begin{equation}
=1+2\sum_{n=1}^{\infty }r^{n}\cos \left[ 2n\pi \left( x-\xi \right) \right] ;
\label{0.49}
\end{equation}%
\begin{equation}
L\left( x,\xi ,\Lambda \right) =\frac{1}{1-2\lambda -\Lambda }%
+2\sum_{n=1}^{\infty }\frac{r^{2n}}{1-2\lambda r^{n}-\Lambda r^{2n}}\cos %
\left[ 2n\pi \left( x-\xi \right) \right] .  \label{0.50}
\end{equation}

Parameter%
\[
\lambda \neq :0;r^{-n};\frac{1}{2}r^{-n};\left( -1\pm \sqrt{2}\right)
r^{-n},\quad n=1,2,\ldots ; 
\]%
parameter%
\[
\mu \neq \mu _{n},\quad n=1,2,\ldots , 
\]%
where $\mu _{n}$ are characteristic numbers of the homogeneous equation%
\[
\psi \left( x\right) =\mu \int\limits_{0}^{1}K\left( x,\xi \right) \psi
\left( \xi \right) d\xi ,\quad x\in \left[ 0,1\right] . 
\]

When the values of $\mu $, $\lambda $ and $r$ are chosen (they can be
corrected later, depending on various arguments), the sequence of
computational procedures is as follows:

- determination of the kernels $K$ of Eq. (\ref{0.42}) from (\ref{0.45}),
using expression (\ref{0.46});

- determination of the free term $F_{1}$ from (\ref{0.44});

- determination of the function $\psi _{1}$ from (\ref{0.42}), with $F\equiv
F_{1}$;

- determination of the function $\rho $ from (\ref{0.48}), using expression (%
\ref{0.49});

- determination of the function $\kappa $ from (\ref{0.47}), using
expression (\ref{0.50});

- determination of the free term $F_{0}$ from (\ref{0.43});

- determination of the function $\psi _{0}$ from (\ref{0.42}), with $F\equiv
F_{0}$;

- determination of the sought function $\psi $ from (\ref{0.41}).

\bigskip

\chapter{A reduction of linear boundary-value and initial-boundary-value
problems to Fredholm integral equations of the first kind}

\section{Ordinary differential equations}

Consider, for example, 
\begin{equation}  \label{5.1}
u^{\prime \prime }-a\left( x\right) u=f\left( x\right) ,\quad x\in \left[ 0,1%
\right] ;
\end{equation}
\begin{equation}  \label{5.2}
u^{\prime }\left( 0\right) =u\left( 1\right) =0,
\end{equation}
where $a\left( x\right) $ and $f\left( x\right) $ are given $L_2$-functions.

From the notation 
\begin{equation}
u^{\prime \prime }\left( x\right) =\psi \left( x\right) ,  \label{5.3}
\end{equation}%
it follows: 
\begin{equation}
u^{\prime }\left( x\right) =\int\limits_{0}^{x}\psi \left( \xi \right) d\xi
+c_{1};  \label{5.4}
\end{equation}%
\begin{equation}
u\left( x\right) =\int\limits_{0}^{x}\left( x-\xi \right) \psi \left( \xi
\right) d\xi +c_{1}x+c_{0},  \label{5.5}
\end{equation}%
where $c_{0}$, $c_{1}$ are the constants of integration.

The substitution of expressions (\ref{5.3}) and (\ref{5.5}) into (\ref{5.1})
leads to a Volterra integral equations of the second kind: 
\begin{equation}
\psi \left( x\right) =a\left( x\right) \int\limits_{0}^{x}\left( x-\xi
\right) \psi \left( \xi \right) d\xi +\left( c_{1}x+c_{0}\right) a\left(
x\right) +f\left( x\right) ,\quad x\in \left[ 0,1\right] ,  \label{5.6}
\end{equation}%
whose the solution is 
\begin{equation}
\psi \left( x\right) =\left( c_{1}x+c_{0}\right) a\left( x\right) +f\left(
x\right) +\int\limits_{0}^{x}Q\left( x,\xi \right) \left[ \left( c_{1}\xi
+c_{0}\right) a\left( \xi \right) +f\left( \xi \right) \right] d\xi ,
\label{5.7}
\end{equation}%
where $Q\left( x,\xi \right) $ is the resolvent of the kernel $a\left(
x\right) \left( x-\xi \right) $.

Taking into account (\ref{5.4}), (\ref{5.5}) and (\ref{5.7}), we find from
the boundary conditions (\ref{5.2}): $c_1=0;$%
\begin{equation}  \label{5.8}
c_0=-\frac{\int\limits_0^1\left( 1-\xi \right) \left[ f\left( \xi \right)
+\int\limits_0^\xi Q\left( \xi ,\zeta \right) f\left( \zeta \right) d\zeta %
\right] d\xi }{1+\int\limits_0^1\left( 1-\xi \right) \left[ a\left( \xi
\right) +\int\limits_0^\xi Q\left( \xi ,\zeta \right) a\left( \zeta \right)
d\zeta \right] d\xi }.
\end{equation}

One can act in a different way: Namely, upon the substitution of expressions
(\ref{5.4}), (\ref{5.5}) into (\ref{5.2}), we get $c_{1}=0;$%
\[
c_{0}=-\int\limits_{0}^{1}\left( 1-\xi \right) \psi \left( \xi \right) d\xi
, 
\]%
and, as a result, 
\begin{equation}
u\left( x\right) =\left[ \int\limits_{0}^{x}\left( x-\xi \right)
-\int\limits_{0}^{1}\left( 1-\xi \right) \right] \psi \left( \xi \right)
d\xi .  \label{5.9}
\end{equation}

In contrast to (\ref{5.6}), the problem reduces to the Fredholm integral
equation of the second kind 
\begin{equation}
\psi \left( x\right) =a\left( x\right) \left[ \int\limits_{0}^{x}\left(
x-\xi \right) -\int\limits_{0}^{1}\left( 1-\xi \right) \right] \psi \left(
\xi \right) d\xi +f\left( x\right) ,\quad x\in \left[ 0,1\right] ,
\label{5.10}
\end{equation}

whose the solution is 
\begin{equation}
\psi \left( x\right) =f\left( x\right) +\int\limits_{0}^{1}Q\left( x,\xi
\right) f\left( \xi \right) d\xi ,  \label{5.11}
\end{equation}%
where $Q\left( x,\xi \right) $ is the resolvent of the kernel\footnote{%
It is assumed that the homogeneous equation (\ref{5.10}) has only a trivial
solution.
\par
{}} 
\[
-a\left( x\right) \left\{ 
\begin{array}{c}
1-\xi ,\quad x<\xi \leq 1; \\ 
1-x,\quad 0\leq \xi \leq x.%
\end{array}%
\right. 
\]

The substitution of (\ref{5.7}) into (\ref{5.5}), taking into account (\ref%
{5.8}), or the substitution of (\ref{5.11}) into (\ref{5.9}), allows us to
find the solution of the problem (\ref{5.1}), (\ref{5.2}). Note that the
outlined approach is substantially indifferent to the order of the
differential equations, the form of initial or boundary conditions and to
the data of the problem.

Analogous transformations are traditionally discussed in courses of the
theory of integral equations (see, e.g., \cite{5.1,5.2}). At the same time,
as far as the solution of applied problems is concerned, the construction of
integral equations of the second kind did not gain sufficient popularity,
which can be characterized as a kind of a paradox. It is rather surprising
in light of rather active attempts of its popularization: see, e.g.,
publications of S. E. Mikeladze, I. A. Birger, and A. N. Golubentsev \cite%
{5.3,5.4,5.5}.

It seems that the reasons for this situation are, on the one hand,
inefficiency of technical means of numerical realization of integral
equations before wide-spread implementation of computers, and, on the other
hand, insufficient popularity of the techniques of the theory of integral
equations among specialists in applied science.

Nonetheless, here is an opinion of G. Wiarda (\cite{5.6}, p. 5): "... an
integral equation substitutes for a corresponding differential equation with
its boundary conditions that, as far as a concrete physical phenomenon is
concerned, necessarily arise with any differential equation. An integral
equation already contains all the elements specifying the physical problem.
One more advantage of integral equations lie in the fact that, in most
cases, we arrive at equations of the same type..., whereas the types of
differential equations, even in closely related problems, often turn out to
be rather different."

\section{An illustration of the procedure of reduction}

Let us turn to the problem of bending of a membrane stretched along a
contour by a uniform load: 
\begin{equation}
\partial _{x}^{2}u+\partial _{y}^{2}u=-1,  \label{5.12}
\end{equation}%
\begin{equation}
u\left( 0,y\right) =u\left( 1,y\right) =0,  \label{5.13}
\end{equation}%
\begin{equation}
u\left( x,0\right) =u\left( x,1\right) =0.  \label{5.14}
\end{equation}

From the notation 
\begin{equation}
\partial _{x}^{2}u\left( x,y\right) =\psi \left( x,y\right) ,  \label{5.15}
\end{equation}%
it follows: 
\begin{equation}
u\left( x,y\right) =\int\limits_{0}^{x}\left( x-\xi \right) \psi \left( \xi
,y\right) d\xi +xg_{11}\left( y\right) +g_{12}\left( y\right) ,  \label{5.16}
\end{equation}%
where $g_{1j}\left( y\right) $ are functions of integration.

In view of (\ref{5.15}), equation (\ref{5.12}) takes the form 
\[
\partial _y^2u\left( x,y\right) =-1-\psi \left( x,y\right) , 
\]
and, respectively, 
\begin{equation}  \label{5.17}
u\left( x,y\right) =-\frac 12y^2-\int\limits_0^y\left( y-\eta \right) \psi
\left( x,\eta \right) d\eta +yg_{21}\left( x\right) +g_{22}\left( x\right) ,
\end{equation}
where $g_{2j}\left( y\right) $ are also functions of integration.

The substitution of expressions (\ref{5.16}), (\ref{5.17}) into the boundary
conditions (\ref{5.13}) and (\ref{5.14}), respectively, allows us to
determine $g_{12}=g_{22}=0;$ 
\[
g_{11}\left( y\right) =-\int\limits_0^1\left( 1-\xi \right) \psi \left( \xi
,y\right) d\xi ;\quad g_{21}\left( x\right) =\frac 12+\int\limits_0^1\left(
1-\eta \right) \psi \left( x,\eta \right) d\eta . 
\]

As a result, expressions (\ref{5.16}) and (\ref{5.17}), respectively, take
the form 
\begin{equation}
u\left( x,y\right) =\left[ \int\limits_{0}^{x}\left( x-\xi \right)
-x\int\limits_{0}^{1}\left( 1-\xi \right) \right] \psi \left( \xi ,y\right)
d\xi ;  \label{5.18}
\end{equation}%
\begin{equation}
u\left( x,y\right) =\frac{1}{2}y\left( 1-y\right) -\left[ \int%
\limits_{0}^{y}\left( y-\eta \right) -y\int\limits_{0}^{1}\left( 1-\eta
\right) \right] \psi \left( x,\eta \right) d\eta .  \label{5.19}
\end{equation}

Eliminating $u$ from these expressions, we get a Fredholm integral equation
of the first kind: 
\[
\left[ \int\limits_{0}^{x}\left( x-\xi \right) -x\int\limits_{0}^{1}\left(
1-\xi \right) \right] \psi \left( \xi ,y\right) d\xi 
\]%
\begin{equation}
+\left[ \int\limits_{0}^{y}\left( y-\eta \right) -y\int\limits_{0}^{1}\left(
1-\eta \right) \right] \psi \left( x,\eta \right) d\eta =\frac{1}{2}y\left(
1-y\right) .  \label{5.20}
\end{equation}

Thus, a principal difference from the one-dimensional case consists in the
reduction of the problem (\ref{5.12})-(\ref{5.14}) to an ill-posed one.
However, here we will be interested not in the determination of the function 
$\psi $ satisfying Eq. (\ref{5.20}) (just note that the algorithms of
sections 5.4, 5.6 and 6.3 apply to it as well) but in the universality of
the procedure of transformation.

Indeed, let the domain of the problem be different from the canonical one,
and let, for example the second condition (\ref{5.13}) have the form $%
u\left( \gamma ,y\right) =0$, where $x=\gamma \left( y\right) $ a certain
single-valued function. Instead of (\ref{5.18}), we have 
\[
u\left( x,y\right) =\left[ \int\limits_{0}^{x}\left( x-\xi \right)
-x\int\limits_{0}^{\gamma \left( y\right) }\left[ \gamma \left( y\right)
-\xi \right] \right] \psi \left( \xi ,y\right) d\xi , 
\]%
and, from a computational point of view, any differences are absent. For the
transition to an ordinary procedure of the evaluation of the integral on a
rectangular domain, it is sufficient to employ a non-orthogonal mapping of
the type $x=\gamma \bar{x}$, $y=\bar{y}$.

It is not difficult to notice that each of expressions (\ref{5.18}) and (\ref%
{5.19}) satisfy identically the pair of boundary conditions (\ref{5.13}) and
(\ref{5.14}), respectively. The rest of the conditions are fulfilled
approximately, depending on the accuracy of the determination of $\psi $. At
the same time, the solution can be represented in the form that satisfies
identically both the conditions (\ref{5.13}) and (\ref{5.14}): 
\[
U_{1}\left( x,y\right) =u_{1}\left( x,y\right) -\left( 1-y\right)
u_{1}\left( x,0\right) -yu_{1}\left( x,1\right) ; 
\]%
\[
U_{2}\left( x,y\right) =u_{2}\left( x,y\right) -\left( 1-x\right)
u_{2}\left( 0,y\right) -xu_{2}\left( 1,y\right) . 
\]

Here, the functions $u_{1}$, $u_{2}$ are determined by (\ref{5.18}) and (\ref%
{5.19}), respectively.

The norm of the error of closure of the values of $u_1\left( x,y\right) $ or 
$U_1\left( x,y\right) $ allows us to estimate the error of the approximate
solution: 
\[
\delta =\frac{2\left\| U_1\left( x,y\right) -U_2\left( x,y\right) \right\| }{%
\left\| U_1\left( x,y\right) +U_2\left( x,y\right) \right\| }. 
\]

However, if instead of (\ref{5.13}) the conditions 
\[
\partial _{x}u\left( 0,y\right) =\partial _{x}u\left( 1,y\right) =0 
\]%
would be imposed, they could not be satisfied by the expression for the
derivative%
\[
\partial _{x}u\left( x,y\right) =\int\limits_{0}^{x}\psi \left( \xi
,y\right) d\xi +g_{11}\left( y\right) 
\]%
that follows from (\ref{5.15}).

Nevertheless, this complication can be easily overcome by the use, in
particular, of the relation%
\[
\partial _{x}^{2}u+\beta u=\psi , 
\]%
where $\beta $ is a constant that allows us to retain both the functions of
integration $g_{1j}\left( y\right) $.

Let us turn to an equivalent formulation of the problem (\ref{5.12})-(\ref%
{5.14}):%
\begin{equation}
\partial _{x}^{2}u_{1}+\partial _{y}^{2}u_{2}=-1;\quad u_{1}\left(
x,y\right) =u_{2}\left( x,y\right) ,  \label{5.21}
\end{equation}%
\begin{equation}
u_{1}\left( 0,y\right) =u_{1}\left( 1,y\right) =u_{2}\left( x,0\right)
=u_{2}\left( x,1\right) =0,  \label{5.22}
\end{equation}%
using a representation of the solution of the type 
\[
u_{1}\left( x,y\right) =\int\limits_{0}^{x}k_{1}\left( x,y,\xi \right) \psi
_{1}\left( \xi ,y\right) d\xi +\sum_{j=1}^{2}\mu _{1j}\left( x\right)
g_{1j}\left( y\right) ; 
\]%
\[
u_{2}\left( x,y\right) =\int\limits_{0}^{y}k_{2}\left( x,y,\eta \right) \psi
_{2}\left( x,\eta \right) d\eta +\sum_{j=1}^{2}\mu _{2j}\left( y\right)
g_{2j}\left( x\right) . 
\]

We assume that the kernels are given and satisfy the conditions 
\[
k_1\left( x,y,x\right) =k_1\left( x,y,y\right) =0; 
\]
\begin{equation}  \label{5.23}
\partial _xk_1\left( x,y,x\right) \neq 0;\quad \partial _yk_2\left(
x,y,y\right) \neq 0,\quad x,y\in \left[ 0,1\right] ;
\end{equation}
$\mu _{1j}\left( x\right) $, $\mu _{2j}\left( y\right) $ are also given; $%
g_{1j}\left( y\right) $, $g_{2j}\left( x\right) $ are to be determined from
the boundary conditions as discussed above.

Let us set $\mu _{11}=x$, $\mu _{21}=y$, $\mu _{12}=\mu _{22}=1$. In this
case, under the conditions (\ref{5.22}), we get: 
\begin{equation}
u_{1}\left( x,y\right) =\left[ \int\limits_{0}^{x}k_{1}\left( x,y,\xi
\right) -x\int\limits_{0}^{1}k_{1}\left( 1,y,\xi \right) \right] \psi
_{1}\left( \xi ,y\right) d\xi ;  \label{5.24}
\end{equation}%
\begin{equation}
u_{2}\left( x,y\right) =\left[ \int\limits_{0}^{y}k_{2}\left( x,y,\eta
\right) -y\int\limits_{0}^{1}k_{2}\left( x,1,\eta \right) \right] \psi
_{2}\left( x,\eta \right) d\eta ,  \label{5.25}
\end{equation}%
and, respectively, 
\begin{equation}
\partial _{x}k_{1}\left( x,y,x\right) \psi _{1}\left( x,y\right)
+\int\limits_{0}^{x}\partial _{x}^{2}k_{1}\left( x,y,\xi \right) \psi
_{1}\left( \xi ,y\right) d\xi =\partial _{x}^{2}u_{1}\left( x,y\right) ;
\label{5.26}
\end{equation}%
\begin{equation}
\partial _{y}k_{2}\left( x,y,y\right) \psi _{2}\left( x,y\right)
+\int\limits_{0}^{y}\partial _{y}^{2}k_{2}\left( x,y,\eta \right) \psi
_{2}\left( x,\eta \right) d\eta =\partial _{y}^{2}u_{2}\left( x,y\right) .
\label{5.27}
\end{equation}

Let, in addition to the conditions (\ref{5.23}), $\partial
_{x}^{2}k_{1}\left( x,y,\xi \right) $ and $\partial _{y}^{2}k_{2}\left(
x,y,\eta \right) $ be $L_{2}$-kernels. Here, (\ref{5.26}), (\ref{5.27}) are
Volterra integral equations of the second kind with respect to the functions 
$\psi _{1}$, $\psi _{2}$, whose solutions, by general theory, exist and are
unique. Therefore, the representations (\ref{5.24}) and (\ref{5.25})
correspond to the physical content of the problem (\ref{5.21}), (\ref{5.22}).

In (\ref{5.23}), we can set 
\[
k_{1}\left( x,y,\xi \right) =\left( x-\xi \right) k_{1}^{\prime }\left(
x,y,\xi \right) ;\quad k_{2}\left( x,y,\eta \right) =\left( y-\eta \right)
k_{2}^{\prime }\left( x,y,\eta \right) , 
\]%
where 
\[
k_{1}^{\prime }\left( x,y,x\right) \neq 0;\quad k_{2}^{\prime }\left(
x,y,y\right) \neq 0,\quad x,y\in \left[ 0,1\right] , 
\]%
using these expressions to refract the a priori information about the
solution in order to smooth the sought functions $\psi _{i}$ and, in
general, to simplify the procedure of calculations. It is clear that this
point is important for more complicated problems with different kinds of
singularities of the behavior of the solutions, and we just outline it here.

The substitution of $\partial _{x}^{2}u_{1}$ and $\partial _{y}^{2}u_{2}$
from (\ref{5.26}), (\ref{5.27}) into (\ref{5.21}) leads to a system of
integral equations 
\begin{equation}
\psi _{2}\left( x,y\right) =-\frac{1}{\partial _{y}k_{2}\left( x,y,y\right) }%
\int\limits_{0}^{y}\partial _{y}^{2}k_{2}\left( x,y,\eta \right) \psi
_{2}\left( x,\eta \right) d\eta +F\left( x,y,\psi _{1}\right) ,  \label{5.28}
\end{equation}%
where 
\[
F\left( x,y,\psi _{1}\right) =-\frac{1}{\partial _{y}k_{2}\left(
x,y,y\right) }\left[ 1+\partial _{x}k_{1}\left( x,y,x\right) \psi _{1}\left(
x,y\right) \right. 
\]%
\[
\left. +\int\limits_{0}^{x}\partial _{x}^{2}k_{1}\left( x,y,\xi \right) \psi
_{1}\left( \xi ,y\right) d\xi \right] ; 
\]%
\[
\left[ \int\limits_{0}^{x}k_{1}\left( x,y,\xi \right)
-x\int\limits_{0}^{1}k_{1}\left( 1,y,\xi \right) \right] \psi _{1}\left( \xi
,y\right) d\xi 
\]%
\begin{equation}
-\left[ \int\limits_{0}^{y}k_{2}\left( x,y,\eta \right)
-y\int\limits_{0}^{1}k_{2}\left( x,1,\eta \right) \right] \psi _{2}\left(
x,\eta \right) d\eta =0,\quad x,y\in \left[ 0,1\right] .  \label{5.29}
\end{equation}

From Eq. (\ref{5.28}), we find 
\begin{equation}  \label{5.30}
\psi _2\left( x,y\right) =F\left( x,y,\psi _1\right) +\int\limits_0^yQ\left(
x,y,\eta \right) F\left( x,\eta ,\psi _1\right) d\eta ,
\end{equation}
where $Q\left( x,y,\eta \right) $ is the resolvent of the kernel $-\partial
_y^2k_2\left( x,y,\eta \right) /\partial _yk_2\left( x,y,y\right) $.

The substitution of expression (\ref{5.30}) into (\ref{5.29}) allows us to
obtain a Fredholm integral equation of the first kind with respect to the
function $\psi _{1}\left( x,y\right) $. Clearly, the above reduction scheme
is more cumbersome compared to that based on the formulation of the problem
in the standard interpretation (\ref{5.12})-(\ref{5.14}). At the same time,
one may discern in it some iteration elements that result from the fact that
(\ref{5.28}) is a Volterra integral equation of the second kind with respect
to both $\psi _{1}$ and $\psi _{2}$.

The reduction procedure applies also to differential equations of other
types. As an illustration we consider the simplest problem of thermal
conductivity: 
\begin{equation}
\partial _{t}u-\partial _{x}^{2}u_{2}=0,  \label{5.31}
\end{equation}%
\begin{equation}
u\left( x,0\right) =u_{0}\left( x\right) ;\quad u\left( 0,t\right) =u\left(
1,t\right) =0.  \label{5.32}
\end{equation}

From $\psi =\partial _x^2u$, equation (\ref{5.31}) and conditions (\ref{5.32}%
), we get 
\[
u\left( x,t\right) =\left[ \int\limits_0^x\left( x-\xi \right)
-x\int\limits_0^1\left( 1-\xi \right) \right] \psi \left( \xi ,t\right) d\xi
; 
\]
\[
u\left( x,t\right) =\int\limits_0^t\psi \left( x,\eta \right) d\eta
+u_0\left( x\right) . 
\]

Accordingly, 
\[
\left[ \int\limits_{0}^{x}\left( x-\xi \right) -x\int\limits_{0}^{1}\left(
1-\xi \right) \right] \psi \left( \xi ,t\right) d\xi
-\int\limits_{0}^{t}\psi \left( x,\eta \right) d\eta =u_{0}\left( x\right)
;\quad x,y\in \left[ 0,1\right] . 
\]

In order to make an analogous reduction of the problem of bending of a
rectangular plate of variable stiffness $D$, fixed along a contour \cite{5.7}%
, 
\[
D\Delta \Delta u+2\partial _{x}D\partial _{x}\Delta u+2\partial
_{y}D\partial _{y}\Delta u+\Delta D\Delta u 
\]%
\begin{equation}
-\left( 1-\nu \right) \left( \partial _{x}^{2}D\partial _{y}^{2}u-2\partial
_{xy}D\partial _{xy}u+\partial _{y}^{2}D\partial _{x}^{2}u\right) =q,
\label{5.33}
\end{equation}%
\begin{equation}
\partial _{x}^{n}u\left( 0,y\right) =\partial _{x}^{n}u\left( a,y\right)
=\partial _{y}^{n}u\left( x,0\right) =\partial _{y}^{n}u\left( x,b\right)
=0,\quad n=0,1,  \label{5.34}
\end{equation}%
where $\Delta =\partial _{x}^{2}+\partial _{y}^{2}$; $\nu $ is the Poisson
coefficient; $q\left( x,y\right) $ is the intensity of the transverse load,
we can set 
\begin{equation}
u\left( x,y\right) =\int\limits_{0}^{x}k\left( x,y,\xi \right) \psi \left(
\xi ,y\right) d\xi +\sum_{j=1}^{4}x^{j-1}g_{1j}\left( y\right) .
\label{5.35}
\end{equation}

Here, 
\[
\partial _{x}^{n}k_{1}\left( x,y,x\right) =0,\quad n=0,1,2;\quad \partial
_{x}^{3}k_{2}\left( x,y,y\right) \neq 0,\quad x\in \left[ 0,a\right] ;\,y\in %
\left[ 0,b\right] , 
\]%
and the functions $g_{1j}\left( y\right) $ are intended to satisfy the
conditions (\ref{5.34}) for $x=0$, $x=a$. The second representation of the
solution via $\psi \left( x,y\right) $ is determined by means of the
substitution of (\ref{5.35}) into Eq. (\ref{5.33}) and four-fold integration
over the variable $y$. The appearing functions $g_{2j}\left( x\right) $
allow us to satisfy the conditions (\ref{5.34}) for $y=0$, $y=b$. After
that, $u\left( x,y\right) $ is eliminated from the representation of the
solution..

Note that with the help of $k\left( x,y,\eta \right) $ one can easily
satisfy conditions at isolated points inside the considered domain, e.g., $%
u\left( x_{i},y_{i}\right) =0$. The procedure of the reduction also applies
to mixed boundary conditions (a change of the type along a side) and to the
case of a connection of plates. Analogously, three-dimensional problems of
mathematical physics can also be reduced to Fredholm integral equations of
the first kind.

\section{Universality and analogous approaches}

Thus, a comparatively elementary method of the reduction of linear
boundary-value and initial-boundary-value problems to Fredholm integral
equations of the first kind is rather universal from the point of view of
its realizations as far as the following aspects are concerned:

- the order and structure of differential equations;

- the form of boundary conditions;

- the availability of variable coefficients;

- the form of the domain;

- the dimensionality of the problem.

In this situation, all the information about a concrete problem is
transferred into a functional equation, whose solution does not require any
conditions on the contour of the domain, which poses a substantial
advantage. Thus, its solution can be sought in the form of a series of a
system of coordinate elements intended exclusively to ensure the efficiency
of the procedure of the numerical realization.

However, the problem obtained as a result of transformations is incorrect,
hence its numerical realization requires adequate methods. At the same time,
in applications, the solution of such a problem can be acceptably
approximated by a series with the number of terms that does not affect the
stability of the numerical algorithms. Therefore, one can hardly explain the
absence of interest to a systematic use of this procedure, especially in the
period before the general orientation at the discretization of problems of
mathematical modeling.

One may state that special literature did not point out the existence of a
formalized method of the reduction of practically arbitrary
initial-boundary-value problems to Fredholm integral equations of the first
kind. At the same time, there a number of examples of applications of
analogous transformations in rather particular situations. As a rule, they
were a given physical interpretation that considerably disguised the
generality of this approach.

Thus, Yu. V. Repman used as a function closely related to $\psi $ boundary
forces of a plate of a canonical configuration that allowed one to satisfy
conditions on an internal contour of complex configuration \cite{5.8}. L. A.
Rozin has developed a method of separation that admits a reduction of the
problems of calculations of membranes to systems of Fredholm integral
equations of the first kind for the forces of interaction of isolated bars (%
\cite{5.9}, section 9). Some publications point out the advantages of the
approximation of higher-order derivative of differential equations with
respect to one of the variables that, compared to numerical differentiation,
are much more accurate. However, as a rule, no comments were made on an
actual transition to ill-posed problems (see, e.g., \cite{5.10}).

Some problems for differential equations, and, in particular, the following
one: 
\[
\partial _{xy}u=a\partial _{x}u+b\partial _{y}u+cu+f, 
\]%
where $a$, $b$, $c$ and $f$ are given functions of the variables $x$ and $y$%
, can be reduced directly to Volterra and Fredholm integral equations of the
second kind with respect to the higher-order derivative ($\psi =\partial
_{xy}u$). These issues are studied in detail by G. M\"{u}ntz \cite{5.11}. Of
considerable interest is the fact, established by this author, that
analogous transformations cannot be extended to the case of the simplest
equation of the elliptic type.

\section{A connection to the algorithm of section 6.4}

The Fredholm integral equation of the first kind that arises as a result of
the reduction of two-dimensional boundary-value (initial-value) problems,
can be represented in the form 
\begin{equation}
\int\limits_{0}^{1}\tau _{1}\left( x,y,\xi \right) \psi \left( \xi ,y\right)
d\xi +\int\limits_{0}^{1}\tau _{2}\left( x,y,\eta \right) \psi \left( x,\eta
\right) d\eta =f\left( x,y\right) ,\quad x,y\in \left[ 0,1\right] ,
\label{5.36}
\end{equation}%
where $\tau _{1}\left( x,y,\xi \right) $, $\tau _{2}\left( x,y,\eta \right) $
and $f\left( x,y\right) $ are given functions; $\psi \left( x,y\right) $ has
to be determined.

Under the assunption that the function satisfying (\ref{5.36}) exists and
unique, it is represented in the form%
\begin{equation}
\psi \left( x,y\right) =\psi _{0}\left( x,y\right) +\psi _{1}\left(
x,y\right) ,  \label{5.37}
\end{equation}%
where $\psi _{0}\left( x,y\right) $ and $\psi _{1}\left( x,y\right) $ are
solutions of the Fredholm integral equation og the second kind%
\[
\psi \left( x,y\right) =\mu \left[ \int\limits_{0}^{1}N\left( x,y,\xi
\right) \psi \left( \xi ,y\right) d\xi +\int\limits_{0}^{1}M\left( x,y,\eta
\right) \psi \left( x,\eta \right) d\eta \right. 
\]%
\begin{equation}
+\left. \int\limits_{0}^{1}d\xi \int\limits_{0}^{1}T\left( x,y,\xi ,\eta
\right) \psi \left( \xi ,\eta \right) d\eta \right] +F\left( x,y\right)
,\quad x,y\in \left[ 0,1\right] ,  \label{5.38}
\end{equation}%
with the free term, respectively,%
\begin{equation}
F\left( x,y\right) \equiv F_{0}\left( x,y\right) =\lambda
\int\limits_{-1}^{0}H\left( x,\xi ,\lambda \right) \kappa \left( \xi
,y\right) d\xi ;  \label{5.39}
\end{equation}%
\begin{equation}
F\left( x,y\right) \equiv F_{1}\left( x,y\right) =-\mu \left[ f\left(
x,y\right) +\int\limits_{0}^{1}H\left( x,\xi ,\lambda \right) f\left( \xi
,y\right) d\xi \right] .  \label{5.40}
\end{equation}

In Eq. (\ref{5.38}), the kernels are given by%
\[
N\left( x,y,\xi \right) =\tau _{1}\left( x,y,\xi \right) +\lambda
\int\limits_{0}^{1}H\left( x,\zeta ,\lambda \right) \tau _{1}\left( \zeta
,y,\xi \right) d\zeta ; 
\]%
\begin{equation}
M\left( x,y,\eta \right) =\tau _{2}\left( x,y,\eta \right) ;\quad T\left(
x,y,\xi ,\eta \right) =\lambda H\left( x,\xi ,\lambda \right) \tau
_{2}\left( \xi ,y,\eta \right) .  \label{5.41}
\end{equation}

Here and above,%
\begin{equation}
H\left( x,\xi ,\lambda \right) =\frac{1}{1-2\lambda }+2\sum_{n=1}^{\infty }%
\frac{r^{n}}{1-2\lambda r^{n}}\cos \left[ 2n\pi \left( x-\xi \right) \right]
,  \label{5.42}
\end{equation}%
where the parameter is $0<r<1$;%
\begin{equation}
\kappa \left( x,y\right) =\rho \left( x,y\right) +\Lambda
\int\limits_{-1}^{0}L\left( x,\xi ,\Lambda \right) \rho \left( \xi ,y\right)
d\xi .  \label{5.43}
\end{equation}

In this expansion, $\Lambda =\lambda ^{2}$;%
\begin{equation}
\rho \left( x,y\right) =-\lambda \int\limits_{0}^{1}h\left( x,\xi \right)
\psi _{1}\left( \xi ,y\right) d\xi ,  \label{5.44}
\end{equation}%
where%
\[
h\left( x,\xi \right) =\frac{1-r^{2}}{1-2r\cos \left[ 2n\pi \left( x-\xi
\right) \right] +r^{2}} 
\]%
\begin{equation}
=1+2\sum_{n=1}^{\infty }r^{n}\cos \left[ 2n\pi \left( x-\xi \right) \right] ;
\label{5.45}
\end{equation}%
\begin{equation}
L\left( x,\xi ,\Lambda \right) =\frac{1}{1-2\lambda -\Lambda }%
+2\sum_{n=1}^{\infty }\frac{r^{2n}}{1-2\lambda r^{n}-\Lambda r^{2n}}\cos %
\left[ 2n\pi \left( x-\xi \right) \right] .  \label{5.46}
\end{equation}

The parameter%
\[
\lambda \neq :0;r^{n};\frac{1}{2}r^{-n};\left( -1\pm \sqrt{2}\right)
r^{-n},\quad n=1,2,\ldots ; 
\]%
the parameter%
\[
\mu \neq \mu _{n},\quad n=1,2,\ldots , 
\]%
where $\mu _{n}$ are the characteristic numbers of the homogenous equation%
\[
\psi \left( x,y\right) =\mu \left[ \int\limits_{0}^{1}N\left( x,y,\xi
\right) \psi \left( \xi ,y\right) d\xi +\int\limits_{0}^{1}M\left( x,y,\eta
\right) \psi \left( x,\eta \right) d\eta \right. 
\]%
\[
+\left. \int\limits_{0}^{1}d\xi \int\limits_{0}^{1}T\left( x,y,\xi ,\eta
\right) \psi \left( \xi ,\eta \right) d\eta \right] ,\quad x,y\in \left[ 0,1%
\right] . 
\]

After the choice of the values of $\mu $, $\lambda $ and $r$ has been made
(note that they can be corrected afterwards, depending on different
situayions), the sequence of the computational procedures is as follows:

- determination of the kernels of Eq. (\ref{5.38}), $N$, $M$ and $T$, from (%
\ref{5.41}), with the use of expression (\ref{5.42});

- determination of the free term $F_{1}$ from (\ref{5.40});

- determination of the function $\psi _{1}$ from (\ref{5.38}), with $F\equiv
F_{1}$;

- determination of the function $\rho $ from (\ref{5.44}), with the use of
expression (\ref{5.45});

- determination of the function $\kappa $ from (\ref{5.43}), with the use of
expression (\ref{5.46});

- determination of the free term $F_{0}$ from (\ref{5.39});

- determination of the function $\psi _{0}$ from (\ref{5.38}), with $F\equiv
F_{0}$;

- determination of the sought function $\psi $ from (\ref{5.37}).

As can be seen, the algorithm of section 6.4 applies to the solution of the
two-dimensional Fredholm integral equation of the first kind without any
substantial changes. In this case, the variable $y$ plays the role of a
parameter.

\section{A verification of the solvability of boundary-value problems}

In the above consideration, we have assumed that the function $\psi \left(
x,y\right) $ satisfying the Fredholm integral equation of the firs kind (\ref%
{5.36}) in the space $L_{2}$ exists and is unique. Nonetheless, by formal
use of the computational relations of section 7.4, one can "find" $\psi
\left( x,y\right) $ also in those cases when Eq. (\ref{5.36}) has no
solution at all or has a variety of solutions. In the first case, the
function $\psi \left( x,y\right) $ being substituted into (\ref{5.36})
cannot satisfy this equation.

Indeed, the function thus obtained is senseless because the construction of
the algorithm (see section 6.3) was based on the assumption that the
function satisfying Eq. (\ref{4.1}) existed, and what is more, the free term 
$f\left( x\right) $ was interpreted as a result of previously performed
integration. However, on the other hand, if the function $\psi \left(
x,y\right) $ found by means of the algorithm of section 7.4 does not satisfy
Eq. (\ref{5.36}) upon substitution, it implies that this equation is
insolvable.

So, what have we got in the end? The unpleasant properties of the Fredholm
integral equations of the first kind, expounded on above, can be rather
efficiently employed to verify the solvability of boundary-value
(initiall-value) problems. Indeed, they are easily reduced to
two-dimensional (or of higher dimension) Fredholm integral equations of the
first kind, which was discussed in section 7.2. Consequently, after the
realization of the algorithm of section 7.4, we are left only with a
necessity to verify whether the obtained solution satisfies an equation of
the type (\ref{5.36}).

For comparatively simple problems of the previous sections, such a
verification is not very important; however, a lot of investigations are
concerned with the adequacy of the problem (\ref{5.33}), (\ref{5.34}) with
regard to the description of the bending $u\left( x,y\right) $ at the
corners of a rectangular plate. Of interest is another issue: one of the
most important problems of numerical simulations is, as a matter of fact, a
formulation of problems that implies construction of differential or
integro-differential equations. In this regard, the Fredholm integral
equation of the first kind (after a reduction to it of a certain posed
problem) may serve as a filter discarding invalid versions!

This short subsection seems to be important. Its brevity results from the
fact that it is based on the material given above.

\chapter{Other classes of problems}

\section{The initial-boundary-value problem for the Korteweg-de Vries
equation}

Let us assume that the problem 
\begin{equation}
\partial _{t}u-6u\partial _{x}u+\partial _{x}^{3}u=0,  \label{6.1}
\end{equation}%
\begin{equation}
u\left( x,0\right) =u_{0}\left( x\right) ;\quad u\left( 0,t\right)
=u_{1}\left( t\right) ;\quad \partial _{x}u\left( 0,t\right) =u_{2}\left(
t\right) ;\quad u\left( 1,t\right) =u_{3}\left( t\right) ,  \label{6.2}
\end{equation}%
has a unique solution in the space $L_{2}$ for given functions $u_{0}\left(
x\right) $; $u_{i}\left( t\right) $, $i=1,2,3$.

There is no general theory that would allow us to make a priori judgements
about the solvability of the problems of this type. Results of numerical
simulations as well as solutions of specially simplified equations near the
boundary (see \cite{6.1}, section 10) may prove to be the main tool of
refinement on physical models.\footnote{%
In this regard, the arguments of section 7.5 \ may prove to be useful.
\par
{}}

Using the procedure of the previous section, we can reduce the problem (\ref%
{6.1}), (\ref{6.2}) to an integral equation of the first kind with respect
to 
\[
\psi \left( x,t\right) =\partial _{x}^{3}u\left( x,t\right) , 
\]%
which yields 
\begin{equation}
u\left( x,t\right) =\frac{1}{2}\int\limits_{0}^{x}\left( x-\xi \right)
^{2}\psi \left( \xi ,t\right) d\xi +\frac{1}{2}x^{2}g_{3}\left( t\right)
+xg_{2}\left( t\right) +g_{1}\left( t\right) ,  \label{6.3}
\end{equation}%
with the functions determined from the boundary conditions: 
\[
g_{1}\left( t\right) =u_{1}\left( t\right) ;\quad g_{2}\left( t\right)
=u_{2}\left( t\right) ; 
\]%
\[
g_{3}\left( t\right) =2\left[ u_{3}\left( t\right) -u_{2}\left( t\right)
-u_{1}\left( t\right) \right] -\int\limits_{0}^{1}\left( 1-\xi \right)
^{2}\psi \left( \xi ,t\right) d\xi . 
\]

Substitution into (\ref{6.3}) leads to the expression 
\[
u\left( x,t\right) =\frac 12\left[ \int\limits_0^x\left( x-\xi \right)
^2-x^2\int\limits_0^1\left( 1-\xi \right) ^2\right] \psi \left( \xi
,t\right) d\xi 
\]
\begin{equation}  \label{6.4}
+x^2u_3\left( t\right) +x\left( 1-x\right) u_2\left( t\right) +\left(
1-x^2\right) u_1\left( t\right) .
\end{equation}

Now we rewrite Eq. (\ref{6.1}) in the form 
\begin{equation}  \label{6.5}
\partial _tu=6u\partial _xu-\partial _x^3u.
\end{equation}

The substitution of (\ref{6.4}) into the right-hand side of (\ref{6.5}) and
integration from $0$ to $t\,$ under the initial condition (\ref{6.2}) allows
us to determine 
\begin{equation}
u\left( x,t\right) =6\int\limits_{0}^{t}\left[ 6u\left( x,\eta \right)
\partial _{x}u\left( x,\eta \right) -\partial _{x}^{3}u\left( x,\eta \right) %
\right] d\eta +u_{0}.  \label{6.6}
\end{equation}

The elimination of $u\left( x,t\right) $ from (\ref{6.4}), (\ref{6.6}) leads
to an equation of the form 
\begin{equation}
\left( A\psi \right) \left( x,t\right) =f\left( x,t\right) ,\quad x,t\in
\Omega :0\leq x,t\leq 1,  \label{6.7}
\end{equation}%
where $A$ is a nonlinear integral operator, and the function $f$ depends on
the data of the problem.

In order to determine the function $\psi \left( x,t\right) $, we can employ
the algorithm of section 7.4 (the variable $y$ is replaced by $t$). This
function will satisfy a nonlinear integral equation of the second kind. By
the contraction mapping theorem, for small absolute values of the parameter $%
\mu $, its solution can be found by means of simple iterations \cite{6.2}.

\section{A boundary-value problem for a substantially nonlinear differential
equation}

Here, we discuss nonlinearity related to higher-order derivatives. As an
example, consider Monge-Amp\`{e}re's equation : 
\begin{equation}
\partial _{x}^{2}u\partial _{y}^{2}u-\left( \partial _{xy}u\right)
^{2}=s_{1}\partial _{x}^{2}u+s_{2}\partial _{y}^{2}u+s_{3}\partial _{xy}u+q,
\label{6.8}
\end{equation}%
where $s_{i}$, $i=1,2,3\,$ and $q$, in general, depend on the variables $x$, 
$y$, the sought function $u\left( x,y\right) $ and its first derivatives $%
\partial _{x}u$, $\partial _{y}u$ \cite{6.3}.

Let us assume that $s_{i}=s_{i}\left( x,y\right) $, $q=q\left( x,y\right) $
and 
\begin{equation}
u\left( 0,y\right) =u\left( 1,y\right) =u\left( x,0\right) =u\left(
x,1\right) =0.  \label{6.9}
\end{equation}%
We also assume that the solution of the problem in $L_{2}$ exists and is
uniqe. Using the notation%
\[
\partial _{x}^{2}u\left( x,y\right) =\psi _{1}\left( x,y\right) ;\quad
\partial _{y}^{2}u\left( x,y\right) =\psi _{2}\left( x,y\right) ; 
\]%
\[
\partial _{xy}u\left( x,y\right) =\psi \left( x,y\right) ,\quad x,y\in
\Omega :0\leq x,y\leq 1, 
\]%
taking into account (\ref{6.9}), we get 
\[
u\left( x,y\right) =\left[ \int\limits_{0}^{x}\left( x-\xi \right)
-x\int\limits_{0}^{1}\left( 1-\xi \right) \right] \psi _{1}\left( \xi
,y\right) d\xi ; 
\]%
\[
u\left( x,y\right) =\left[ \int\limits_{0}^{y}\left( y-\eta \right)
-y\int\limits_{0}^{1}\left( 1-\eta \right) \right] \psi _{2}\left( x,\eta
\right) d\eta ; 
\]%
\[
u\left( x,y\right) =\int\limits_{0}^{x}d\xi \int\limits_{0}^{y}\psi \left(
\xi ,\eta \right) d\eta . 
\]

Upon the substitution of these expressions into (\ref{6.8}) and the
elimination of the function $u$, we reduce the problem to the following
system of equations: 
\[
\psi _{1}\left( x,y\right) \psi _{2}\left( x,y\right) -\psi ^{2}\left(
x,y\right) =s_{1}\left( x,y\right) \psi _{1}\left( x,y\right) 
\]%
\begin{equation}
+s_{2}\left( x,y\right) \psi _{2}\left( x,y\right) +s_{3}\left( x,y\right)
\psi \left( x,y\right) +q\left( x,y\right) ;  \label{6.10}
\end{equation}%
\begin{equation}
\left[ \int\limits_{0}^{x}\left( x-\xi \right) -x\int\limits_{0}^{1}\left(
1-\xi \right) \right] \psi _{1}\left( \xi ,y\right) d\xi
-\int\limits_{0}^{x}d\xi \int\limits_{0}^{y}\psi \left( \xi ,\eta \right)
d\eta =0;  \label{6.11}
\end{equation}%
\begin{equation}
\left[ \int\limits_{0}^{y}\left( y-\eta \right) -y\int\limits_{0}^{1}\left(
1-\eta \right) \right] \psi _{2}\left( x,\eta \right) d\eta
-\int\limits_{0}^{x}d\xi \int\limits_{0}^{y}\psi \left( \xi ,\eta \right)
d\eta =0,\quad x,y\in \Omega .  \label{6.12}
\end{equation}

Two-fold differentiation of Eqs. (\ref{6.11}), (\ref{6.12}) with respect to $%
x$ and $y$ yields, respectively,%
\[
\psi _{1}\left( x,y\right) =\int\limits_{0}^{y}\partial _{x}\psi \left(
x,\eta \right) d\eta ;\quad \psi _{2}\left( x,y\right)
=\int\limits_{0}^{x}\partial _{y}\psi \left( \xi ,y\right) d\xi . 
\]

Equation (\ref{6.10}) takes the form 
\[
\left[ \int\limits_{0}^{y}\partial _{x}\psi \left( x,\eta \right) d\eta %
\right] \left[ \int\limits_{0}^{x}\partial _{y}\psi \left( \xi ,y\right)
d\xi \right] -\psi ^{2}\left( x,y\right) =s_{1}\left( x,y\right)
\int\limits_{0}^{y}\partial _{x}\psi \left( x,\eta \right) d\eta 
\]%
\[
+s_{2}\left( x,y\right) \int\limits_{0}^{x}\partial _{y}\psi \left( \xi
,y\right) d\xi +s_{3}\left( x,y\right) \psi \left( x,y\right) +q\left(
x,y\right) ,\quad x,y\in \Omega , 
\]%
and after integration in the limits $0,x$ and $0,y$ reduces to the
following: 
\begin{equation}
\left( A\psi \right) \left( x,y\right) =f\left( x,y\right) ,\quad x,y\in
\Omega ,  \label{6.13}
\end{equation}%
where $A$ is a corresponding nonlinear operator; 
\[
f\left( x,y\right) =\int\limits_{0}^{x}d\xi \int\limits_{0}^{y}q\left( \xi
,\eta \right) d\eta . 
\]

The above implies the boundedness of the derivatives $\partial _{x}s_{1}$, $%
\partial _{y}s_{2}$. A possible way of the solution of this equation is
discussed in section 8.1.

\section{Nonlinearity of the boundary condition}

Consider a typical problem of the irradiation of an infinite plate with a
thermally insulated surface into a medium whose absolute temperature is
equal to zero \cite{6.4}:

\begin{equation}
\partial _{t}u-a\partial _{x}^{2}u=0,  \label{6.14}
\end{equation}%
\begin{equation}
u\left( x,0\right) =u_{0}\left( x\right) ;\quad p\partial _{x}u\left(
0,t\right) +u^{m}\left( 0,t\right) =0;\quad \partial _{x}u\left( 1,t\right)
=0.  \label{6.15}
\end{equation}

Here, $u\left( x,t\right) $ is the temperature gradient; $u_{0}\left(
x\right) $ is a given function; $a$ is the temperature conductivity; $%
p=\lambda /\alpha $, with $\lambda $, $\alpha $ being the
thermal-conductivity and the heat-transfer coefficients, respectively; $m$
is a parameter.

Introduce the notation 
\begin{equation}  \label{6.16}
\partial _x^2u\left( x,y\right) =\psi \left( x,t\right) ,
\end{equation}
which leads to 
\[
u\left( x,t\right) =\int\limits_0^x\left( x-\xi \right) \psi \left( \xi
,t\right) d\xi +xg_1\left( t\right) +g_2\left( t\right) , 
\]
where $g_i\left( t\right) $ are functions of integration.

The boundary conditions (\ref{6.15}) yield 
\[
g_{1}\left( t\right) =-\int\limits_{0}^{1}\psi \left( \xi ,t\right) d\xi
;\quad pg_{1}\left( t\right) +g_{2}^{m}\left( t\right) =0, 
\]%
and, accordingly, 
\[
u\left( x,t\right) =\left[ \int\limits_{0}^{x}\left( x-\xi \right)
-x\int\limits_{0}^{1}\right] \psi \left( \xi ,t\right) d\xi +\left[
p\int\limits_{0}^{1}\psi \left( \xi ,t\right) d\xi \right] ^{\frac{1}{m}}. 
\]

Using (\ref{6.14}), (\ref{6.16}) and taking into account the initial
condition (\ref{6.15}), we get 
\[
u\left( x,t\right) =a\int\limits_{0}^{t}\psi \left( x,\eta \right) d\eta
+u_{0}\left( x\right) , 
\]%
and the problem is reduced to the solution of the nonlinear integral
equation of the first kind (\ref{6.7}), where 
\[
A_{\bullet }=\left[ \int\limits_{0}^{x}\left( x-\xi \right)
-x\int\limits_{0}^{1}\right] _{\bullet }d\xi +\left(
p\int\limits_{0}^{1}\left. {}\right. _{\bullet }d\xi \right) ^{\frac{1}{m}%
}-a\int\limits_{0}^{t}\left. {}\right. _{\bullet }d\eta ; 
\]%
\[
f\left( x,t\right) =u_{0}\left( x\right) . 
\]

\section{A small parameter by the highest-order derivative of the
differential equation of the problem}

As an illustration of general considerations, we consider the problem of
heat transport induced by the processes of thermal conduction and convection
(the first and the second terms of the equation, respectively) \cite{6.5}:

\begin{equation}  \label{6.17}
\partial _tu=\epsilon \partial _x^2u+\beta \partial _xu.
\end{equation}

Here, $\beta >0$ is a constant; $\epsilon \,$ is a small parameter, 
\begin{equation}
u\left( x,0\right) =0;\quad u\left( 1,t\right) =0;\quad u\left( 0,t\right)
=u_{1}\left( t\right) ,  \label{6.18}
\end{equation}%
with $u_{1}\left( t\right) $ being a given $L_{2}$-function.

The notation (\ref{6.16}) under the boundary conditions (\ref{6.18}) leads
to 
\begin{equation}
u\left( x,t\right) =\left[ \int\limits_{0}^{x}\left( x-\xi \right)
-x\int\limits_{0}^{1}\left( 1-\xi \right) \right] \psi \left( \xi ,t\right)
d\xi .  \label{6.19}
\end{equation}

The integration of (\ref{6.17}) in the limits $0,t$ with the use of (\ref%
{6.19}) and of the initial condition (\ref{6.18}) yields 
\[
u\left( x,t\right) =\int\limits_{0}^{t}\left[ \epsilon \psi \left( x,\eta
\right) +\beta \int\limits_{0}^{x}\psi \left( \xi ,\eta \right) d\xi \right]
d\eta +u_{1}\left( t\right) . 
\]

The elimination of the function $u$ from these relations leads to Eq. (\ref%
{6.7}), where 
\[
A=A^{\prime }+A^{\prime \prime }, 
\]%
\[
A_{\bullet }^{\prime }=\epsilon \int\limits_{0}^{t}\left. {}\right.
_{\bullet }d\eta ;\quad A_{\bullet }^{\prime \prime }=\beta
\int\limits_{0}^{x}d\xi \int\limits_{0}^{t}\left. {}\right. _{\bullet }d\eta
-\left[ \int\limits_{0}^{x}\left( x-\xi \right) -x\int\limits_{0}^{1}\left(
1-\xi \right) \right] _{\bullet }d\xi ; 
\]%
\[
f\left( x,t\right) =-u_{1}\left( t\right) . 
\]

The algorithm of section 7.4 allows us to reduce the problem to a Fredholm
integral equation of the second kind of the form 
\begin{equation}
s\left( x,t\right) =\mu \left[ \left( \epsilon R_{1}+R_{2}\right) s\right]
\left( x,t\right) +f\left( x,t\right) ,  \label{6.20}
\end{equation}%
where $R_{1}$ and $R_{2}$ are corresponding integral operators; $s\left(
x,t\right) \equiv \psi \left( x,t\right) $.

As a result of the expansion \cite{6.6} 
\[
s\left( x,t\right) =\sum_{m=0}^{\infty }\epsilon ^{m}s_{m}\left( x,t\right)
, 
\]%
we get a sequence of recursion relations 
\[
s_{0}\left( x,t\right) =\mu \left( R_{2}s_{0}\right) \left( x,t\right)
+f\left( x,t\right) ; 
\]%
\[
s_{1}\left( x,t\right) =\mu \left( R_{2}s_{1}\right) \left( x,t\right) +\mu
\left( R_{1}s_{0}\right) \left( x,t\right) ; 
\]%
\[
\ldots 
\]%
\[
s_{m+1}\left( x,t\right) =\mu \left( R_{2}s_{m+1}\right) \left( x,t\right)
+\mu \left( R_{1}s_{m}\right) \left( x,t\right) 
\]%
that are canonical Fredholm integral equations of the second kind.

It follows from the above that the proposed approach is rather efficient in
the problems of mathematical physics with a singular perturbation, whose
numerical realization, as a rule, meets with considerable difficulties (see,
in particular, \cite{6.7}). Indeed, we managed to transform the singular
perturbation (\ref{6.17}) into the regular one (\ref{6.20}), which
facilitated a radical simplification of the problem.\footnote{%
Singular and regular perturbations affect,respectively, main and dependent
terms of the operators.}

\section{Equations of a mixed type}

Boundary-value problems for equations of this type are characterized by
complexity of the investigation into the issues of existence and uniqueness
(see \cite{6.8}). As a consequence, one has to consider such equations on
rather special domains, which restricts the field of practical applications.

Leaving this issue be, only for the sake of an illustration of the procedure
of reduction, we turn to well-known Tricomi's equation 
\begin{equation}
y\partial _{x}^{2}u+\partial _{y}^{2}u=0  \label{6.21}
\end{equation}%
that belongs both to the hyperbolic and elliptical types for $y<0$ and $y>0$%
, respectively. As an for example, we employ the following boundary
conditions: 
\begin{equation}
u\left( 0,y\right) =u\left( 1,y\right) =u\left( x,-1\right) =0;\quad u\left(
x,1\right) =\nu \left( x\right) ,  \label{6.22}
\end{equation}%
where the function $\nu \left( x\right) $ is such that $\nu \left( 0\right)
=\nu \left( 1\right) =0$.

From the notation 
\begin{equation}
\partial _{x}^{2}u\left( x,y\right) =\psi \left( x,y\right) ,  \label{6.23}
\end{equation}%
by (\ref{6.22}), it follows: 
\[
u\left( x,y\right) =\left[ \int\limits_{0}^{x}\left( x-\xi \right)
-x\int\limits_{0}^{1}\left( 1-\xi \right) \right] \psi \left( \xi ,y\right)
d\xi . 
\]

Two-fold integration of Eq. (\ref{6.21}) in the limits $-1$, $y$ under the
conditions (\ref{6.23}) and (\ref{6.22}) yields the expression 
\[
u\left( x,y\right) =-\left[ \int\limits_{-1}^{y}\left( y-\eta \right) -\frac{%
1+y}{2}\int\limits_{-1}^{1}\left( 1-\eta \right) \right] \eta \psi \left(
x,\eta \right) d\eta +\frac{1}{2}\left( 1+y\right) \nu \left( x\right) . 
\]

The problem reducea to a Fredholm integral equation of the second kind (\ref%
{6.13}) on the domain $\Omega :0\leq x\leq 1,-1\leq y\leq 1$ with the
operator 
\[
A_{\bullet }=\left[ \int\limits_{0}^{x}\left( x-\xi \right)
-x\int\limits_{0}^{1}\left( 1-\xi \right) \right] _{\bullet }d\xi +\left[
\int\limits_{-1}^{y}\left( y-\eta \right) -\frac{1+y}{2}\int\limits_{-1}^{1}%
\left( 1-\eta \right) \right] \eta _{\bullet }d\eta 
\]%
and the free term 
\[
f\left( x,y\right) =\frac{1}{2}\left( 1+y\right) \nu \left( x\right) . 
\]

Note that the so-called condition of "matching" on the line of parabolic
degeneracy $y=0$, imposed on the solution of Eq. (\ref{6.21}) (\cite{6.8},
p. 27), is fulfilled in a natural way: 
\[
\lim_{y\rightarrow +0}u\left( x,y\right) =\lim_{y\rightarrow -0}u\left(
x,y\right) ,\quad x\in \left[ 0,1\right] ; 
\]%
\[
\lim_{y\rightarrow +0}\partial _{y}u\left( x,y\right) =\lim_{y\rightarrow
-0}\partial _{y}u\left( x,y\right) ,\quad x\in \left[ 0,1\right] . 
\]

As in the previous subsection, this situation results from the fact that the
singularity of the problem is transferred from the main term of the relevant
operator to the dependent one.

\section{The inverse problem of the restoration of the coefficient of the
differential equation}

Small oscillations in the transverse direction of a stretched string of
variable density are described by the equation 
\begin{equation}  \label{6.24}
\partial _t^2u=a\left( x\right) \partial _x^2u.
\end{equation}

Here, $x$, $t$ are dimensionless coordinates; 
\[
a\left( x\right) =NT^2/\rho \left( x\right) l^2, 
\]
with $N$ being the tension, $\rho \left( x\right) $ the density of the
material, $2l$ the length of the string, $T$ the time interval.

We assume that the ends of the string are fixed, whereas its density and the
oscillations are symmetric with respect to the coordinate $x=0$. The
corresponding boundary conditions have the form 
\begin{equation}  \label{6.25}
\partial _xu\left( 0,t\right) =u\left( 1,t\right) =0.
\end{equation}

We also employ the following initial conditions: 
\begin{equation}  \label{6.26}
u\left( x,0\right) =u_0\left( x\right) ;\quad \partial _tu\left( x,0\right)
=0.
\end{equation}

The coefficient $a\left( x\right) $ is to be determined from (\ref{6.24})-(%
\ref{6.26}) for given $u_{0}\left( x\right) $, $N$, $l$, $T$ and additional
information on the oscillations of the middle cross-section of the string:

\begin{equation}
u\left( 0,t\right) =\nu \left( t\right) .  \label{6.27}
\end{equation}%
We assume that the conditions ensuring the existence and uniqueness of the
solution of the considered problem (\cite{6.9}, section 4) are fulfilled.

By analogy with what was done many times before, using the notation (\ref%
{6.16}) and (\ref{6.24})-(\ref{6.26}), we find 
\begin{equation}
u\left( x,y\right) =\left[ \int\limits_{0}^{x}\left( x-\xi \right)
-\int\limits_{0}^{1}\left( 1-\xi \right) \right] \psi \left( \xi ,y\right)
d\xi ;  \label{6.28}
\end{equation}%
\[
u\left( x,y\right) =a\left( x\right) \int\limits_{0}^{t}\left( t-\eta
\right) \psi \left( x,\eta \right) d\eta +u_{0}\left( x\right) . 
\]

By eliminating $u\left( x,t\right) $, we obtain an equation of the type (\ref%
{6.7}). The substitution of (\ref{6.28}) into (\ref{6.27}) leads to the
integral equation 
\[
\left( A^{\prime }\psi \right) \left( t\right) =f^{\prime }\left( t\right)
,\quad t\in \left[ 0,1\right] , 
\]%
where 
\[
A_{\bullet }^{\prime }=\int\limits_{0}^{1}\left( 1-\xi \right) _{\bullet
}d\xi ;\quad f^{\prime }\left( t\right) =-\nu \left( t\right) . 
\]

The procedure of the so posed system of equations can be viewed in the
context of supplementing the algorithm of section 7.4 by iterations with the
function $a$.

\section{The problem of the Stefan type}

Consider the classical model \cite{6.10}: 
\begin{equation}
\partial _{t}u=\partial _{x}^{2}u,\quad 0<x<\gamma \left( t\right) ;\quad
0<t\leq 1,  \label{6.29}
\end{equation}%
\begin{equation}
u\left( x,0\right) =u_{0}\left( x\right) ;\quad u\left( 0,t\right) =u\left(
\gamma \left( t\right) ,t\right) =0,\quad u_{0}\left( 0\right) =0.
\label{6.30}
\end{equation}

On the moving boundary that separates the phases an additional condition is
imposed: 
\begin{equation}  \label{6.31}
\alpha \partial _xu\left( \gamma \left( t\right) ,t\right) =\gamma ^{\prime
}\left( t\right) ,\quad \gamma \left( 0\right) =\gamma _0,
\end{equation}
where $\gamma _0>0$; the constant $\alpha $ can be both positive and
negative; $\gamma ^{\prime }\left( t\right) =d\gamma \left( t\right) /dt$.

Thus, the data of the problem are $u_0\left( x\right) $, $\alpha $ and $%
\gamma _0$; the functions $u\left( x,t\right) $ and $\gamma \left( t\right) $
are to be determined.

In Eqs. (\ref{6.29})-(\ref{6.31}), we make a non-orthogonal mapping 
\begin{equation}
\bar{x}=x/\gamma \left( t\right) ,\quad \bar{t}=t  \label{6.32}
\end{equation}%
on a canonical domain $\Omega :0\leq \bar{x},\bar{t}\leq 1$. We get: 
\begin{equation}
\partial _{\bar{t}}u-\left[ \bar{x}\gamma ^{\prime }\left( \bar{t}\right)
/\gamma \left( \bar{t}\right) \right] \partial _{\bar{x}}u=\partial _{\bar{x}%
}^{2}u,  \label{6.33}
\end{equation}%
\begin{equation}
u\left( \bar{x},0\right) =0;\quad u\left( 0,\bar{t}\right) =u\left( 1,\bar{t}%
\right) =0;  \label{6.34}
\end{equation}%
\begin{equation}
\alpha \partial _{\bar{x}}u\left( 1,\bar{t}\right) =\gamma ^{\prime }\left( 
\bar{t}\right) ,\quad \gamma \left( 0\right) =\gamma _{0}.  \label{6.35}
\end{equation}

By analogy with the above, the notation 
\[
\partial _{\bar{x}}^{2}u\left( \bar{x},\bar{t}\right) =\psi \left( \bar{x},%
\bar{t}\right) , 
\]%
conditions (\ref{6.34}) and equation (\ref{6.33}) lead to 
\begin{equation}
u\left( \bar{x},\bar{t}\right) =\left[ \int\limits_{0}^{\bar{x}}\left( \bar{x%
}-\xi \right) -\bar{x}\int\limits_{0}^{1}\left( 1-\xi \right) \right] \psi
\left( \xi ,\bar{t}\right) d\xi ;  \label{6.36}
\end{equation}%
\[
u\left( \bar{x},\bar{t}\right) =\int\limits_{0}^{t}\left\{ \psi \left(
x,\eta \right) +\left[ \bar{x}\gamma ^{\prime }\left( \bar{t}\right) /\gamma
\left( \bar{t}\right) \right] \left[ \int\limits_{0}^{x}-\int\limits_{0}^{1}%
\left( 1-\xi \right) \right] \psi \left( \xi ,\eta \right) d\xi \right\}
d\eta 
\]%
\begin{equation}
+u_{0}\left( x\right) .  \label{6.37}
\end{equation}

The substitution of (\ref{6.36}) into (\ref{6.35}) yields 
\[
\gamma ^{\prime }\left( \bar{t}\right) =\alpha \int\limits_{0}^{1}\xi \psi
\left( \xi ,\bar{t}\right) d\xi , 
\]%
from which we get 
\begin{equation}
\gamma \left( \bar{t}\right) =\alpha \int\limits_{0}^{1}\xi d\xi
\int\limits_{0}^{\bar{t}}\psi \left( \xi ,\eta \right) d\eta +\gamma _{0}.
\label{6.38}
\end{equation}%
Accordingly, in the expression (\ref{6.37}), we have 
\[
\frac{\bar{x}\gamma ^{\prime }\left( \bar{t}\right) }{\gamma \left( \bar{t}%
\right) }=\alpha \bar{x}\int\limits_{0}^{1}\xi \psi \left( \xi ,\bar{t}%
\right) d\xi /\left[ \alpha \int\limits_{0}^{1}\xi d\xi \int\limits_{0}^{%
\bar{t}}\psi \left( \xi ,\eta \right) d\eta +\gamma _{0}\right] . 
\]

The elimination of $u\left( \bar{x},\bar{t}\right) $ from (\ref{6.36}), (\ref%
{6.37}) leads to the integral equation of the first kind (\ref{6.7}). The
function $\bar{\psi}$ determined from this equation should be approximated
by an analytical dependence on $\bar{x}$ in order to make an inverse change
of variables. The sought separation boundary $\gamma \left( \bar{t}\right) $
is determined from the nonlinear integral equation (\ref{6.38}). Then, by (%
\ref{6.36}), using (\ref{6.32}), we can calculate the function $u\left(
x,t\right) $.

\chapter{Conclusions}

Let us summarize the main points of the above consideration. Thus, the
solution of the Fredholm integral equation of the first kind 
\begin{equation}
\left( A\psi \right) \left( x\right) \equiv \int\limits_{0}^{1}k\left( x,\xi
\right) \psi \left( \xi \right) d\xi =f\left( x\right) ,\quad x\in \left[ 0,1%
\right]  \label{1}
\end{equation}%
in the \textquotedblright convenient\textquotedblright\ for the numerical
realization space $L_{2}$ is an ill-posed problem. In the case of the space $%
l_{2}^{\prime }$ that is adequate to the range of the operator $A$, the
situation is different: the data of Eq. (\ref{1}) may, theoretically,
satisfy the conditions of its correctness, but, nevertheless, the solution
will constitute a series that diverges as a result of the accumulation of
errors of calculations.

It should be noted that even a verification of whether the data of (\ref{1})
belongs to the space $l_{2}^{\prime }$ is, in general, infeasible. At the
same time, as an objective factor of incorrectness, there appear the error
of experimental determination of $f\left( x\right) $ and, sometimes,
inaccurate information about the function $k\left( x,\xi \right) $ that
characterizes the system under consideration.

The basis of our work is formed by the suggestion to connect adaptively Eq. (%
\ref{1}) to the space $l_{2}^{\prime }$ by means of a modelling of the error 
$\left( \delta f\right) \left( x\right) $ that arises owing to the smoothing
of information by the procedure of integration. We assume that the function
satisfying this equation exists, is unique, and the condition%
\begin{equation}
\left( \delta f\right) \left( x\right) =0,\quad x\in \left[ 0,1\right] ,
\label{2}
\end{equation}%
reflecting objective smallness of this error compared to $\psi \left(
x\right) $ and the data of the problem, is employed.

Starting by heuristic considerations that were later supported by more firm
arguments, we demonstrated the expediency of the representation of the error
as a difference between the sought function in an explicit form and of the
integral component: 
\begin{equation}
\left( \delta f\right) \left( x\right) =\psi \left( x\right) -\lambda \left(
B\psi \right) \left( x\right) ,\quad x\in \left[ 0,1\right] ,  \label{3}
\end{equation}%
where%
\[
B_{\bullet }=\int\limits_{-1}^{1}h\left( x,\xi \right) _{\bullet }d\xi
,\quad \Psi \left( x\right) =\left\{ 
\begin{array}{c}
\psi \left( x\right) ,\quad x\in \left[ 0,1\right] ;~ \\ 
\varphi \left( x\right) ,\quad x\in \left[ -1,0\right) .%
\end{array}%
\right. 
\]

Of exclusive importance for the whole complex of the transformations,
especially in the first version of their realization, was the representation%
\begin{equation}
h\left( x,\xi \right) =1+2\sum_{n=1}^{\infty }r^{n}\cos \left[ 2n\pi \left(
x-\xi \right) \right] ,\quad 0<r<1,  \label{4}
\end{equation}%
i.e., in the form of Poisson's kernel, that allowed us to satisfy the
condition (\ref{2}) for the case when the function $\psi \left( x\right) $
was harmonic.

By the use of (\ref{2}), (\ref{3}), the formulation of the problem (\ref{1})
was transformed:%
\begin{equation}
\Psi \left( x\right) =\lambda \left( B\Psi \right) \left( x\right) +\left\{ 
\begin{array}{c}
0,\quad x\in \left[ 0,1\right] ;~ \\ 
\kappa \left( x\right) ,\quad x\in \left[ -1,0\right) ;%
\end{array}%
\right.  \label{5}
\end{equation}%
\begin{equation}
\Psi ^{\prime }\left( x\right) =\lambda \left( B\Psi ^{\prime }\right)
\left( x\right) +\left\{ 
\begin{array}{c}
\chi \left( x\right) ,\quad x\in \left[ 0,1\right] ;~ \\ 
0,\quad x\in \left[ -1,0\right) ;%
\end{array}%
\right.  \label{6}
\end{equation}%
where%
\[
\Psi ^{\prime }\left( x\right) =\left\{ 
\begin{array}{c}
\psi \left( x\right) ,\quad x\in \left[ 0,1\right] ;~ \\ 
\varphi ^{\prime }\left( x\right) ,\quad x\in \left[ -1,0\right) ,%
\end{array}%
\right. 
\]%
\begin{equation}
\Psi \left( x\right) =\lambda \left( B\Psi \right) \left( x\right) +\left\{ 
\begin{array}{c}
\mu \left( A\psi \right) \left( x\right) -\mu f\left( x\right) ,\quad x\in 
\left[ 0,1\right] ;~ \\ 
\kappa \left( x\right) ,\quad x\in \left[ -1,0\right) ;%
\end{array}%
\right.  \label{7}
\end{equation}%
\begin{equation}
\Psi ^{\prime }\left( x\right) =\lambda \left( B\Psi ^{\prime }\right)
\left( x\right) +\left\{ 
\begin{array}{c}
\mu \left( A\psi \right) \left( x\right) -\mu f\left( x\right) +\chi \left(
x\right) ,\quad x\in \left[ 0,1\right] ;~ \\ 
0,\quad x\in \left[ -1,0\right) .%
\end{array}%
\right.  \label{8}
\end{equation}

From the point of view of constructiveness of what followed on the basis of (%
\ref{2})-(\ref{4}), a key role was played by the following factors:

- an extension of Eq. (\ref{3}) under the condition (\ref{2}) to $x\in \left[
-1,0\right) $. As a result, there arose the Fredholm integral equation of
the second kind (\ref{5}) with an undefined function $\kappa \left( x\right) 
$, a component of its free term;

- the use of the equation of analogous structure, Eq. (\ref{6}), whose the
free term goes to zero on the second part of the interval of definition;

- an incompletely continuous perturbation of the operator $A$ by consecutive
addition of (\ref{5}) and (\ref{6}) to (\ref{1}). As a result, Eqs. (\ref{7}%
) and (\ref{8}) arose.

The elimination of the function $\chi \left( x\right) $ from (\ref{8}) using
the equation%
\begin{equation}
\Psi _{0}^{\prime }\left( x\right) =\lambda \left( B\Psi _{0}^{\prime
}\right) \left( x\right) +\left\{ 
\begin{array}{c}
\mu \left( A\psi _{0}\right) \left( x\right) +\chi \left( x\right) ,\quad
x\in \left[ 0,1\right] ;~ \\ 
0,\quad x\in \left[ -1,0\right) ,%
\end{array}%
\right.  \label{9}
\end{equation}%
where%
\[
\Psi _{0}^{\prime }\left( x\right) =\left\{ 
\begin{array}{c}
\psi _{0}\left( x\right) ,\quad x\in \left[ 0,1\right] ;~ \\ 
\varphi _{0}^{\prime }\left( x\right) ,\quad x\in \left[ -1,0\right) ,%
\end{array}%
\right. 
\]%
allowed us to obtain%
\begin{equation}
\Psi _{1}^{\prime }\left( x\right) =\lambda \left( B\Psi _{1}^{\prime
}\right) \left( x\right) +\left\{ 
\begin{array}{c}
\mu \left( A\psi _{1}\right) \left( x\right) -\mu f\left( x\right) ,\quad
x\in \left[ 0,1\right] ;~ \\ 
0,\quad x\in \left[ -1,0\right) .%
\end{array}%
\right.  \label{10}
\end{equation}%
Here, 
\begin{equation}
\Psi _{1}^{\prime }\left( x\right) =\left\{ 
\begin{array}{c}
\psi _{1}\left( x\right) =\psi \left( x\right) -\psi _{0}\left( x\right)
,\quad x\in \left[ 0,1\right] ;~ \\ 
\varphi _{1}^{\prime }\left( x\right) =\varphi ^{\prime }\left( x\right)
-\varphi _{0}^{\prime }\left( x\right) ,\quad x\in \left[ -1,0\right) .%
\end{array}%
\right.  \label{11}
\end{equation}

The inversion of the operator $I-\lambda B$ in (\ref{10}) leads to the
Fredholm integral equation of the second kind%
\begin{equation}
\psi _{1}\left( x\right) =\mu \int\limits_{0}^{1}K\left( x,\xi \right) \psi
_{1}\left( \xi \right) d\xi +f_{1}\left( x\right) ,\quad x\in \left[ 0,1%
\right] ,  \label{12}
\end{equation}%
where the kernel and the free term are determined by the data of (\ref{1});
the parameter $\mu \,$, as in analogous cases $\lambda $, must satisfy a
solvability condition. After the determination of $\psi _{1}\left( x\right) $%
, the function $\varphi _{1}^{\prime }\left( x\right) $ is given by
quadratures.

From (\ref{5}) and (\ref{6}), it follows:%
\begin{equation}
\chi \left( x\right) =\lambda \int\limits_{-1}^{0}h\left( x,\xi \right) 
\left[ \varphi \left( \xi \right) -\varphi ^{\prime }\left( \xi \right) %
\right] d\xi .  \label{13}
\end{equation}

A further orientation of the transformations was concentrated on the
determination of the difference in the right-hand side of this equation in
order to turn it into a Fredholm integral equation of the second kind with
respect to $\chi \left( x\right) $. The fact that such a result could be
achieved was by no means obvious, because, although the function $\varphi
^{\prime }\left( x\right) $ was expressed in a simple way via $\chi \left(
x\right) $ [from Eq. (\ref{6})], such a possibility was absent for the
function $\varphi \left( x\right) $.

In the process of attaining the set objective, a stress was put on obtaining
a relation between the solutions of (\ref{7}) and (\ref{8}) that would allow
one of this equation to turn into another. The form of the free terms of
Eqs. (\ref{7}) and (\ref{8}) implies a possibility of such transformations,
that is, a possibility of a "flow" of their nonzero components from one part
of the interval of definition to the other, which opens up a prospect of the
representation of $\psi \left( x\right) $ in two ways, i.e., with and
without the function $\chi \left( x\right) $ in an explicit form.

The actual realization of the above arguments showed that the solutions of
Eqs. (\ref{5})-(\ref{8}) on $x\in \left[ -1,0\right) $ are mutually related
via the functions entering (\ref{9}):%
\begin{equation}
\varphi \left( x\right) =\varphi ^{\prime }\left( x\right) +\varphi
_{0}^{\prime }\left( x\right) ;  \label{14}
\end{equation}%
\begin{equation}
\kappa \left( x\right) =\lambda \int\limits_{0}^{1}h\left( x,\xi \right)
\psi _{0}\left( \xi \right) d\xi .  \label{15}
\end{equation}

From (\ref{11}), (\ref{14}),%
\begin{equation}
\varphi \left( x\right) -\varphi ^{\prime }\left( x\right) =\varphi
_{0}^{\prime }\left( x\right) =\varphi ^{\prime }\left( x\right) -\varphi
_{1}^{\prime }\left( x\right) ,  \label{16}
\end{equation}%
and, after substitution into (\ref{13}), the problem reduced to the solution
of the Fredholm integral equation of the second kind 
\begin{equation}
\chi \left( x\right) =\Lambda \int\limits_{0}^{1}l\left( x,\xi \right) \chi
\left( \xi \right) d\xi +q\left( x\right) ,\quad x\in \left[ 0,1\right] .
\label{17}
\end{equation}

Here, $\Lambda =\lambda ^{2}$;%
\[
l\left( x,\xi \right) =\int\limits_{-1}^{0}h\left( x,\zeta \right) H\left(
\zeta ,\xi ,\lambda \right) d\zeta , 
\]%
with $H\left( x,\xi ,\lambda \right) $ being the resolvent of the kernel $%
h\left( x,\xi \right) $ on $x\in \left[ -1,1\right] $;%
\[
q\left( x\right) =-\lambda \int\limits_{-1}^{0}h\left( x,\xi \right) \varphi
_{1}^{\prime }\left( \xi \right) d\xi . 
\]

The solution to Eq. (\ref{17}) has the form%
\[
\chi \left( x\right) =q\left( x\right) +\Lambda \int\limits_{0}^{1}L\left(
x,\xi ,\Lambda \right) q\left( \xi \right) d\xi , 
\]%
where%
\[
L\left( x,\xi ,\Lambda \right) =\frac{1}{1-2\lambda -\Lambda }%
+2\sum_{n=1}^{\infty }\frac{r^{2n}}{1-2\lambda r^{n}-\Lambda r^{2n}}\cos %
\left[ 2n\pi \left( x-\xi \right) \right] 
\]%
is the resolvent of the kernel $l\left( x,\xi \right) $.

The inversion of the operator $I-\lambda B$ in Eq. (\ref{5}) allowed us to
represent the function as a Fourier series in terms of the elements $\left\{
\cos \left( 2n\pi x\right) ,\sin \left( 2n\pi x\right) \right\} $ whose
coefficients were expressed via the data of (\ref{1}) and depended on the
parameter $r$. Note that on the previous stage of the calculations by (\ref%
{12}) the function $\psi _{1}\left( x\right) $ was determined in an
analogous form.

The solution so obtained was restricted only by the case when the function $%
\psi \left( x\right) $ satisfying Eq. (\ref{1}) was harmonic. However, a
passage to the limit $r\rightarrow 1$ easily removes the problems by
transferring the free term of Eq. (\ref{1}), as well as the function $\psi
\left( x\right) $ satisfying this equation, into the space $L_{2}$.

This is the main point in achieving the final objective of the
transformations. Accordingly, condition (\ref{2}) takes the form%
\[
\left\Vert \delta f\right\Vert _{L_{2}\left( 0,1\right) }=0. 
\]

In general, the transformations seem to be rather transparent. Thus, the
determination of the function $\psi \left( x\right) $ is transformed into
the problem (\ref{5})-(\ref{8}). The obtained function $\Psi ^{\prime
}\left( x\right) $ is a part of the solution of Eq. (\ref{8}) that depends
on $-\mu f\left( x\right) $. From (\ref{14}) and (\ref{11}) the function $%
\varphi _{0}^{\prime }\left( x\right) $ is determined in two ways, which is
reflected by relation (\ref{16}). Hence (\ref{13}) turns into Eq. (\ref{17}%
). Finally, in the obtained solution a passage to the limit with respect to $%
r$ was made.

The following interpretation of the algorithm of the reduction is possible.
First, the transformed formulation of the problem is "deformed" by
eliminating the function $\chi \left( x\right) $ from Eq. (\ref{8}). Then
this "deformation" is adaptively smoothed out by means, which is very
important, of the solution of the Fredholm integral equations of the second
kind (\ref{12}), (\ref{17}) and (\ref{6}).

We have discussed the first version of the solution of the problem. The
second version of its solution is also based on the relations given above.
By the use of (\ref{15}), an analogue of Eq. (\ref{17}) for the function $%
\kappa \left( x\right) $ was obtained. Its solution has the form%
\begin{equation}
\kappa \left( x\right) =q^{\prime }\left( x\right) +\Lambda
\int\limits_{-1}^{0}L\left( x,\xi ,\Lambda \right) q^{\prime }\left( \xi
\right) d\xi ,\quad x\in \left[ -1,0\right) ,  \label{18}
\end{equation}%
where%
\begin{equation}
q^{\prime }\left( x\right) =-\lambda \int\limits_{0}^{1}h\left( x,\xi
\right) \psi _{1}\left( \xi \right) d\xi .  \label{19}
\end{equation}

The problem (\ref{1}) is reduced to a numarical realization of,
consecutively, two Fredholm integral equations of the second kind, namely,
Eq. (\ref{12}) and%
\begin{equation}
\psi _{0}\left( x\right) =\mu \int\limits_{0}^{1}K\left( x,\xi \right) \psi
_{0}\left( \xi \right) d\xi +F_{0}\left( x\right) ,\quad x\in \left[ 0,1%
\right] ,  \label{20}
\end{equation}%
where%
\[
F_{0}\left( x\right) =\lambda \int\limits_{0}^{1}H\left( x,\xi ,\lambda
\right) \kappa \left( \xi \right) d\xi . 
\]%
As a result, its solution is sought in the form%
\[
\psi \left( x\right) =\psi _{0}\left( x\right) +\psi _{1}\left( x\right) , 
\]%
see also (\ref{9}).

Note that the kernels of these equations are the same. It is also shown
that, by the use of (\ref{5}), (\ref{11}), (\ref{18}) and (\ref{19}), the
problem is reduced to a numerical realization of a single Fredhom integral
equation of the second kind with respect tot he function $\psi \left(
x\right) $.

In contrast to the previous version of the solution, there is no need here
to evaluate the Fourier coefficients of the functions $k\left( x,\xi \right) 
$ and $f\left( x\right) $ and perfor the summation of infinite series, which
may be regarded as an advantage. At the same time, universal algorithms are
available for the solution of Eqs. (\ref{12}) and (\ref{20}). In general,
the second version of the solution of the problem is more formalized. As an
advantage of the first version, one should point out a possibility of
obtaining the function $\psi \left( x\right) $ in a convenient, as a rule,
form of a Fourier series.

The principal difference between the two versions lies in the way of
satisfying (\ref{2}), or the equation%
\begin{equation}
\lambda \int\limits_{-1}^{0}h\left( x,\xi \right) \varphi \left( \xi \right)
d\xi =\psi \left( x\right) -\lambda \int\limits_{0}^{1}h\left( x,\xi \right)
\psi \left( \xi \right) d\xi ,\quad x\in \left[ 0,1\right] ,  \label{21}
\end{equation}%
where, for $\psi \left( x\right) \in L_{2}\left( 0,1\right) $, the function $%
\varphi \left( x\right) $ can only be a generalized function. In contrast to
the first version, where, for this reason, the transformations were
performed with the function $\psi \left( x\right) $ that was assumed
harmonic up to the final stages, in the second version, it was implied that
Eq. (\ref{21}) was satisfied in the sense of generalized functions.

Specifically, we employed an equation obtained by applying to (\ref{21}) the
operator%
\[
\int\limits_{0}^{1}h\left( x,\xi \right) _{\bullet }d\xi 
\]%
with respect to to the generalized functions\footnote{%
Note the following characteristic feature: the final result of the
transformations appears to be the same as if, without these transformations,
one postulated the applicability of the theory of Fredholm integral
equations of the second kind to the case when the function $\Psi \left(
x\right) $ from Eq. (\ref{5}) is a generalized function.}%
\[
\hat{\psi}\left( x\right) =\lambda \int\limits_{0}^{1}h\left( x,\xi \right)
\psi \left( \xi \right) d\xi ,\quad \hat{\varphi}\left( x\right) =\lambda
\int\limits_{-1}^{0}h\left( x,\xi \right) \varphi \left( \xi \right) d\xi . 
\]

At the same time, exactly condition (\ref{2}) appears to be absolutely
necessary for the realization of both the first and the second versions of
the solution of the problem. Indeed, equation (\ref{21}) that may be called
"free-lance" cardinally changes the problem (\ref{1}) with regard to the
solvability of the Fredholm integral equation of the first kind. With the
help of (\ref{2}), one essetially removes an inherently insurmountable
problem of an objective mismatch of $f\left( x\right) $ and $R\left(
A\right) $, which is the reason for the problem (\ref{1}) being ill-posed.

The liberation of $f\left( x\right) $ from formal association with $R\left(
A\right) $ by means of (\ref{2}) and (\ref{7}), (\ref{8}) simultaneously
results in the fact that the free term of Eq. (\ref{21}), when considered as%
\[
\left( B^{\prime }\varphi \right) \left( x\right) =\tilde{f}\left( x\right)
,\quad x\in \left[ 0,1\right] , 
\]%
where%
\[
B_{\bullet }^{\prime }=\int\limits_{-1}^{0}h\left( x,\xi \right) _{\bullet
}d\xi ,\quad \tilde{f}\left( x\right) =\frac{1}{\lambda }\psi \left(
x\right) -\int\limits_{0}^{1}h\left( x,\xi \right) \psi \left( \xi \right)
d\xi , 
\]%
becomes functional.

As a consequence, the condition $f\left( x\right) \in R\left( A\right) $
(that is actually infeasible) is replaced by the following:%
\begin{equation}
\tilde{f}\left( x\right) \in R\left( B^{\prime }\right) ,  \label{22}
\end{equation}%
which, in fact, is equivalent to%
\[
f\left( x\right) +\left( \delta f\right) /\mu \in R\left( A\right) . 
\]%
Thus, it proves to be possible to go over from a numerical comparison
between $f\left( x\right) $ and $R\left( A\right) $ just to the question of
the existence of the function function $\psi \left( x\right) $ allowing for
the fulfillment of the condition (\ref{22}).

Moreover, given that (\ref{21}) is a Fredholm integral equation of the
second kind with respect to $\psi \left( x\right) $, in the course of
subsequent transformations, there occurs, in a sense, a readdressing of the
status between $\tilde{f}\left( x\right) $ and $R\left( B^{\prime }\right) $%
. Namely, the range of the operator $B^{\prime }$ manifests itself as the
free term, and the problem essentially reduces to finding the function $\psi
\left( x\right) $ from it. Here, the fact that $R\left( A\right) $ is not
closed does not play any role.

Note the following: as a result of (\ref{21}), the determination of the
function $\psi \left( x\right) $ satisfying Eq. (\ref{1}) was carried out,
figuratively, by "materialized pressing" with regard to the validity of
seemingly abstract Banach's theorem on the inverse operator. Specifically,
this is done by the identity operator from $I-\lambda B$ by ensuring the
entering of the above-mentioned function in an explicit form.

It is shown that wide classes of problems of numerical simulation are easily
reduced to Fredholm integral equations of the first kind. After that, the
procedure of correct formulation and of constructive realization, discussed
for a one-dimensional case, is directly extended to them. Therefore, a
differentiation between direct and inverse formulations of problems of
mathematical physics to a certain extent loses significance. We have also
proposed a method of verification of the solvability of problems formulated
in terms of partial differential equations.

In light of the above, we can draw a conclusion that, if the phenomenon
(process) admits an adequate description by methods of numerical
simulations, the restoration of its underlying cause or of different
parameters from an objectively sufficient volume of additional information
does not pose principal difficulties, because the corresponding problems can
be well-posed. From this point of view, an analysis of actually observed
events, including multi-factor social-economic and ecological processes, can
be done with much larger efficiency.

Maybe, it would be reasonable to suggest that, in general, the process of
the understanding of the World is much simpler than a wide audience usually
supposes it to be under the influence of the sphere of applied science that,
at present, armed with means of electronic processing of information,
constitutes, in fact, a natural monopoly with an almost dominant role of
commercial component and, correspondingly, a systematic drive for investment?

Thus, colossal means are invested in problems of the restoration of
dependencies from empirical data and, in particular, in remote probing of
the surface of the Earth by spacecraft. What is actually realized is a
search for minimally and maximally acceptable values of the parameter $%
\alpha $ in the integral equation of the type 
\[
\alpha \psi \left( x\right) +\int\limits_0^1k\left( x,\xi \right) \psi
\left( \xi \right) d\xi =f\left( x\right) ,\quad x\in \left[ 0,1\right] . 
\]

The essence lies in the necessity to establish a balance between
computational and, respectively, financial abilities of the solution of an
almost degenerate algebraic problem and an approximation to the "exact"
formulation that is associated with the factor of incorrectness for $\alpha
=0$.

In this regard, we note that, of course, it would be incorrect to suppose
that problems in science are altogether absent or that one can develop,
irrespective of the circumstances, efficient means to overcome these
problems. However, in our opinion, complications of principal character are
inherent, in the first place, to direct formulations of some problems, that
is, to the construction of mathematical models of insufficiently studied
processes and phenomena.

It is clear that the solution of some classes of inverse problems of
numerical simulation may also pose substantial difficulties, but,
nevertheless, the wide-spread dogma that the procedure of the restoration of
the cause from the consequence is ill-posed, in general, seems to be
manifestly erroneous.

J. Hadamard's statement that the problems that adequately describe real
processes are well-posed is an ingenious idea, whose constructive
development allows one to attain a qualitatively higher level of the
potential of methods of numerical simulations.

March 2, 2005.

\bigskip

E-mail: eperchik@bk.ru

\end{document}